\documentclass{aa}
\usepackage{graphicx}
\usepackage{txfonts}
\usepackage[bf,nooneline]{subfigure}
\usepackage{dcolumn}
	\renewcommand{\thesubtable}{\thetable\alph{subtable}}
	\makeatletter
	\renewcommand{\p@subtable}{}
	\renewcommand{\@thesubtable}{Table~\thesubtable.\hskip\subfiglabelskip}
	\makeatother

	\makeatletter
	\renewcommand{\p@subfigure}{}
	\renewcommand{\@thesubfigure}{}
	\makeatother

	\newcolumntype{+}{D{+}{\,\pm\,}{4,4}}

\begin{document}

   \title{New galactic open cluster candidates from DSS and 2MASS imagery
		\thanks{Tables 2e, 4a and 4b are only available in electronic form at the CDS via anonymous ftp to
		cdsarc.u-strasbg.fr (130.79.125.5) or via cdsweb.u-strasbg.fr/Abstract.html}
   		}

   \author{M. Kronberger\inst{1}
          \and
		  P. Teutsch\inst{1}\fnmsep\inst{2}
		  \and
		  B. Alessi\inst{1}
		  \and
		  M. Steine\inst{1}
		  \and
		  L. Ferrero\inst{1}
		  \and
		  K. Graczewski\inst{1}
		  \and
		  M. Juchert\inst{1}
		  \and
		  D. Patchick\inst{1}
		  \and
		  D. Riddle\inst{1}
		  \and
		  J. Saloranta\inst{1}
		  \and
		  M. Schoenball\inst{1}
		  \and
		  C. Watson\inst{1}
		  }

   \offprints{P. Teutsch, \email{Philipp.Teutsch@uibk.ac.at}}

   \institute{Deepskyhunters Collaboration\\
              \email{deepskyhunters@yahoo.com}
		\and
             Institut f\"ur Astrophysik, Leopold-Franzens-Universit\"at Innsbruck, Austria\\
             \email{Philipp.Teutsch@uibk.ac.at}
             }
   \date{Received 17 August 2005 / Accepted 05 October 2005}

   \abstract{
   An inspection of the DSS and 2MASS images of selected Milky Way regions has led
   to the discovery of 66 stellar groupings
   whose morphologies, color-magnitude diagrams, and stellar density
   distributions suggest that these objects are possible open clusters that
   do not yet appear to be listed in any catalogue. For 24 of
   these groupings, which we consider to be the most likely to be candidates,
   we provide extensive descriptions on the basis of 2MASS photometry and
   their visual impression on DSS and 2MASS. Of these cluster candidates, 9
   have fundamental parameters
   determined by fitting the color-magnitude diagrams with solar metallicity
   Padova isochrones. An additional 10
   cluster candidates have distance moduli and
   reddenings derived from $K$ magnitudes and $(J-K)$ color indices of
   helium-burning red clump stars. As an addendum, we also provide a list of a
   number of apparently unknown galactic and extragalactic objects that
   were also discovered during the survey.
   }
	\titlerunning{New galactic open cluster candidates}
   \maketitle


\section{Introduction}

   In the most up-to-date catalogues of galactic star clusters, 
   Dias~et~al.~(\cite{dias}) (hereafter DAML02) and
   Bica~et~al.~(\cite{bica03a}) (hereafter BDB03), a total of 1875 open
   clusters, open cluster candidates, and stellar groups visible in the visual
   or infrared spectral bands are listed. Despite the large number of
   objects included in these catalogues, they are apparently still far from
   complete, illustrated by the fact that several hundred new
   clusters and cluster candidates have been added to the list of known
   stellar groupings since these catalogues were published. 
   
   This large number of new discoveries is mostly the result of
   the release of the Two Micron All Sky Survey
   (Skrutskie~et~al.~\cite{skrutskie}, hereafter 2MASS). The images and
   photometric data obtained from this survey in the $J$, $H$, and
   $K_\mathrm{S}$ spectral bands have proved to be fertile ground in the
   search for new open clusters and have offered new possibilities for the
   analysis of newly detected and already known cluster candidates. 
   Important to mention in this respect are the following publications: 
   Bica~et~al.~(\cite{bica03c}), who reports 3 new open cluster candidates in
   the Cygnus X region, detected by visually inspecting 2MASS and
   Digitized Sky Survey (hereafter DSS) images; Bica~et~al.~(\cite{bica03b})
   and Dutra~et~al.~(\cite{dutra03}), who carried out a survey of infrared star
   clusters and stellar groups around the central positions of optical and
   radio nebulae and found 167 and 179 previously unknown objects,
   respectively; and Ivanov~et~al.~(\cite{ivanov}) and 
   Borissova~et~al.~(\cite{borissova}), who used a search algorithm based on
   finding peaks in the apparent stellar surface density, leading to the
   detection of an additional 14 clusters via 2MASS. 
   
   Other important publications include: 
   Alessi~et~al.~(\cite{alessi}), who discovered 11
   previously unknown cluster candidates in the solar neighborhood using
   astrometric and photometric data provided by
   the Tycho-2 catalogue (H{\o}g~et~al.~\cite{hog}); 
   Bica~et~al.~(\cite{bica04}), who reported 3 new open clusters found on maps
   obtained from the Guide Star Catalogue and images from the DSS; and Drake~(\cite{drake}), who discovered 8
   clusters by searching for density fluctuations within the USNO A2.0 catalogue.

   The recent discovery of new galactic open clusters in optical
   wavelengths 
   by Bica~et~al.~(\cite{bica03c} and \cite{bica04}) and Drake~(\cite{drake}) 
   is remarkable in that as it illustrates
   that it is still possible to detect unknown clusters on the DSS,
   even though the systematic searches for faint clusters on the
   Palomar and ESO/SERC Schmidt plates -- which led to the Berkeley open
   cluster list (Setteducati~\&~Weaver~\cite{setteducati}) and to the
   individual lists which generated the ESO catalogue
   (Lauberts~\cite{lauberts}) -- were carried out very carefully. This is also
   underlined by Pfleiderer~et~al.~(\cite{pfleiderer}) and
   Saurer~et~al.~(\cite{saurer}), who reported the discovery of a number of
   faint clusters apparently not detected during these surveys.

   As a first result of a systematic survey of selected Milky Way fields based on the
   visual inspection of DSS and 2MASS images, we announce in this paper the discovery of 66 previously
   unknown (or, at least, unlisted) clusters and cluster candidates.

   A discussion of the search technique and a description of the methods used
   to analyze the objects is given in the following section. Section~3 is dedicated to the presentation and
   discussion of the most interesting cluster
   candidates, and it provides a list of basic and fundamental parameters
   derived for these objects. Finally, in 
   Section~4, concluding remarks are given.


\section{Search strategy and analysis}

   A stellar field in which an open cluster is present is characterized by a number of specific features which reveal
   its existence:

   \begin{enumerate}
      \item The presence of an open cluster in a star field leads to increased stellar density when compared with
	  the neighboring fields.
   
      \item The cluster members show similar radial velocities and proper motions.
   
      \item Due to their common origin, the member stars of an open cluster have similar ages and chemical
	  compositions.
   \end{enumerate}

   Starting with these considerations, we carried out a systematic survey of several Milky Way fields near the
   galactic plane using red, blue, and infrared First and Second Generation DSS images extracted from the ESO Online
   Digitized Sky Survey facility\footnote{http://archive.eso.org/dss/dss}. As target fields for this survey, we
   favored Milky Way regions close to the galactic equator and characterized by a distinct lack of catalogued open
   clusters, as well as fields with evidence of active star formation. The DSS images of these regions were then
   examined in order to search for fields containing aggregates with a morphology similar to open clusters, i.e.
   fields with a stellar density significantly higher than the surrounding ones. A number of reflection and emission
   nebulae visible on the DSS images were also investigated, and to examine whether they contained any embedded
   clusters or
   stellar groups, we extracted additional $J$, $H$, and $K_\mathrm{S}$ images of these objects from the 2MASS
   facility\footnote{http://irsa.ipac.caltech.edu}. Finally, the exact position and apparent angular diameter of each
   cluster candidate were measured using the Aladin Interactive Sky Atlas provided by the CDS,
   Strasbourg\footnote{http://aladin.u-strasbg.fr/aladin.gml}.


   \subsection{Color-magnitude diagrams (CMDs)}

      As an increased stellar density alone is not definite proof that the observed stellar field contains an open
      cluster ( as it can also be caused by chance agglomerations of stars and absorption holes), we obtained a
	  photometric analysis of
      the candidate fields using data from the 2MASS Point Source Catalogue (Cutri~et~al.~\cite{cutri}) to get further
      clarity on their nature and their properties.
   
      As a first step, we generated preliminary $[J, (J-H)]$ and $[K_\mathrm{S}, (J-K_\mathrm{S})]$ CMDs for every
      candidate. To minimize possible contamination, we selected only those sources with accurate photometric
	  measurements.
      This selection was done by checking the read flag ($rd\_flg$) of the sources provided in every band, which 
      indicates how the magnitude of the source was determined and therefore contains information on the quality of
	  the
      photometric measurement. Based on this criteria, we restricted the data base to: sources with $rd\_flg = 2$
	  (point
      spread-function fitting) or $rd\_flg = 1$ (aperture photometry); sources with a read flag of 0, 4, 6, or 9 in
	  any
	  of the specific bands, indicating either that the source could not be detected in this band 
	  or that there was a bad photometric measurement,
	  were omitted from the CMDs. In addition, we examined all other flags of the selected sources given in the 2MASS
	  Point Source Catalogue, such as the quality flag ($ph\_qual$) and the extended source contamination flag
	  ($gal\_contam$), to minimize the influence of further sources of error.
   
      The CMDs of the cluster fields were then compared with the CMDs of 4 control fields with the same size as the
      cluster fields and situated N, S, E, and W of the cluster candidate; the center-to-center distance between
      comparison field and cluster field was, in each case, taken as 3 times the radius of the cluster candidate.


   \subsection{Fundamental parameters}

      The majority of the clusters and cluster candidates presented in this publication exhibited certain features in
      their CMDs which enabled us to determine fundamental parameters of these objects.

      For those cluster candidates whose CMDs revealed a well-defined cluster sequence, we determined the distances,
      reddenings, and ages by fitting the cluster sequence with solar metallicity Padova isochrones from
      Girardi~et~al.~(\cite{girardi02}) calculated for the $J$, $H$, and $K_\mathrm{S}$ spectral
      bands\footnote{http://pleiadi.pd.astro.it/isoc\_photosys.01/isoc\_photosys.01.html}. To correct for the
	  influence of
      interstellar reddening, we assumed a value of $R_\mathrm{V} = 3.09$ for the total-to-selective absorption and
	  used
	  the relations $A_\mathrm{J} = 0.282 \cdot A_\mathrm{V}$ and $E(J-H) = 0.33 \cdot E(B-V)$
	  (Rieke~\&~Lebowski~\cite{rieke}).

      In a number of cases, the CMD exhibited a structure resembling a helium-burning red clump
      (hereafter RC). This RC -- a typical feature of intermediate-aged and old open clusters -- is a valuable
	  standard
	  candle and allows a determination of the clusters' distance and reddening. In case the CMD of such a cluster 
	  did not allow obtaining an isochrone fit, we applied
      the approach of Grocholski~\&~Sarajedini~(\cite{grocholski}) (hereafter GS02), who used the mean values
      $m_\mathrm{K,RC}$ and $(J-K)_\mathrm{RC}$ of the stars composing the RC to derive the distance and reddening
      parameters. These values were computed by determining the corresponding values $m_\mathrm{K_S,RC}$ and
      $(J-K_S)_\mathrm{RC}$ of the RC stars and subsequent transformation of the 2MASS $K_\mathrm{S}$ to the standard
      Johnson $K$ band magnitudes using the equations quoted in GS02 (see Table~\ref{TabRCvals}).

      For the absolute magnitude $M_\mathrm{K,RC}$ of the RC, we adopted the value of $M_\mathrm{K,RC} = -1.61 \pm
	  0.03$
      (Alves~\cite{alves}) used in GS02. As color index $(J-K)_\mathrm{0,RC}$, we assumed
	  $(J-K)_\mathrm{0,RC}
      = 0.62 \,\pm\, 0.05$, which was derived from solar metallicity Padova isochrones with $8.5 \leq \log{t} \leq
	  10.0$; this value is close to the mean value of $0.61 \,\pm\, 0.06$ for the clusters quoted in GS02.

      To ensure that the observed feature is indeed an RC and not the brightest part of a highly reddened MS, we
      also computed the ratio $(J-K)_\mathrm{RC}/(J-H)_\mathrm{RC}$ (Table~\ref{TabRCvals}). Assuming moderate
	  reddening $(0 < E(B-V) < 2)$, this ratio has a value of $1.2$ to $1.4$ for RC stars and a value close to
	  $1.6$ for stars of early
	  spectral type; therefore, it provides information concerning the stellar composition of the feature.
   
      It should be noted, however, that the parameters that were derived from $m_\mathrm{K,RC}$ and
	  $(J-K)_\mathrm{RC}$
      should only be taken as estimates, as both $M_\mathrm{K,RC}$ and $(J-K)_\mathrm{0,RC}$ show a dependence on age
	  and
      metallicity (GS02; Girardi~\&~Salaris~\cite{girardi01}); differences from the actual values of up to 30\% can be
      expected. To determine more precise fundamental parameters, deeper studies of the cluster candidates are
	  therefore required.


\setcounter{table}{0}
	
   \begin{table}
   \begin{center}
        \caption[]{mean values of $m_\mathrm{K,RC}$, $(J-K)_\mathrm{RC}$ and $(J-K)_\mathrm{RC}/(J-H)_\mathrm{RC}$ for
					the clusters with RCs.}
         \label{TabRCvals}
		 {\scriptsize
			\begin{tabular}{l+++}
			\hline
            \noalign{\smallskip}
			DSH~ID & \multicolumn{1}{c}{$m_\mathrm{K,RC}$} & \multicolumn{1}{c}{$(J-K)_\mathrm{RC}$} 
			& \multicolumn{1}{c}{$(J-K)_\mathrm{RC}/(J-H)_\mathrm{RC}$}\\
			\noalign{\smallskip}
            \hline
            \noalign{\smallskip}
			\object{DSH~J0347.3+5354	} & 11.75 + 0.06	 &	1.50 + 0.03	 & 1.44 + 0.03\\
			\object{DSH~J0920.5-5251	} & 12.95 + 0.12	 &	1.12 + 0.01	 & 1.44 + 0.02\\
			\object{DSH~J1054.2-6144	} & 13.92 + 0.08	 &	1.09 + 0.03	 & 1.36 + 0.04\\
			\object{DSH~J1323.6-6340	} & 12.85 + 0.05	 &	1.11 + 0.01	 & 1.38 + 0.02\\
			\object{DSH~J1906.8+0935	} & 12.02 + 0.05	 &	1.66 + 0.02	 & 1.45 + 0.02\\
			\object{DSH~J1920.8+1540	} & 12.09 + 0.05	 &	1.72 + 0.01	 & 1.49 + 0.01\\
			\object{DSH~J1940.1+2615	} & 14.06 + 0.07	 &	1.06 + 0.03	 & 1.37 + 0.03\\
			\object{DSH~J1942.7+2951	} & 13.13 + 0.03	 &	0.90 + 0.03	 & 1.37 + 0.04\\
			\object{DSH~J1945.1+2809	} & 13.07 + 0.09	 &	0.91 + 0.01	 & 1.32 + 0.03\\
			\object{DSH~J2126.1+5331	} & 12.77 + 0.06	 &	1.40 + 0.02	 & 1.39 + 0.02\\
			\noalign{\smallskip}
            \hline
        	\end{tabular}
			}
   \end{center}
   \end{table}


	  
   \subsection{Radial density profiles (RDPs)}

      In addition to the CMDs, we created RDPs from 2MASS data by counting stars in concentric rings and dividing
	  the counts by the area of each ring. To avoid misleading results due to spatial variations in the number of
	  faint stars, we applied a cut-off at $H = 15.5$ for optical clusters and at $K_\mathrm{S} = 14.5$ for infrared
	  clusters.
	  These RDPs were used to determine the diameter of each candidate with good precision. They also provided
	  a check on whether the fields of the potential clusters indeed exhibited a significant excess in the number of
	  stars compared to the stellar background density.

	  
\section{Results and discussion}

   \subsection{Open clusters and promising open cluster candidates}

      In Tables~\ref{TabIso}~to~\ref{TabNo}, we provide positions, angular diameters, reddenings, distances,
	  and ages of 24 stellar aggregates visible in optical or infrared wavelengths that we consider as being open
	  clusters or, at least, promising open
      cluster candidates according to the results of the analytical procedures described in Sect. 2. Cluster
	  candidates
      whose CMDs allowed a fit to solar metallicity Padova isochrones are listed in Table~\ref{TabIso}, while
	  candidates with RCs in
      their CMDs are given in Table~\ref{TabRC}. Finally, Table~\ref{TabNo} contains those candidates for which
	  the determination of
      fundamental parameters was found to be impossible. Every cluster candidate has a designation in the 
	  IAU-recognized
      format DSH~Jhhmm.m$\pm$ddmm and its discoverer-ID assigned. Unless otherwise noted, all candidates were
	  discovered
      between June 2003 and December 2004; the discovery was in each case announced to the DSH
      newsgroup\footnote{http://groups.yahoo.com/group/deepskyhunters}. One of the described cluster candidates
	  (\object{DSH~J0718.1-1734} = \object{Teutsch~49}) turned out to be
	  independently discovered by Drake~(\cite{drake}); it is listed therein as \object{DC~2}. 
	   
      Figures~\ref{fig1}~to~\ref{fig3} show the DSS images of all optical cluster candidates extracted from the
	  CADC facility. The RDPs of the clusters are given in Figures~\ref{fig4}~to~\ref{fig6}. 
      Finally, in Figures~\ref{fig7}~to~\ref{fig9}, the results of the photometric analysis of the clusters
	  are presented. In each case, the
      radius of the extraction area equals the visually determined cluster radius $R_\mathrm{vis}$. To increase the
      contrast between cluster features and background scatter in the CMDs, we used different grey-tones, with
	  lighter tones indicating a larger distance from the cluster center.

   
   \subsection{Extended notes}

      \subsubsection{Isochrone-fitted cluster candidates}
		
	
	\setcounter{table}{2}
	\setcounter{subtable}{0}

	\begin{table*}
	\begin{center}
		\subtable[][Positions, angular diameters, reddenings, distances, and ages of the open cluster candidates with
 					CMDs fitted with Padova solar metallicity isochrones.]{
         \label{TabIso}
        	\begin{tabular}{llllc++++}
			\noalign{\smallskip}
			\noalign{\smallskip}
			\noalign{\smallskip}
            \hline
            \noalign{\smallskip}
            DSH~ID & discoverer~ID & $RA~J2000$ & $Dec~J2000$ & $D_\mathrm{vis}$ &
			\multicolumn{1}{c}{$E(B-V)$} & \multicolumn{1}{c}{$(m-M)_0$} & \multicolumn{1}{c}{distance} 
			& \multicolumn{1}{c}{age}\\
			&&&& arcmin &&& \multicolumn{1}{c}{kpc} & \multicolumn{1}{c}{$\log{t}$}\\
            \noalign{\smallskip}
            \hline
            \noalign{\smallskip}
			\object{DSH~J0355.3+5823} & \object{Juchert~9} & $03~55~21.0$ & $+58~23~30$ & $3.0$ & 0.79 + 0.06
			& 13.20 + 0.30 & 4.4 + 0.6 & \multicolumn{1}{c}{$< 7.6$} \\
			\object{DSH~J0528.3+3446} & \object{Kronberger~1} & $05~28~21.0$ & $+34~46~30$ & $1.6$ & 0.52 + 0.06
			& 11.40 + 0.20 & 1.9 + 0.2 & \multicolumn{1}{c}{$< 7.5$} \\			
			\object{DSH~J0553.8+2649} & \object{Teutsch~51} & $05~53~51.9$ & $+26~49~47$ & $3.6$ & 1.06 + 0.09
			& 12.60 + 0.20 & 3.3 + 0.3 & 8.90 + 0.10 \\
			\object{DSH~J1453.4-6028} & \object{Teutsch~80} & $14~53~25.6$ & $-60~28~57$ & $3.4$ & 1.61 + 0.06
			& 11.95 + 0.60 & 2.5 + 0.6 & 8.10 + 0.15 \\
			\object{DSH~J1704.3-4204} & \object{Teutsch~84} & $17~04~20.1$ & $-42~04~24$ & $4.0$ & 1.12 + 0.15
			& 11.75 + 0.25 & 2.2 + 0.3 & 9.00 + 0.10 \\
			\object{DSH~J1930.2+1832} & \object{Teutsch~42} & $19~30~13.1$ & $+18~32~09$ & $1.2$ & 2.55 + 0.09
			& 11.00 + 0.20 & 1.6 + 0.2 & 7.50 + 0.25 \\		
			\object{DSH~J1933.9+1831} & \object{Kronberger~79} & $19~33~55.0$ & $+18~31~12$ & $2.1$ 
			& 1.52 + 0.06^\mathrm{a} & 12.80 + 0.30^\mathrm{a} & 3.6 + 0.4^\mathrm{a} 
			& 7.40 + 0.30^\mathrm{a} \\
			&&&&& 1.30 + 0.06^\mathrm{a} & 12.20 + 0.30^\mathrm{a} & 2.7 + 0.4^\mathrm{a} 
			& 8.35 + 0.10^\mathrm{a} \\
			\object{DSH~J2002.3+3518} & \object{ADS~13292~Cl} & $20~02~23.3$ & $+35~18~41$ & $1.1$ & 0.58 + 0.03
			& 11.00 + 0.20 & 1.6 + 0.2 & \multicolumn{1}{c}{$< 7.0$} \\
			\object{DSH~J2027.7+3604} & \object{Teutsch~30} & $20~27~43.0$ & $+36~04~32$ & $3.2$ & 1.24 + 0.03
			& 11.05 + 0.15 & 1.6 + 0.1 & \multicolumn{1}{c}{$< 6.9$} \\
			\noalign{\smallskip}
            \hline
         \end{tabular}
		}
		\begin{list}{}{}
		\item[Note:] $^\mathrm{a}$ see Sect. 3.2.1.
		\end{list}
		\end{center}

	\end{table*}
	  

         \object{DSH~J0355.3+5823} = \object{Juchert~9}
		    is a small and slightly irregularly shaped cluster with 15 brighter stars and an unknown number of fainter
			members. It is well separated from the Milky Way background (Fig.~\ref{fig1}). The center of the
            cluster is dominated by a pair of equally bright stars, which is listed in the Washington Catalogue of
			Double Stars (Worley~\&~Douglass~\cite{worley}; hereafter WDS) as \object{Stein~2015}. As can be seen from
			the CMD shown in Fig.~\ref{fig7}, the cluster stars align along a well-defined MS. A giant branch is not
			evident, implying that the cluster is a rather young object. As an upper limit, we found 
			$\log{t} = 7.6$.\\

         \object{DSH~J0528.3+3446} = \object{Kronberger~1}
		 	is an irregularly shaped, N-S elongated cluster situated $22\arcmin$ N of the HII region \object{IC~417}
			and $4.7\arcmin$ N of the bright K0 star \object{HD~35\,742}. The brightest cluster member,
			\object{LS~+34~33} (Reed~\cite{reed}), is a luminous star of spectral type B, which is also listed as a
			double star with the designation \object{Scheiner~272} in the WDS. The CMD of \object{Kronberger~1}
			(Fig.~\ref{fig7}) exhibits a sparsely populated, but fairly well-defined cluster
			sequence that lacks any evolutionary features and shows a gap between $J = 11.5$ and $J = 13$. The solar
			metallicity isochrone fit of this feature gives a distance of $d = 1.9  \,\pm\, 0.2 ~\mathrm{kpc}$ 
			and a reddening of $E(B-V) = 0.52 \,\pm\, 0.06$. As in the case of \object{Juchert~9}, an exact age cannot
			be determined; the upper limit appears to be $\log{t} = 7.5$.\\

   		 The three cluster candidates
         \object{DSH~J0553.8+2649} = \object{Teutsch~51}, 
		 \object{DSH~J1453.4-6028} = \object{Teutsch~80} and
		 \object{DSH~J1704.3-4204} = \object{Teutsch~84}
            are cluster candidates that are situated in Milky Way parts that appear to be heavily obscured by
			interstellar 
			dust; it is therefore possible that these objects are actually regions of lower extinction and not
			physical groups of stars. Nonetheless, in all cases the CMD (Fig.~\ref{fig7}) supports the idea of a real
			clustering, as they all reveal well-defined structures that resemble cluster sequences and are not visible
			in the CMDs of the comparison fields. \\

         \object{DSH~J1930.2+1832} = \object{Teutsch~42} 
		    is a compact cluster candidate that is located $17\arcmin$ N of the HII region \object{S~82}
			(Sharpless \cite{sharpless}) in Sagitta. It appears as a tight group dominated by a triangle of brighter
			stars (Fig.~\ref{fig1}). As its color excess of $E(B-V) = 2.55 \pm 0.09$ implies, it is a highly reddened
			object. One of the remarkable things about this cluster is the presence of the X-ray source
			\object{2E~1928.0+1825} = \object{2RXP~J193013.8+183216} (Harris~et~al.~\cite{harris}; ROSAT~\cite{rosat})
			in its field. Whether there is indeed a physical connection between this object and the cluster requires
			deeper study and will not be discussed in this publication; we note, however, the short distance from the
			cluster center of $12 \,\pm\, 10$ arcseconds (ROSAT~\cite{rosat}).\\

         \object{DSH~J1933.9+1831} = \object{Kronberger~79}
            is a compact object of irregular morphology that contrasts very well with the Milky Way background 
			(Fig.~\ref{fig1}). This is also indicated by its RDP (Fig.~\ref{fig4}). With a sharp central peak and a
			steady decline with increasing distance from the center, it has a shape that is typical 
			of open clusters with a strong central condensation. Photometric analysis of the cluster field reveals a
			well defined cluster sequence with a MS, a possible
			turn-off at about $J = 12$ and $(J-H) = 0.4$, and a few possible giant stars positioned around $J = 11.2$
			and $(J-H) = 0.9$ (Fig.~\ref{fig7}). Assuming that these stars are indeed members of
			\object{Kronberger~79}, distance, age, and color
			excess are found to be $d = 2.7 \,\pm\, 0.4 ~\mathrm{kpc}$, $E(B-V) = 1.30 \,\pm\, 0.06$, and $\log{t} =
			8.35 \,\pm\, 0.1$. Another solution, excluding the giant stars, yields 
			$d = 3.6 \,\pm\, 0.4 ~\mathrm{kpc}$, $E(B-V) = 1.52 \,\pm\, 0.06$, and $\log{t} = 7.4 \,\pm\, 0.3$.\\

         \object{DSH~J2002.3+3518} = \object{ADS~13292} Cl
		 	is a very compact group with a morphology similar to the Trapezium Cluster in the Orion Nebula 
			(Fig.~\ref{fig1}). Although it was noted as a multiple star as early as 1894 by
			Thomas Espin, who catalogued it as \object{Espin~202} (WDS), and listed as potential trapezium system \#67
			by Ambartsumian~(\cite{ambartsumian}), it was not recognized as a possible cluster until recently when it
			was independently discovered by B.~Skiff and P.~Teutsch (Archinal~\&~Hynes~\cite{archinal}). A recent
			study of trapezium systems by Abt~\&~Corbally~(\cite{abt}) (hereafter AC00) revealed the group indeed to
			be a physical system and identified 6 probable cluster members, all of them within
			$0.55\arcmin$ from the cluster center. However, as AC00 considered just the brightest stars of the system
			and ignored the fainter group members the actual number of cluster members may therefore be much higher.
			From a solar metallicity isochrone fit of the CMD (Fig.~\ref{fig7}), we found a distance modulus $(m-M)_0$
			of $11.0 \,\pm\, 0.2$ and a color
			excess $E(B-V) = 0.58 \,\pm\, 0.06$. This is in good agreement with the
			values quoted in AC00, which are $(m-M)_0 = 10.9$ and $E(B-V) = 0.61$, respectively. The cluster appears
			to be a very young object, with an age of probably less than $10~\mathrm{Myr}$.\\


\setcounter{table}{2}
\setcounter{subtable}{1}

   \begin{table*}
	\begin{center}	
		\subtable[][Positions, angular diameters, reddenings and distances of the open cluster candidates with RCs.]{
         \label{TabRC}
 			\begin{tabular}{llllc+++}
			\noalign{\smallskip}
			\noalign{\smallskip}
			\noalign{\smallskip}
            \hline
            \noalign{\smallskip}
            DSH~ID & discoverer~ID & $RA~J2000$ & $Dec~J2000$ & $D_\mathrm{vis}$ &
			\multicolumn{1}{c}{$E(B-V)$} & \multicolumn{1}{c}{$(m-M)_0$} & \multicolumn{1}{c}{distance}\\
			&&&& arcmin &&& \multicolumn{1}{c}{kpc}\\
            \noalign{\smallskip}
            \hline
            \noalign{\smallskip}
			\object{DSH~J0347.3+5354} & \object{Juchert~11} & $03~47~18.0$ & $+53~54~35$ & $5.0$ & 1.68 + 0.15 
			& 12.75 + 0.15 & 3.6 + 0.2 \\				
			\object{DSH~J0920.5-5251} & \object{Teutsch~48} & $09~20~31.8$ & $-52~51~06$ & $2.2$ & 0.96 + 0.11
			& 14.25 + 0.20 & 7.0 + 0.6 \\	
			\object{DSH~J1054.2-6144} & \object{Kronberger~39} & $10~54~13.6$ & $-61~44~16$ & $0.8$ & 0.90 + 0.15
			& 15.20 + 0.15 & 11.1 + 0.8\\	
			\object{DSH~J1323.6-6340} & \object{Teutsch~79} & $13~23~38.8$ & $-63~40~10$ & $2.0$ & 0.95 + 0.12
			& 14.15 + 0.10 & 6.7 + 0.4\\			
			\object{DSH~J1906.8+0935} & \object{Alessi~56} & $19~06~52.1$ & $+09~35~00$ & $2.2$ & 2.00 + 0.13
			& 12.95 + 0.10 & 3.9 + 0.2\\				
			\object{DSH~J1920.8+1540} & \object{Alessi~57} & $19~20~53.8$ & $+15~40~36$ & $2.5$ & 2.11 + 0.11
			& 12.95 + 0.10 & 3.9 + 0.2\\	
			\object{DSH~J1940.1+2615} & \object{Kronberger~31} & $19~40~11.0$ & $+26~15~48$ & $1.3$ & 0.84 + 0.15
			& 15.40 + 0.15 & 11.9 + 0.8\\				
			\object{DSH~J1942.7+2951} & \object{Teutsch~43} & $19~42~46.9$ & $+29~51~20$ & $1.3$ & 0.54 + 0.15
			& 14.55 + 0.10 & 8.1 + 0.4\\		
			\object{DSH~J1945.1+2809} & \object{Kronberger~4} & $19~45~11.4$ & $+28~09~40$ & $1.3$ & 0.56 + 0.12
			& 14.50 + 0.15 & 7.9 + 0.6\\
			\object{DSH~J2126.1+5331} & \object{Kronberger~81} & $21~26~08.9$ & $+53~31~58$ & $3.5$ & 1.50 + 0.13
			& 13.85 + 0.15 & 5.9 + 0.4\\
			\noalign{\smallskip}
            \hline
			\end{tabular}
		}
	\end{center}
	\end{table*}


	     \object{DSH~J2027.7+3604} = \object{Teutsch~30}
   			is a fairly poor and loose group of stars associated with the faint HII region \object{LBN~198}
			(Lynds~\cite{lynds65}) and located immediately E of a long, thin streak of interstellar matter that runs
			in N-S direction (Fig.~\ref{fig1}). The stellar aggregate is positioned around the B star
			\object{BD+35~4126}, which is also
			listed in the WDS as a quadruple star with the designation \object{Espin~2193}. As its position within a
			HII region suggests, this cluster appears to be a very young object, with $\log{t} \leq 6.9$.

      \subsubsection{cluster candidates with RCs}

   		 \object{DSH~J1323.6-6340} = \object{Teutsch~79} 
		 	is a faint cloud of stars not well separated from the background Milky Way and with an obvious dark
			spot in the immediate center (Fig.~\ref{fig2}). Its CMD (Fig.~\ref{fig8}) reveals a distinct RC with
			similarities to the structure that is observed in
			the CMD of the cluster \object{C~1426-607} = \object{Pismis~19} (DAML02) (Fig.~\ref{fig10}). 
			From the position of
			the RC in the CMD, distance and reddening of this cluster candidate are found to be 
			$d = 6.7 \,\pm\, 0.4 ~\mathrm{kpc}$ and $E(B-V) = 0.95 \,\pm\, 0.12$.\\

         \object{DSH~J2126.1+5331} = \object{Kronberger~81}
            is a rich cluster of stars, appearing as a dense cloud of extremely faint stars on red and infrared 
			DSS images (Fig.~\ref{fig2}); on blue DSS images, the cluster is invisible, indicating strong interstellar
			absorption in the cluster field. As in the case of \object{Teutsch~42}, an X-ray source is located in the
			immediate vicinity of the
			cluster, about $2.0\arcmin$ distant from the cluster center and less than $10\arcsec$ from the star
			\object{TYC~3966-892-1}; its designation in the ROSAT all-sky Bright Source Catalogue
			(Voges~et~al.~\cite{voges}) is \object{1RXS~J212622.1+533224}.
   			The CMD of the cluster (Fig.~\ref{fig8}) differs significantly from the CMDs of the comparison fields, as
			it completely lacks a distinct population of blue stars but instead reveals a conspicuous feature on the
			red side of the diagram at $(J-H) \approx 1$, which is, in all probability, the well-populated giant
			branch of the cluster, with the RC positioned at $J \approx 14$. This is especially distinct in the
			innermost $0.55\arcmin$ of the cluster, where almost all stars detected by 2MASS align along this feature.
			All this is a strong argument in favor of the reality
			of this object, as the presence of a dust hole, which might have a similar optical appearance, would 
			result in a shift of the stellar distribution to the blue side of the CMD. Moreover, the
			$(J-K)_\mathrm{RC}/(J-H)_\mathrm{RC}$ value
			of $1.39 \,\pm\, 0.02$ that is determined for this feature (see Table~\ref{TabRCvals}) indicates that it
			is indeed composed of giant
			stars and not a population of reddened main-sequence stars. As a comparison with the cluster
			\object{NGC~2158} (whose giant branch is of comparable appearance) shows, the lack of an MS in the CMD of
			\object{Kronberger~81} is
			probably due to the fact that the MS stars are too faint to be detected (Fig.~\ref{fig11}). Assuming that
			both objects have
			similar properties in respect to age and stellar composition, the brightness of the turn-off can be
			estimated as $\approx 16$ in $J$ band and would therefore be at the limit of detection for 2MASS.
			As follows from $m_\mathrm{K,RC}$ and $(J-K)_\mathrm{RC}$, distance and reddening of
			\object{Kronberger~81} are found to be
			$d = 5.9 \,\pm\, 0.4 ~\mathrm{kpc}$ and $E(B-V) = 1.50 \,\pm\, 0.13$; like several other clusters
			presented in this paper, it appears to be a member of the Perseus Arm of the Galaxy.\\

         The cluster candidates 
		 \object{DSH~J0347.3+5354} = \object{Juchert~11}, 
		 \object{DSH~J0920.5-5251} = \object{Teutsch~48}, 
		 \object{DSH~J1054.2-6144} = \object{Kronberger~39}, 
		 \object{DSH~J1906.8+0935} = \object{Alessi~56}, 
		 \object{DSH~J1920.8+1540} = \object{Alessi~57}, 
		 \object{DSH~J1940.1+2615} = \object{Kronberger~31}, 
		 \object{DSH~J1942.7+2951} = \object{Teutsch~43}, and 
		 \object{DSH~J1945.1+2809} = \object{Kronberger~4} 
		 	all appear as faint but distinct concentrations of stars (Fig.~\ref{fig2}), and all
			have similar CMDs with distinct clumps of red giant stars present that are not evident in the CMDs of
			the comparison fields (Fig.~\ref{fig8}). The RDPs (Fig.~\ref{fig5}) are in all cases consistent with the
			assumption of an open cluster in the field, but most of them lack the very
			distinct over-density of stars evident in the RDPs of the majority of the cluster and
			cluster candidates presented here due to the faintness of the majority of stars composing
			these aggregates, which puts them below the cut-off we defined in Sect. 2, . The
			distances $d$ of these clusters vary between $3.6 \,\pm\, 0.2 ~\mathrm{kpc}$ for \object{Juchert~11} and
			$11.9 \,\pm\, 0.8 ~\mathrm{kpc}$ for \object{Kronberger~31} while reddenings $E(B-V)$ are between 
			$0.54 \,\pm\, 0.15$ (\object{Teutsch~43}) and $2.11 \,\pm\, 0.11$ (\object{Alessi~57}).


\setcounter{table}{2}
\setcounter{subtable}{2}

   \begin{table*}
	\begin{center}
		\subtable[][Positions and angular diameters of open cluster candidates without fundamental parameters
	  				determined.]{
         \label{TabNo}
        	\begin{tabular}{llllc}
			\noalign{\smallskip}
			\noalign{\smallskip}
			\noalign{\smallskip}
            \hline
            \noalign{\smallskip}
            DSH~ID & discoverer~ID & $RA~J2000$ & $Dec~J2000$ & $D_\mathrm{vis}$\\
			&&&& arcmin\\
            \noalign{\smallskip}
            \hline
            \noalign{\smallskip}
			\object{DSH~J0718.4-1734} & \object{Teutsch~49} & $07~18~26.8$ & $-17~34~40$ & $2.4$\\							
			\object{DSH~J1052.8-5927} & \object{Teutsch~31} & $10~52~50.1$ & $-59~27~53$ & $1.0$\\							
			\object{DSH~J1822.6-1443} & \object{Kronberger~25} & $18~22~39.9$ & $-14~43~41$ & $1.0$\\							
			\object{DSH~J2003.1+3158} & \object{Kronberger~54} & $20~03~07.9$ & $+31~58~01$ & $0.9$\\						
			\object{DSH~J2111.8+5222} & \object{Kronberger~80} & $21~11~50.5$ & $+52~22~48$ & $0.9$\\	
			\noalign{\smallskip}
            \hline
         \end{tabular}
		} 
		\end{center}
   		
   \end{table*}
   

	\subsubsection{Cluster candidates without fundamental parameters}
	
		\object{DSH~J0718.1-1734} = \object{Teutsch~49}
			is a fairly rich cluster of faint stars that is separated very well from the Milky Way background
			(Fig.~\ref{fig3}). It is identical with the object \object{DC~2} in Drake~(\cite{drake}). The
			 visually
			estimated diameter of this cluster is $2.4\arcmin$, but the RDP (Fig.~\ref{fig6}) suggests that outliners
			extend as far as $2.2\arcmin$ from the cluster center. A similar cluster radius ($2.4\arcmin$) is
			found by Drake~(\cite{drake}), who furthermore notes a King model core radius of $0.55\arcmin$ for
			 this
			cluster. The cluster CMD (Fig.~\ref{fig9}) reveals a well-defined MS at $(J-H) \approx 0.6$, which implies
			strong reddening $E(B-V)$ in the range of $1.5 - 2.0$. A giant branch is not evident in the CMD,
			indicating that the stellar aggregate is rather young.\\

   		\object{DSH~J1052.8-5927} = \object{Teutsch~31}
   			is a very compressed star-cluster candidate with a diameter of 1.2 arcminutes in the vicinity of the
			\object{$\eta$~Carinae} Nebula \object{NGC~3372}. The CMD of this candidate (Fig.~\ref{fig9}) contrasts
			well with the comparison fields and reveals traces of the MS at about $(J-H) = 0.4$, but it is not defined
			well enough to determine any fundamental parameters.\\

		\object{DSH~J1822.6-1443} = \object{Kronberger~25}
   			appears as an extremely dense cluster of stars in the field of the HII region \object{S~48}
			(Sharpless~\cite{sharpless}), with the
			clusters \object{C~1819-146} and \object{C~1820-146} (Kharchenko~\&~Schilbach~\cite{kharchenko}; hereafter
			KS95) situated at distances of 10 and 5.7 arcminutes due NW and N, respectively. The tightly packed
			central core of
			the cluster measures about 20 arcseconds across and contains at least 8 stars, while the overall diameter
			of the group appears to be about $1\arcmin$ (Fig.~\ref{fig3}).
   
   			Due to the high stellar density in the cluster center, which causes the photometric
			measurements of the inner most stars to be greatly influenced by blending, the CMD (Fig.~\ref{fig9}) is
			ill-defined, with just traces of the MS visible. For that reason, a direct determination of the cluster
			parameters was impossible.
			However, it has to be noted that \object{Kronberger~25} is shown in Fig.~12 of KS95 (therein
			dotted as a single
			star) to share the proper motion of the nearby cluster complex; this implies that this newly discovered
			cluster
			might in fact be associated with this complex and, therefore, be situated at a similar distance
			($d_\mathrm{C~1819-146} = 1.99 ~\mathrm{kpc}$; $d_\mathrm{C~1820-146} = 2.13 ~\mathrm{kpc}$; from KS95).\\

   		\object{DSH~J2003.1+3158} = \object{Kronberger~54}
   			is a small cluster of stars that is well separated from the Milky Way background. It is situated
			$4.1\arcmin$ E of the bright star \object{HD~190\,227} and $3.6\arcmin$ N of the possible planetary nebula
			\object{CoMaC~3} (Acker~et~al.~\cite{acker}). 
			   
   			The CMD of \object{Kronberger~54} (Fig.~\ref{fig9}) shows practically all cluster stars aligned along a
			well-defined, although fairly broad MS; a giant branch is not present, indicating that this group is of
			rather young age. Unfortunately, the scatter of the MS, together with the obvious lack of stars with 
			$J > 15$ in the cluster field, prevents a reliable determination of the fundamental parameters of this
			cluster.\\

   		\object{DSH~J2111.8+5222} = \object{Kronberger~80}
   			is also a likely galactic cluster. It is situated in a part of the Milky Way with fairly strong
			interstellar
			obscuration. Very small ($D_\mathrm{vis} = 0.9\arcmin$) and dense, it appears as a conspicuous object on
			the DSS image shown in Fig.~\ref{fig3}.\\
   

   \setcounter{table}{2}
   \setcounter{subtable}{3}
   
   \begin{table*}
   \begin{center}
   		\subtable[][Positions and angular diameters of additional open cluster candidates.]{
         \label{TabOthers}
 			{\scriptsize
        	\begin{tabular}{llllcl}
            \noalign{\smallskip}
			\noalign{\smallskip}
			\noalign{\smallskip}
            \hline
            \noalign{\smallskip}
            DSH~ID & discoverer~ID & $RA~J2000$ & $Dec~J2000$ & $D_\mathrm{vis}$ &
			notes\\
			&&&& arcmin\\
            \noalign{\smallskip}
            \hline
            \noalign{\smallskip}
			\object{DSH~J0016.3+5957} & \object{Juchert-Saloranta~1} & $00~16~20.5$ & $+59~57~43$ & $5.0$\\
			\object{DSH~J0033.1+6507} & \object{Patchick~78} & $00~33~10.2$ & $+65~07~03$ & $1.8$\\							
			\object{DSH~J0207.3+6015} & \object{Riddle~4} & $02~07~22.7$ & $+60~15~25$ & $4.0$
			& brightest~star~\object{HD~236\,946}~(spectrum~K2;~Brodskaya~\cite{brodskaya})~probable~non-member\\
			\object{DSH~J0247.7+6158} & \object{Teutsch~162} & $02~47~42.3$ & $+61~58~29$ & $1.7$
			& \object{[BDS2003]~57}~(Bica~et~al.~\cite{bica03b})~2.5\arcmin~SW;~involved~in~
			  \object{Sh2-193}~(Sharpless~\cite{sharpless})\\
			\object{DSH~J0455.7+3646} & \object{Teutsch~5} & $04~55~43.2$ & $+36~46~55$ & $3.2$\\
			\object{DSH~J0539.4+3320} & \object{Teutsch~1} & $05~39~29.3$ & $+33~20~46$ & $2.0$\\
			\object{DSH~J0541.3+3914} & \object{Teutsch~2} & $05~41~21.8$ & $+39~14~26$ & $1.4$
			& \object{PN~G170.7+04.6}~(Alter et al. \cite{alter})~42\arcsec~distant~from~the~cluster~center\\						
			\object{DSH~J0542.7+3057A} & \object{Teutsch~45} & $05~42~45.9$ & $+30~57~33$ & $2.7$
			& involved~in~nebulosity~\object{DSH~J0542.7+3057B}~(see Appendix)\\						
			\object{DSH~J0544.3+2848} & \object{Teutsch~10} & $05~44~23.0$ & $+28~48~54$ & $5.5$
			& situated~within~the~NE~part~of~\object{Sh2-240}~(Sharpless~\cite{sharpless})\\
			\object{DSH~J0604.1+3129} & \object{Kronberger~60} & $06~04~10.1$ & $+31~29~44$ & $1.8$
			& IR cluster or group; embedded~in~reflection~nebula~\object{GN~06.00.8}~(Magakian~\cite{magakian})\\
			\object{DSH~J0625.4+1351} & \object{Teutsch~11} & $06~25~24.4$ & $+13~51~59$ & $2.3$
			& \object{DSH~J0625.6+1336}~=~\object{Teutsch~12}~16\arcmin~S\\
			\object{DSH~J0625.6+1336} & \object{Teutsch~12} & $06~25~40.3$ & $+13~36~25$ & $4.2$
			& \object{DSH~J0625.4+1351}~=~\object{Teutsch~11}~16\arcmin~N\\
			\object{DSH~J0629.4+0910} & \object{Alessi~53} & $06~29~24.5$ & $+09~10~39$ & $2.8$ 
			& = \object{DC~7} (Drake~\cite{drake})\\	
			\object{DSH~J0643.0+0140} & \object{Alessi~15} & $06~43~04.0$ & $+01~40~19$ & $0.9$
			& \object{DSH~J0643.9+0124}=~\object{Teutsch~13}~20.6\arcmin~SE;~\object{DSH~J0643.5+0210}~=~
			\object{Alessi~16}~31\arcmin~NNE\\						
			\object{DSH~J0643.5+0210} & \object{Alessi~16} & $06~43~35.0$ & $+02~10~24$ & $0.6$ 
			& embedded~in~nebulosity;~\object{DSH~J0643.0+0140}~=~\object{Alessi~15}~31\arcmin~SSW\\						
			\object{DSH~J0643.9+0124} & \object{Teutsch~13} & $06~43~55.0$ & $+01~24~09$ & $2.5$
			& \object{DSH~J0643.0+0140}~=~\object{Alessi~15}~20.6\arcmin~NW\\
			\object{DSH~J0720.9-2251} & \object{Juchert~12} & $07~20~56.7$ & $-22~52~00$ & $2.8$
			& 1.4\arcmin~SE~of~\object{HD~57\,573}\\						
			\object{DSH~J0758.3-3446} & \object{Kronberger~85} & $07~58~20.8$ & $-34~46~11$ & $1.4$
			& 1.6\arcmin~NW~of~reflection~nebula~\object{Bran~89}~(Magakian~\cite{magakian})\\						
			\object{DSH~J0807.1-3603} & \object{Teutsch~50} & $08~07~08.1$ & $-36~03~48$ & $3.2$\\							
			\object{DSH~J1108.6-6042} & \object{Teutsch~143a} & $11~08~40.6$ & $-60~42~50$ & $0.6$ 
			& surrounding~S~Dor~star~\object{Wray~15-751}~(van Genderen~\cite{vangenderen})\\						
			\object{DSH~J1153.2-6236} & \object{Teutsch~77} & $11~53~15.5$ & $-62~36~32$ & $1.5$ 
			& \object{HD~309\,205}~2.8\arcmin~NNE\\						
			\object{DSH~J1713.2-3942} & \object{Teutsch~85} & $17~13~13.8$ & $-39~42~22$ & $0.6$\\							
			\object{DSH~J1907.5+0617} & \object{Juchert~3} & $19~07~33.0$ & $+06~17~10$ & $3.0$\\							
			\object{DSH~J1911.1+1450} & \object{Riddle~15} & $19~11~09.2$ & $+14~50~04$ & $0.8$
			& partly~embedded~in~nebulosity\\
			\object{DSH~J1922.5+1240} & \object{Juchert~1} & $19~22~32.0$ & $+12~40~00$ & $3.2$
			& at~N~edge~of~\object{LDN~684}~(Lynds~\cite{lynds62})\\
			\object{DSH~J1925.2+1356} & \object{Kronberger~13} & $19~25~14.9$ & $+13~56~44$ & $1.5$
			& \object{C1922+136}~=~\object{King~25}~18\arcmin~SW;~\object{DSH~J1926.0+1945}~=~
			\object{Teutsch~26}~16\arcmin~SE\\
			\object{DSH~J1937.3+1841} & \object{Teutsch~27} & $19~37~23.2$ & $+18~41~50$ & $1.1$
			& \object{DSH~J1933.9+1831}~=~\object{Kronberger~79}~53\arcmin~WSW\\
			\object{DSH~J1958.1+3053} & \object{Kronberger~52} & $19~58~07.9$ & $+30~53~18$ & $2.3$
			& 8.6\arcmin~SW~of~\object{HD~189\,395}\\						
			\object{DSH~J1959.5+4918} & \object{Patchick~89} & $19~59~33.0$ & $+49~18~45$ & $1.4$
			& = \object{DC~6} (Drake~\cite{drake})\\							
			\object{DSH~J2001.2+3336} & \object{Toepler~1} & $20~01~17.6$ & $+33~36~54$ & $2.1$ 
			& HII~region~\object{M~1-97}~(Minkowski \cite{minkowski})
			3.3\arcmin~SE;~see~also~Toepler~(\cite{toepler})\\						
			\object{DSH~J2006.5+3534} & \object{Kronberger~28} & $20~06~31.7$ & $+35~34~34$ & $1.3$
			& in~field~of~\object{C2004+356}~=~\object{NGC~6871}\\
			\object{DSH~J2015.0+3355} & \object{Patchick~75} & $20~15~01.5$ & $+33~55~43$ & $1.5$\\							
			\object{DSH~J2017.9+3645} & \object{Kronberger~74} & $20~17~57.2$ & $+36~45~37$ & $0.3$
			& IR~cluster~or~group;~embedded~in~nebulosity;~
			HII~region~\object{Sh2-104}~(Sharpless~\cite{sharpless})~W\\						
			\object{DSH~J2020.7+4112} & \object{Kronberger~58} & $20~20~47.8$ & $+41~12~17$ & $0.2$
			& IR~cluster~or~group;~embedded~in~nebulosity;~\object{V1318~Cyg}~cluster~10\arcmin~NNW\\						
			\object{DSH~J2023.9+3636} & \object{Kronberger~57} & $20~23~57.6$ & $+36~36~17$ & $2.8$\\							
			\object{DSH~J2028.2+3506} & \object{Teutsch~28} & $20~28~16.5$ & $+35~06~50$ & $1.3$\\							
			\object{DSH~J2033.8+4008} & \object{Kronberger~59} & $20~33~49.9$ & $+40~08~53$ & $1.6$
			& IR~cluster~or~group;~embedded~in~nebulosity;~associated~with~\object{IRAS~20319+3958}\\
			\object{DSH~J2035.1+5318} & \object{Patchick~103} & $20~35~08.3$ & $+53~18~00$ & $1.3$\\
			\object{DSH~J2121.7+5036} & \object{Teutsch~144} & $21~21~43.9$ & $+50~36~36$ & $2.0$\\							
			\object{DSH~J2247.4+5946} & \object{Teutsch~54} & $22~47~24.1$ & $+59~46~54$ & $1.8$\\							
			\object{DSH~J2347.9+6259} & \object{Teutsch~23} & $23~47~54.2$ & $+62~59~50$ & $1.8$\\							
			\object{DSH~J2353.1+6247} & \object{Kronberger~55} & $23~53~09.5$ & $+62~47~12$ & $1.0$
			& IR~cluster\\
			\noalign{\smallskip}
            \hline
        \end{tabular}}}
		\end{center}
   \end{table*}


	
	\subsection{Other candidates}

   		Table~\ref{TabOthers} lists further stellar agglomerations reported to the DSH newsgroup between June 2003 and
		December 2004, which we consider to be possible open 
		clusters from the analytical procedures described in Sect. 2. As in Tables~\ref{TabIso} to~\ref{TabNo}, 
		positions and angular diameters of the cluster candidates are given, and a DSH designation and the
		discoverer ID are assigned. It should be noted that 2 of the objects presented in this section
		(\object{DSH~J0629.4+0910} = \object{Alessi~53} and \object{DSH~J1959.5+4918} = \object{Patchick~89}) turned
		out to be described by Drake~(\cite{drake}) as objects \object{DC~6} and \object{DC~7}, respectively;
		nevertheless, as in the case of \object{Teutsch~49} (see Sect. 3.2.3), we decided to include these objects
		as they have been discovered independently.
		 
   		Although the evidence that the objects listed in Table~\ref{TabOthers} are indeed genuine open clusters is
		weaker than for the candidates presented in Tables~\ref{TabIso} to~\ref{TabNo}, their morphologies, CMDs and
		RDPs are good arguments in favor of their being clusters. 
		However, as with the cluster candidates described in the previous section, deeper studies are
		necessary to further clarify their nature.

   		An additional list of 174 stellar fields with evidence that an open cluster might be involved
		(=~Table~2e) is available via the CDS. As in Table~\ref{TabOthers}, the following information is given:
		DSH ID, discoverer ID, $RA~J2000$, $DE~J2000$, and the visual diameter $D_\mathrm{vis}$.


\section{Conclusions}

   An inspection of several Milky Way fields using DSS and 2MASS $JHK_\mathrm{S}$ imagery has led to the discovery of
   66
   stellar aggregates that are thought to be open clusters or, at least, promising open cluster candidates due to
   their appearance, their stellar density distributions, and their CMDs. For 9 of these objects, we were able to
   determine preliminary distances, reddenings, and ages by fitting their CMDs with solar metallicity Padova
   isochrones
   and found distances $d = 1.6 - 4.6 ~\mathrm{kpc}$, reddenings $E(B-V) = 0.52 - 2.55$, and ages ranging from
   less than $10~\mathrm{Myr}$ to $\approx 1~\mathrm{Gyr}$. Additionally, distances and reddenings were derived for 10
   clusters by using their helium-burning red clump stars as standard candles, with the values between
   $3.6 ~\mathrm{kpc}$ and $11.9 ~\mathrm{kpc}$ found for the distances and between $0.54$ and $2.11$ for the
   reddenings.
   Furthermore, we note the presence of X-ray sources in the immediate vicinity of 2 clusters presented here,
   \object{Teutsch~42} and \object{Kronberger~81}; however, the question whether these are indeed physically connected
   with the clusters or not cannot be answered from the available data and therefore shall be the object of a deeper
   investigation.
   
   The high number of newly discovered cluster candidates, as presented in this paper, can be seen as a sign of the
   low completeness level of even the most up-to-date catalogues. Considering that no systematic all-sky survey for
   open clusters has been carried out yet on 2MASS or on the Second Palomar Observatory Sky Survey images, it can be
   assumed that the chances are high that a number of unknown galactic clusters detectable in optical and infrared
   wavelengths is still awaiting discovery. It should be noted, however, that further studies are required to confirm
   that the objects presented here are true clusters and to determine their precise fundamental parameters.


	\newcolumntype{.}{D{.}{.}{2,2}}

	\setcounter{table}{2}
   	\begin{table*}
		\caption[]{Positions and angular diameters of apparently uncatalogued galactic and extragalactic objects.}
       	\label{TabNeb}
		{\scriptsize
        	\begin{tabular}{llll..l}
			\hline
            \noalign{\smallskip}
			DSH~ID & discoverer~ID & $RA~J2000$ & $Dec~J2000$ & \multicolumn{1}{c}{$D_\mathrm{max}$} &
			\multicolumn{1}{c}{$D_\mathrm{min}$} & notes\\
			&&&& \multicolumn{1}{c}{arcmin} & \multicolumn{1}{c}{arcmin}\\
            \noalign{\smallskip}
            \hline
            \noalign{\smallskip}
			\object{DSH~J0540.7+3144} & \object{Teutsch~PN~J0540.7+3144} & $05~40~44.6$ & $+31~44~32$ & 1.8 & 1.8
			& PN;~independent~IPHAS~discovery~(Frew~\cite{frew05d})\\
			\object{DSH~J0542.7+3057B} & \object{Teutsch~GN~J0542.7+3057} & $05~42~44.3$ & $+30~57~57$ & 2.1 & 2.1
			& HII~region?~Associated~with~\object{DSH~J0542.7+3057A}~=~\object{Teutsch~45}\\		
			\object{DSH~J0600.5+2141} & \object{Teutsch~GN~J0600.5+2141} & $06~00~34.8$ & $+21~41~11$ & 1.2 & 1.1
			& compact~HII~region~or~PN\\	
			\object{DSH~J0646.4+0828} & \object{Riddle~PN~J0646.4+0828} & $06~46~24.7$ & $+08~29~02$ & 1.1 & 1.0
			& confirmed~spectroscopically~as~PN~(Frew~\cite{frew05a})\\
			\object{DSH~J1405.9-6040} & \object{Riddle~GN~J1405.9-6040} & $14~05~58.5$ & $-60~40~14$ & 1.7 & 1.1
			& reflection~nebula?\\
			\object{DSH~J1714.8-1415} & \object{Patchick~2} & $17~14~48.9$ & $-14~15~54$ & 0.17 & 0.17
			& small~disk~0.6\arcmin~N~of~\object{SAO~160\,398}; planetary nebula?\\
			\object{DSH~J1909.9+1204} & \object{Teutsch~GN~J1909.9+1204} & $19~09~54.9$ & $+12~04~52$ & 0.2 & 0.2
			& reddish~circular~nebulosity;~planetary~nebula?\\
			\object{DSH~J1919.5+4445} & \object{Patchick~5} & $19~19~30.6$ & $+44~45~44$ & 2.6 & 1.6
			& probable~bipolar~PN~(Block~\cite{block};~Frew~\cite{frew05d})\\
			\object{DSH~J1940.6+2930} & \object{Kronberger~PN~J1940.6+2930} & $19~40~40.4$ & $+29~30~09$ & 0.4 &
			0.4 & misclass.~as~galaxy~by~Roman~et~al.~(\cite{roman});~PN~(Frew~\cite{frew05d})\\
			\object{DSH~J1941.1+1908} & \object{Kronberger~GN~J1941.1+1908} & $19~41~07.7$ & $+19~08~26$ & 0.6 &
			0.3 & irregular~nebulosity~surrounding~2~stars;~classification~uncertain\\
			\object{DSH~J1942.4+2145} & \object{Kronberger~PN~J1942.4+2145} & $19~42~26.1$ & $+21~45~23$ & 0.35 & 0.3
			& small~bipolar~nebulosity;~PN?\\
			\object{DSH~J1944.9+2245} & \object{Kronberger~PN~J1944.9+2245} & $19~44~59.1$ & $+22~45~49$ & 4.0 & 4.0
			& PN~(Frew~\cite{frew05d});~\object{Lan~21}~(Lanning~\&~Meakes~\cite{lanning})~assoc.\\
			\object{DSH~J1947.0+2930} & \object{Patchick~1} & $19~47~02.7$ & $+29~30~26$ & 0.21 & 0.18
			& confirmed~spectroscopically~as~PN~(Frew~\cite{frew05d})\\
			\object{DSH~J1957.3+2639} & \object{Teutsch~PN~J1957.3+2639} & $19~57~22.3$ & $+26~39~06$ & 2.3 & 1.5
			& confirmed~spectroscopically~as~PN~(Frew~\cite{frew05c});~bipolar?\\
			\object{DSH~J2002.4+5533} & \object{Patchick~4} & $20~02~28.4$ & $+55~33~18$ & 3.5 & 3.5
			& nebulosity~with~very~blue~central~star;~PN?~(Frew~\cite{frew05d})\\
			\object{DSH~J2009.6+4114} & \object{Patchick~6} & $20~09~40.9$ & $+41~14~43$ & 0.7 & 0.7
			& probable~PN~(Frew~\cite{frew05d})\\
			\object{DSH~J2046.1+5257} & \object{Patchick~3} & $20~46~10.5$ & $+52~57~06$ & 0.5 & 0.5
			& nebulosity~with~very~blue~central~star;~planetary~nebula?\\	
			\object{DSH~J2051.0+7222} & \object{Schoenball~GN~J2051.0+7222} & $20~51~05.0$ & $+72~22~27$ & 2.5 &
			2.0 & evenly~illuminated~elliptical~nebula;~reflection~nebula?\\
			\object{DSH~J2055.4+3903} & \object{Teutsch~PN~J2055.4+3903} & $20~55~27.3$ & $+39~03~57$ & 0.35 & 0.25
			& small~bipolar~nebulosity;~planetary~nebula?\\
			\object{DSH~J2120.0+5141} & \object{Patchick~7} & $21~20~00.1$ & $+51~41~05$ & 0.08 & 0.08
			& very~small~emission-line~object;~planetary~nebula?\\
			\object{DSH~J2122.0+5504} & \object{Kronberger~PN~J2122.0+5504} & $21~22~01.0$ & $+55~04~30$ & 0.33 &
			0.33 & nebulosity~consisting~of~2~opposing arcs;~planetary~nebula?\\			
			\object{DSH~J2308.8+1712} & \object{Riddle~GX~J2308.8+1712} & $23~08~51.4$ & $+17~12~36$ & 1.8 & 1.5 
			& bright~galaxy; variable~star~\object{DY~Pegasi} superimposed\\			
			\object{DSH~J2327.2+6509} & \object{Teutsch~PN~J2327.2+6509} & $23~27~13.8$ & $+65~09~15$ & 0.3 & 0.3
			& probable~planetary~nebula~(Frew \cite{frew05b})\\
			\noalign{\smallskip}
            \hline
         \end{tabular}}
       
   \end{table*}



\section{Addendum}

   During our survey for new open cluster candidates on DSS and 2MASS, we noted a number of galactic nebulae and
   galaxies that were found not to be listed in any current database. The most impressive of these objects are listed
   in Table~\ref{TabNeb}, with a more extensive list (Tables 4a and 4b) available via the CDS. All objects are
   tentatively classified on the basis of
   their brightness and morphology on DSS-2 blue, red, and infrared images. However, further detailed investigations
   are needed to establish their true nature.


\begin{acknowledgements} 
   This publication made use of data products from the Two Micron All Sky Survey, which is a joint project of the
   University of Massachusetts and the Infrared Processing and Analysis Center/California Institute of Technology,
   funded by the National Aeronautics and Space Administration and the National Science Foundation. Extensive use was
   also made of the SIMBAD and WEBDA databases, the ESO Online Digitized Sky Survey facility, and the Aladin
   Interactive Sky Atlas. Furthermore, special thanks go to: A.~Block of the NOAO, USA, and D.~J.~Frew of the
   Macquarie
   University, Australia, for providing extra information on some of our planetary nebulae candidates;
   to W. S. Dias of the University of S\~ao Paulo; B.~Skiff of Lowell Observatory; O.~Brazell of the Webb Society; and
   A.~Moitinho of the Observatorio Astron\'omico Nacional, M\'exico, for reading and commenting on the manuscript; and
   to M.~Sinhuber for extensive technical support. Last but not least, the Authors of this study would like to thank
   all DSH members for their enthusiasm and far-reaching support.

   \emph{The DSH collaboration is a group of amateur and professional astronomers founded in June 2003. Its
   intentions are the discovery of deep sky objects not yet published in professional literature and the improvement
   of observational data for already known objects.}
\end{acknowledgements}







\setcounter{figure}{0}
\setcounter{subfigure}{0}
\begin{figure*}

\caption[]{Second generation DSS red images of the cluster candidates presented in
 			Table~\subref{TabIso}. The field of view is in each case $8\arcmin$ $\times$
			$8\arcmin$. North is up and East is left.}
\label{fig1}

\subfigure[][\object{Juchert~9}]{
\includegraphics[width=8.75cm]{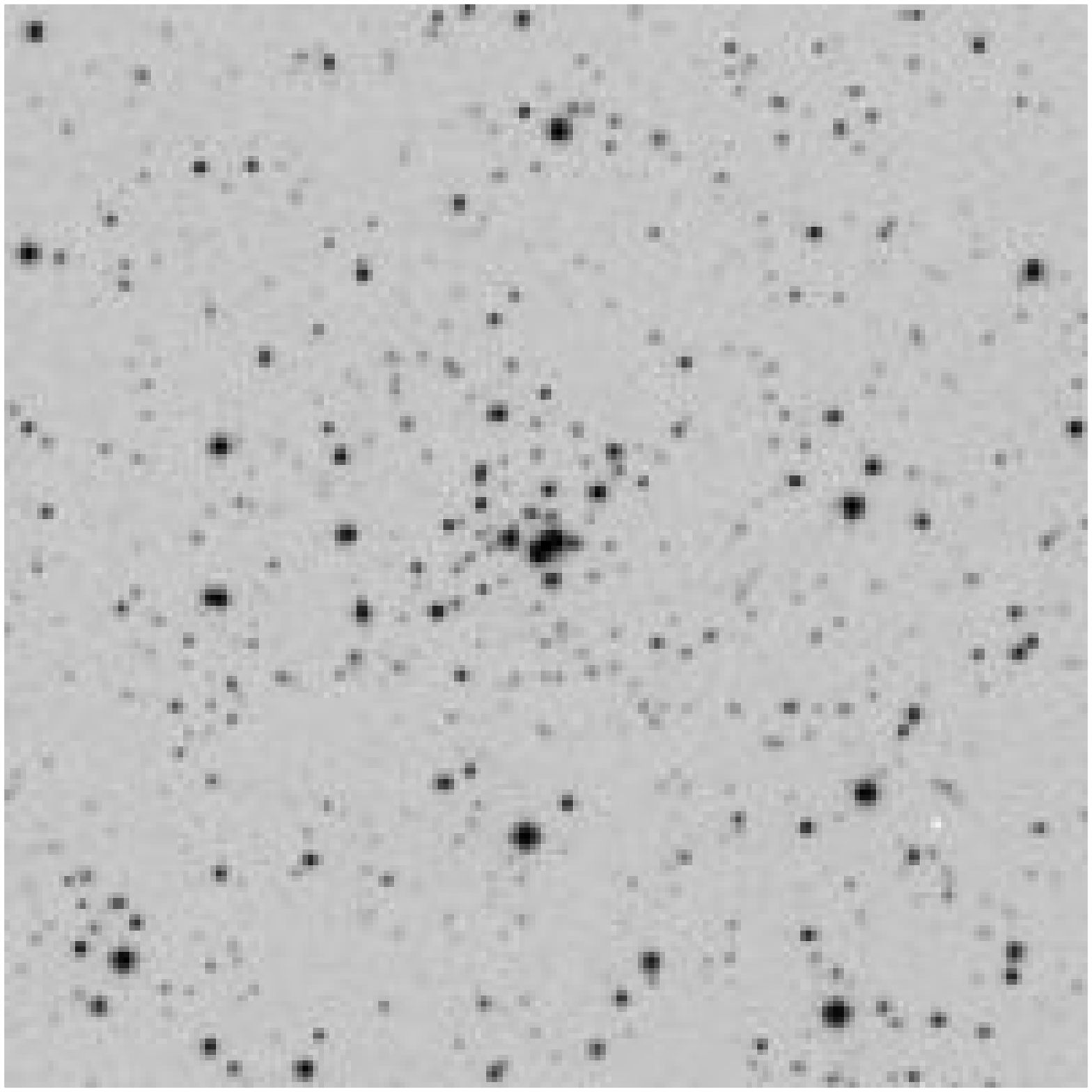}
}
\enskip
\subfigure[][\object{Kronberger~1}]{
\includegraphics[width=8.75cm]{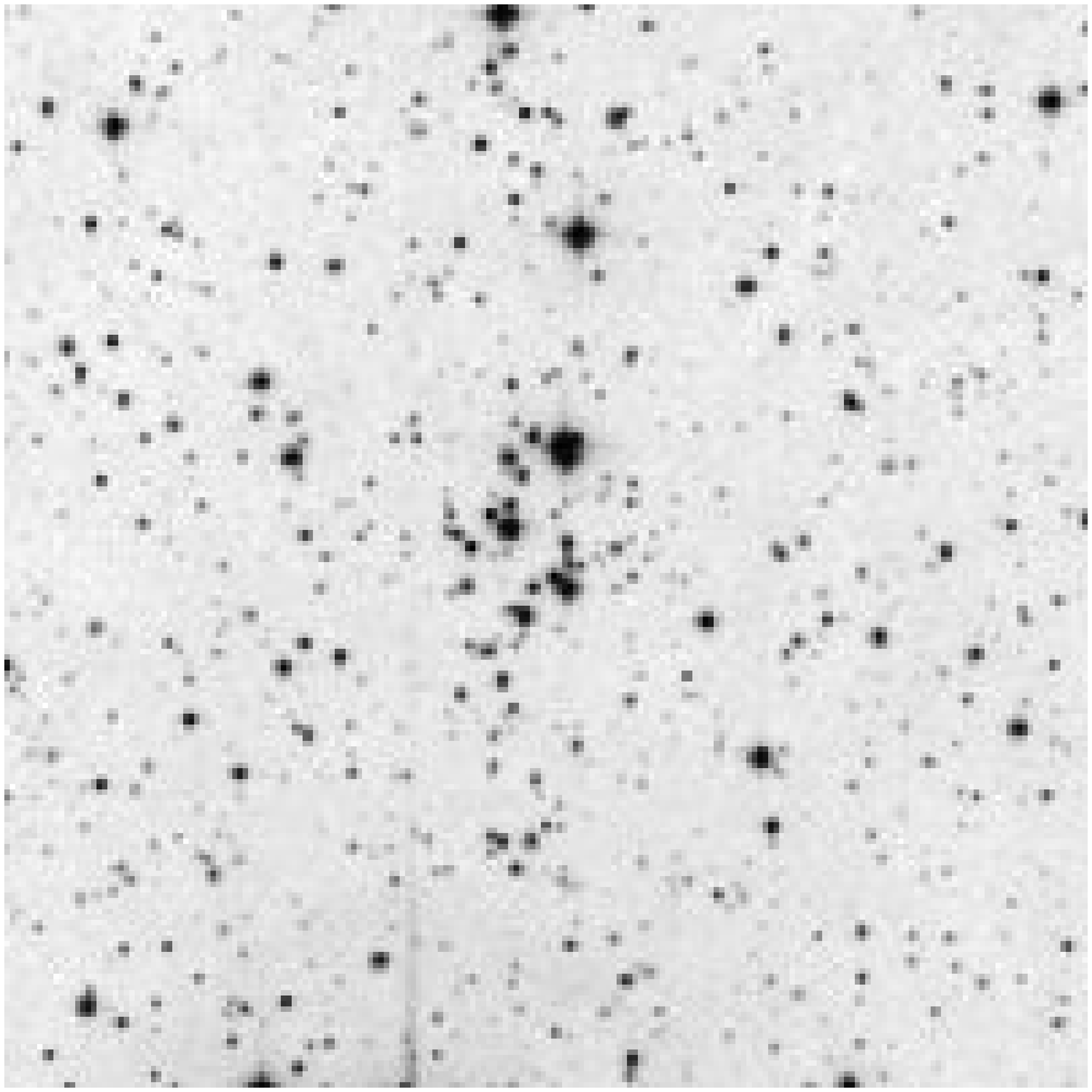}
}

\subfigure[][\object{Teutsch~51}]{
\includegraphics[width=8.75cm]{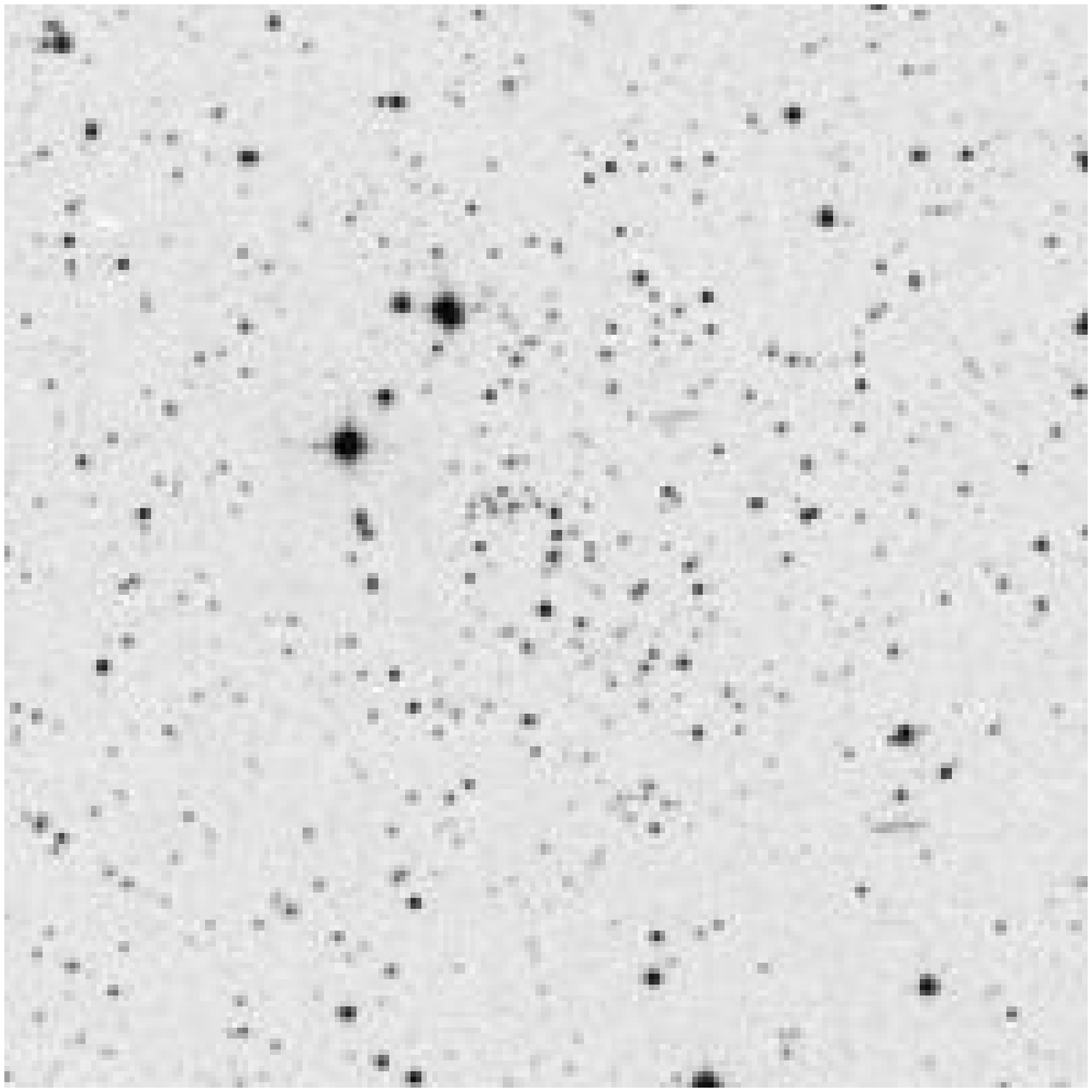}
}
\enskip
\subfigure[][\object{Teutsch~80}]{
\includegraphics[width=8.75cm]{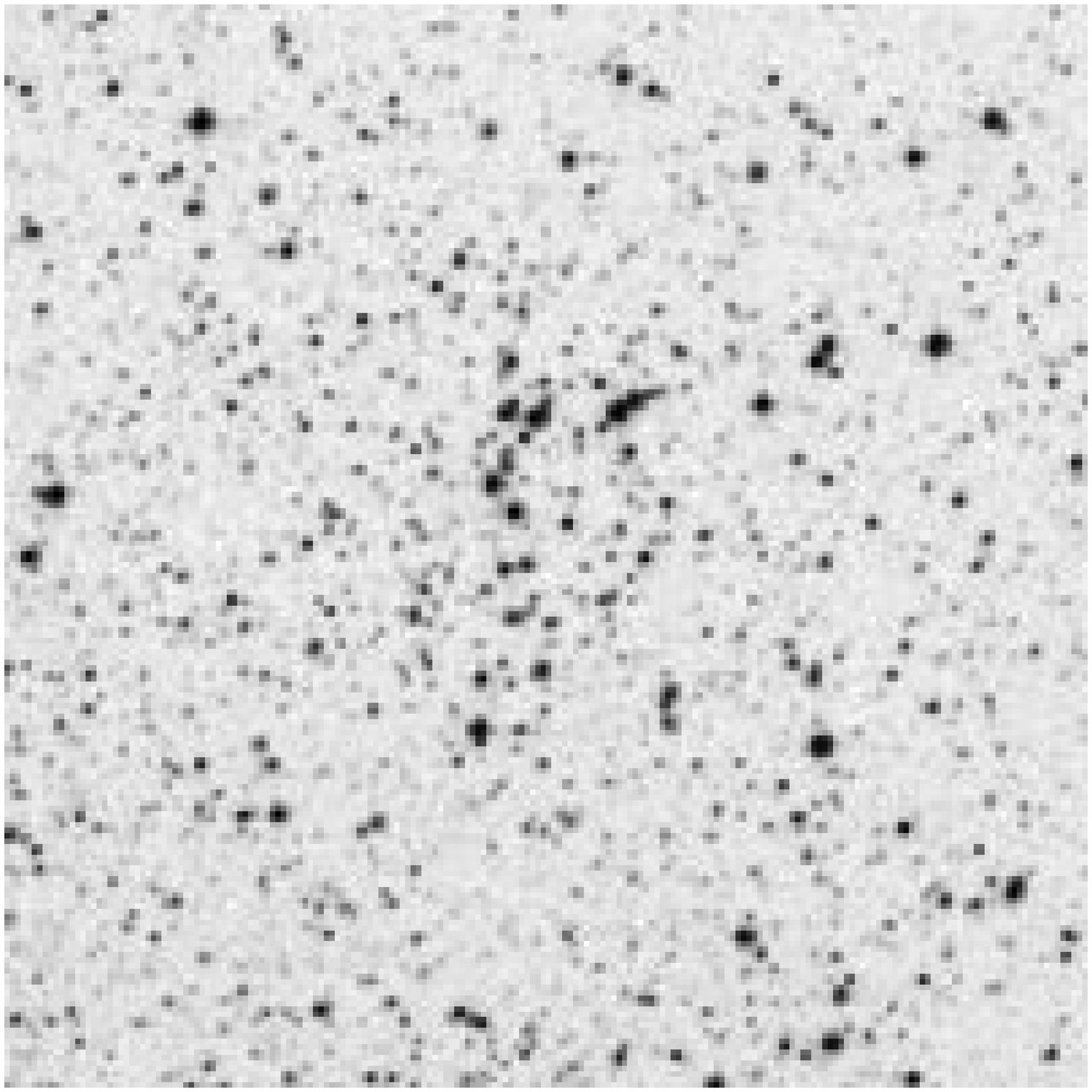}
}

\end{figure*}


\setcounter{figure}{0}
\setcounter{subfigure}{0}
\begin{figure*}
\caption[]{(cont.)}

\subfigure[][\object{Teutsch~84}]{
\includegraphics[width=8.75cm]{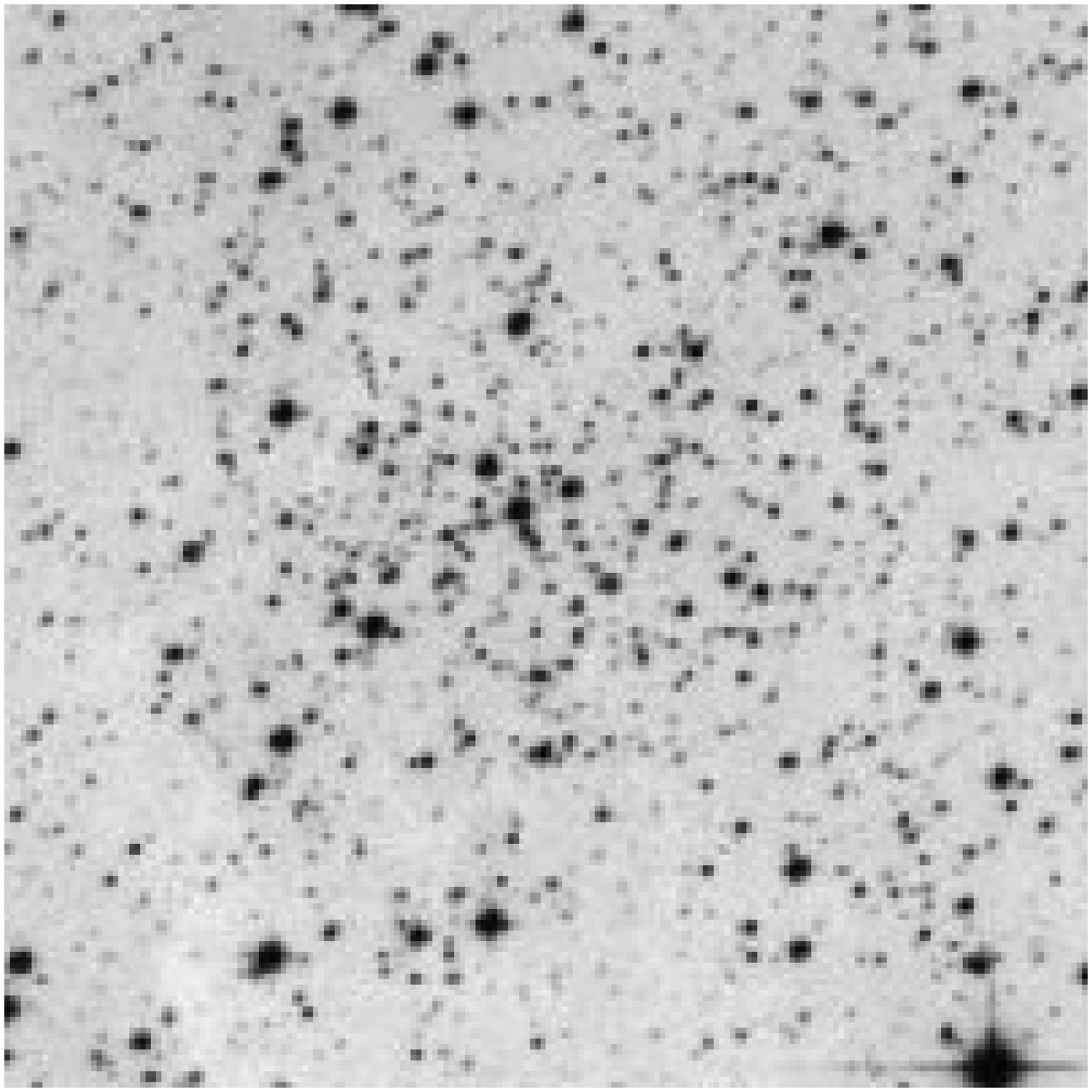}
}
\enskip
\subfigure[][\object{Teutsch~42}]{
\includegraphics[width=8.75cm]{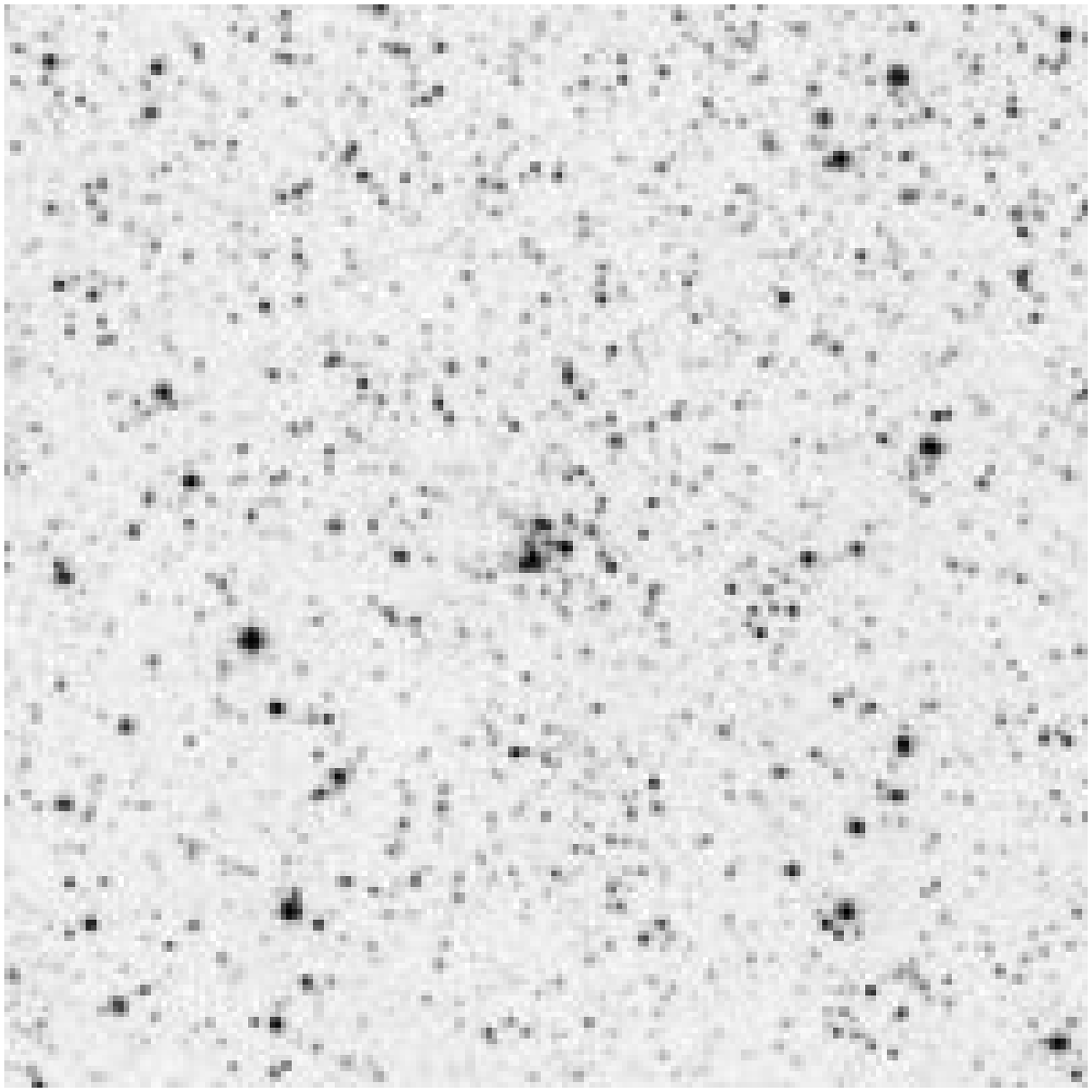}
}

\subfigure[][\object{Kronberger~79}]{
\includegraphics[width=8.75cm]{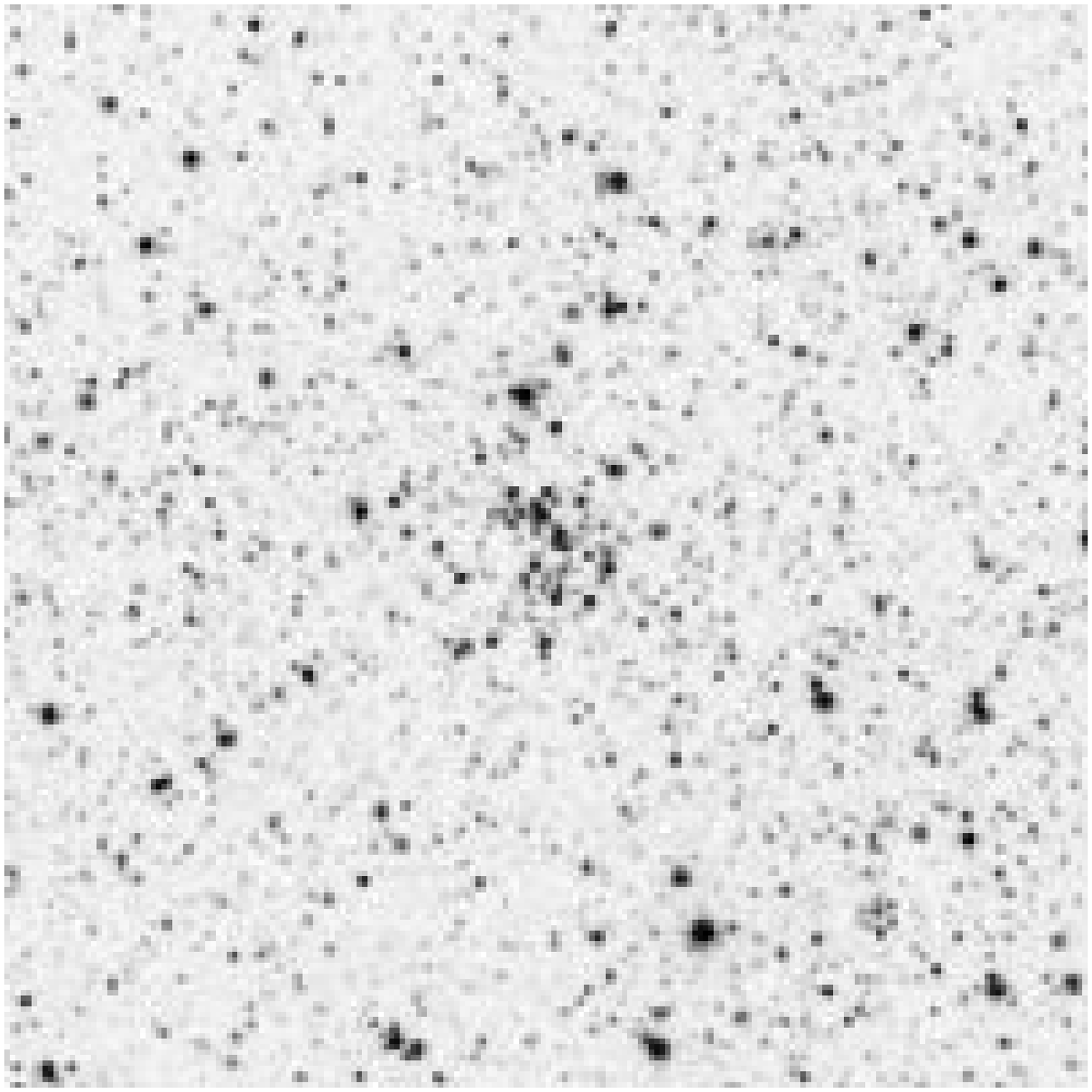}
}
\enskip
\subfigure[][\object{ADS~13292}]{
\includegraphics[width=8.75cm]{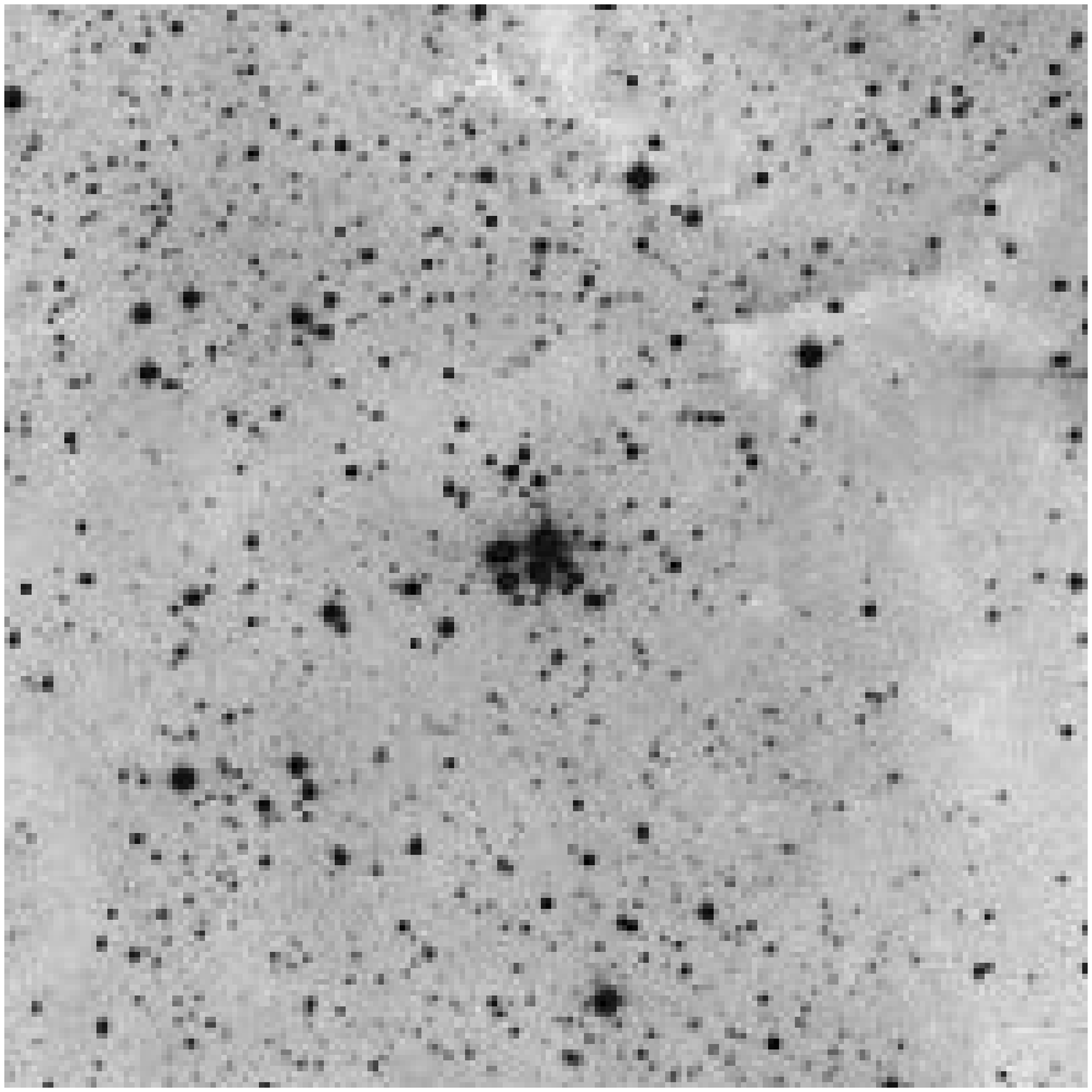}
}

\end{figure*}


\setcounter{figure}{0}
\setcounter{subfigure}{0}
\begin{figure*}
\caption[]{(cont.)}

\subfigure[][\object{Teutsch~30}]{
\includegraphics[width=8.75cm]{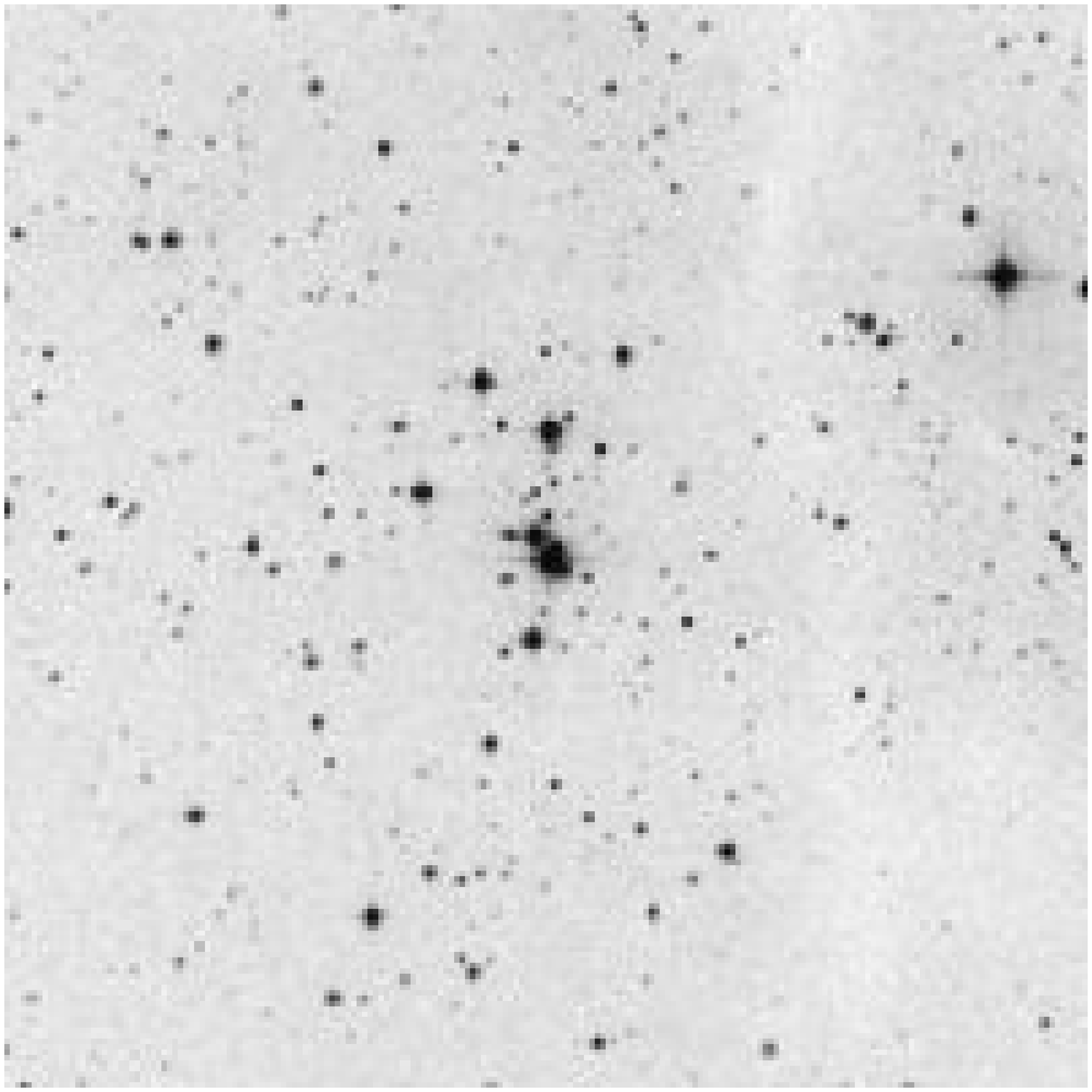}
}

\end{figure*}



\setcounter{figure}{1}
\setcounter{subfigure}{0}
\begin{figure*}
\caption[]{Second generation DSS red images of the cluster candidates presented in
 			Table~\subref{TabRC}. The field of view is in each case $8\arcmin$ $\times$
			$8\arcmin$. North is up and East is left.}
\label{fig2}

\subfigure[][\object{Teutsch~79}]{
\includegraphics[width=8.75cm]{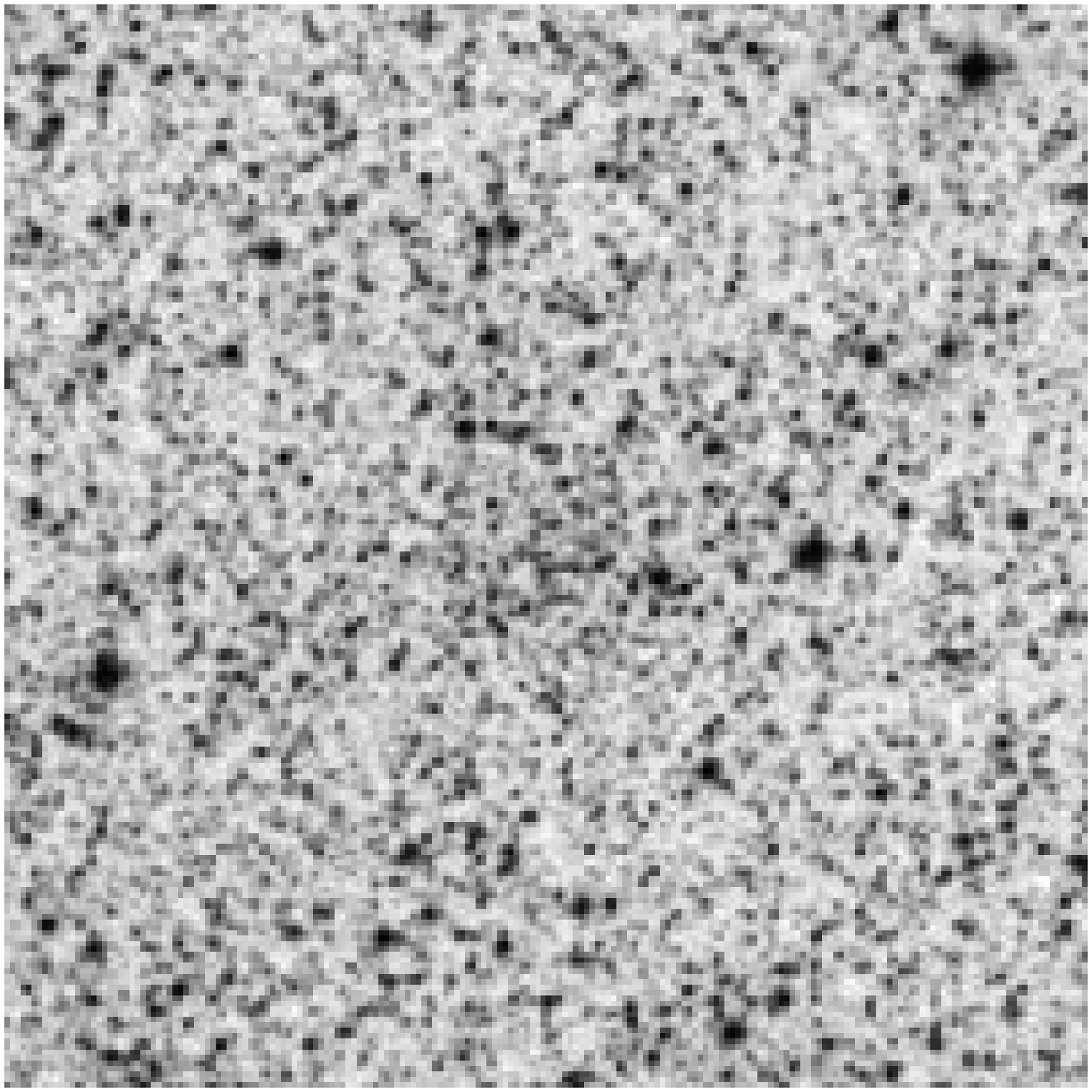}
}
\enskip
\subfigure[][\object{Kronberger~81}]{
\includegraphics[width=8.75cm]{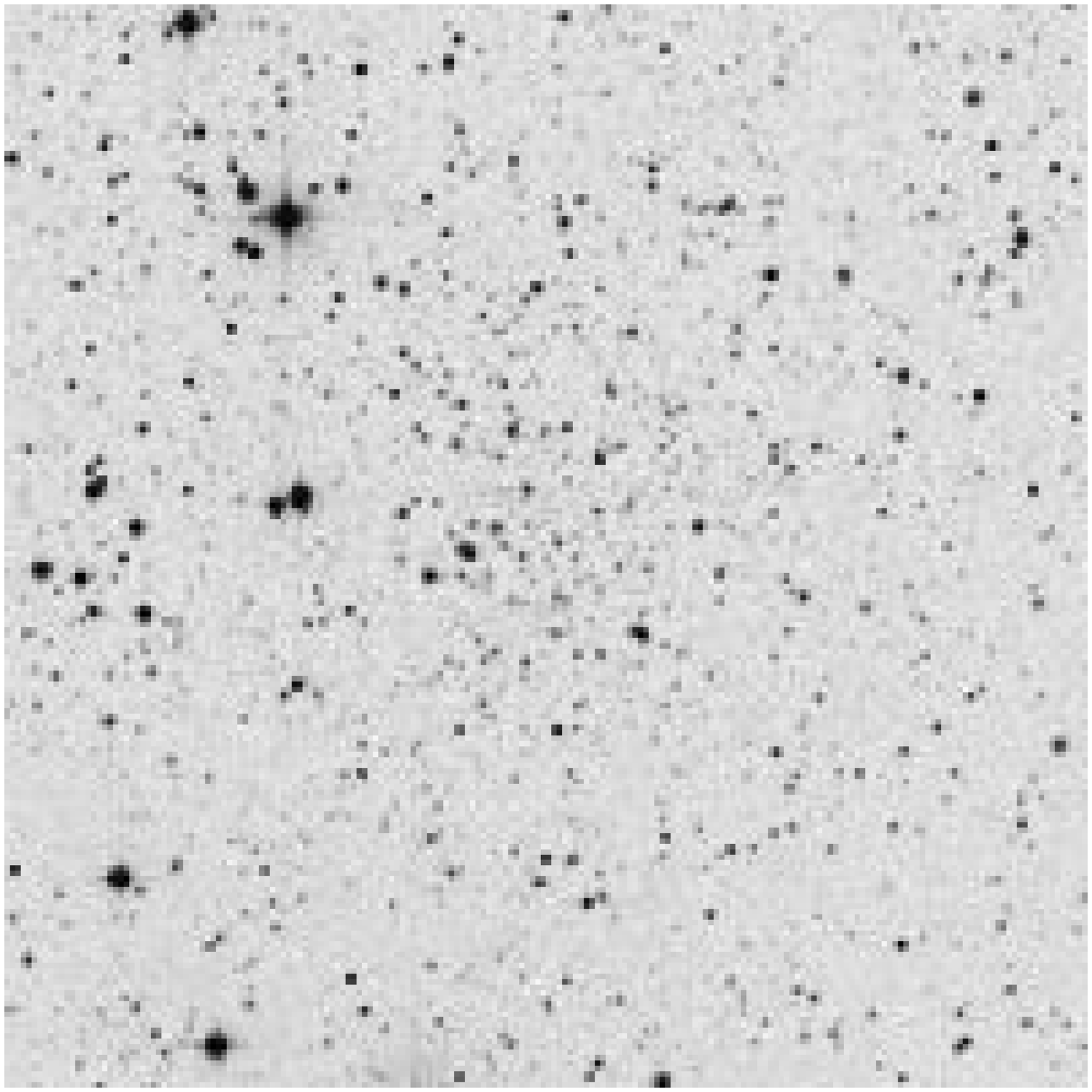}
}

\subfigure[][\object{Juchert~11}]{
\includegraphics[width=8.75cm]{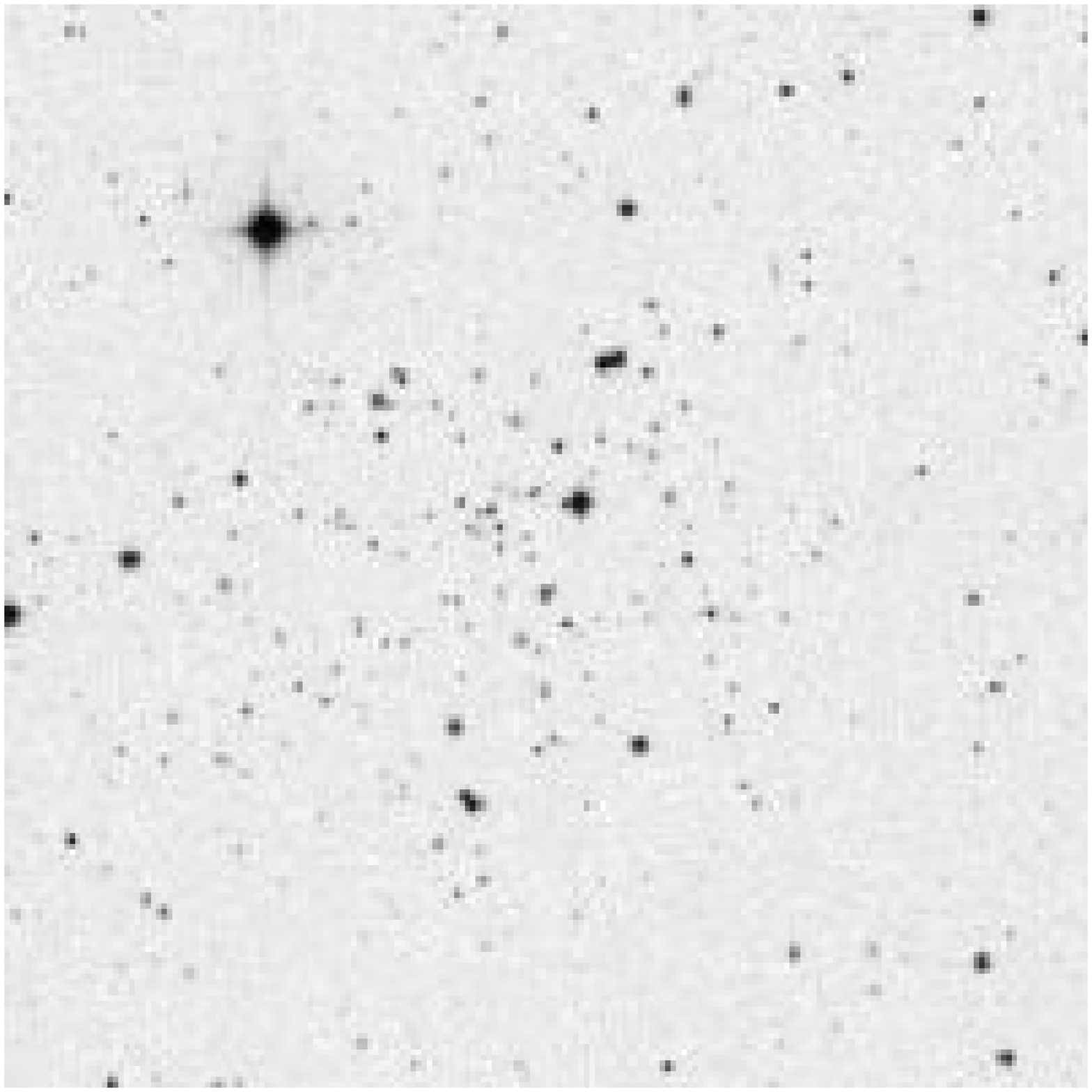}
}
\enskip
\subfigure[][\object{Teutsch~48}]{
\includegraphics[width=8.75cm]{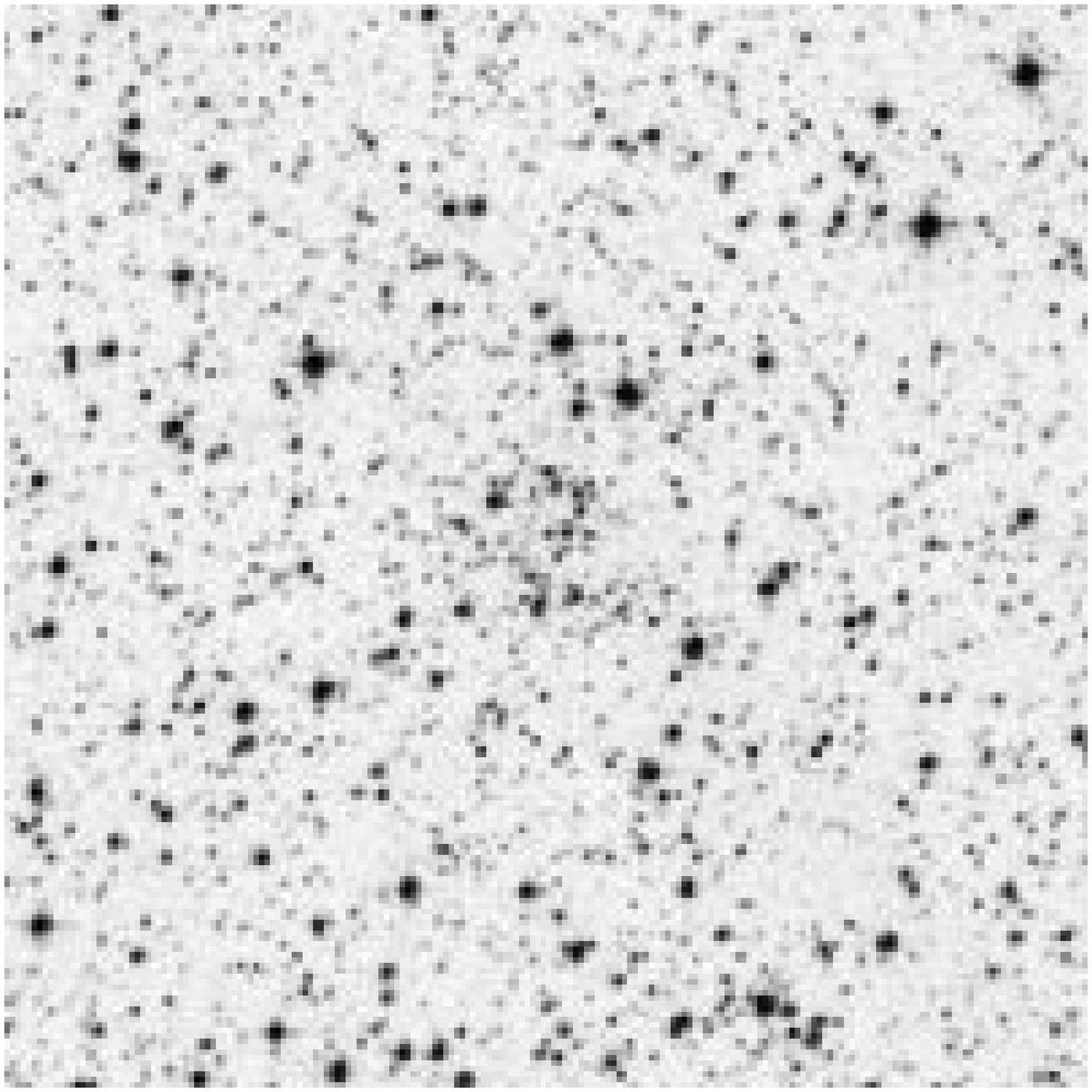}
}

\end{figure*}


\setcounter{figure}{1}
\setcounter{subfigure}{0}
\begin{figure*}
\caption[]{(cont.)}

\subfigure[][\object{Kronberger~39}]{
\includegraphics[width=8.75cm]{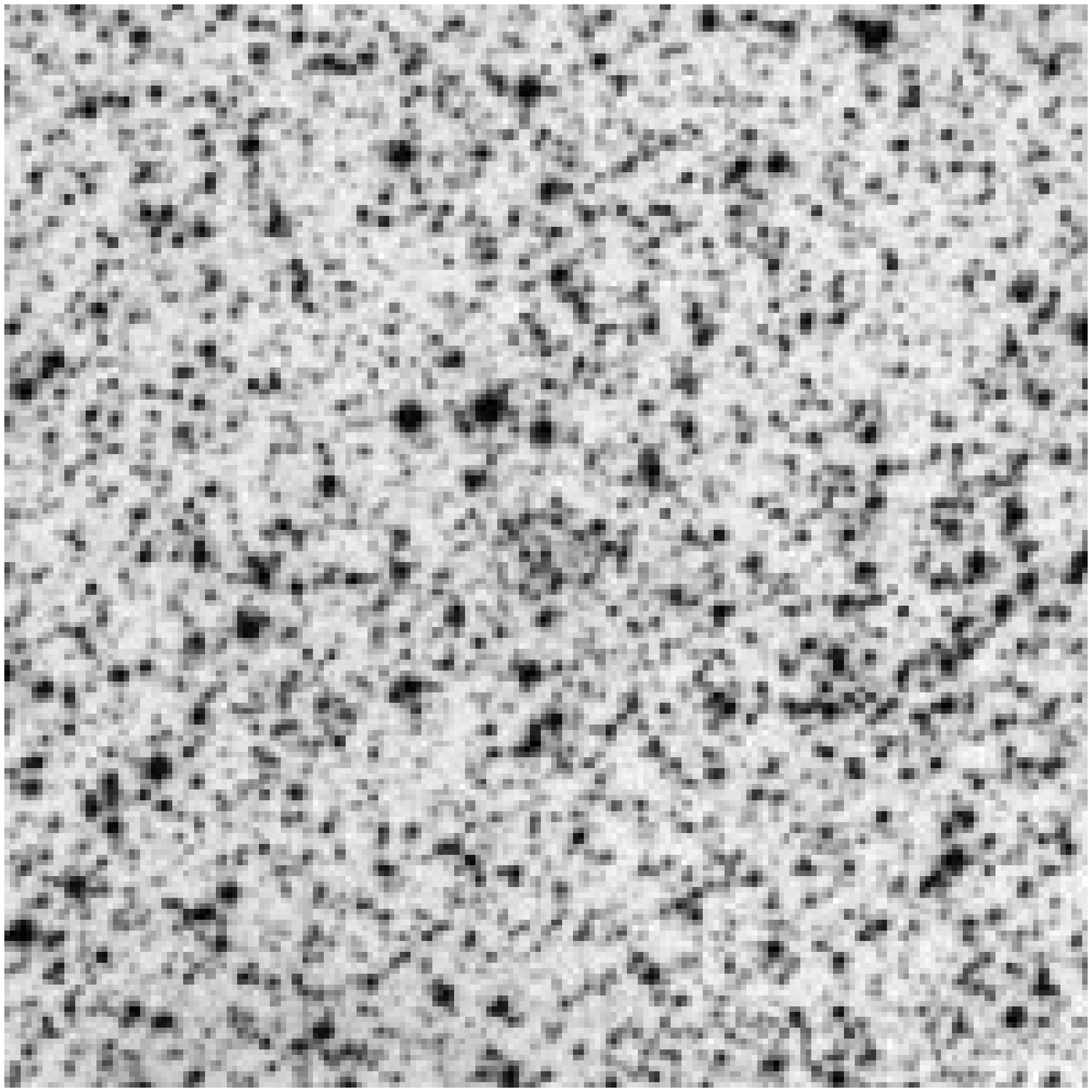}
}
\enskip
\subfigure[][\object{Alessi~56}]{
\includegraphics[width=8.75cm]{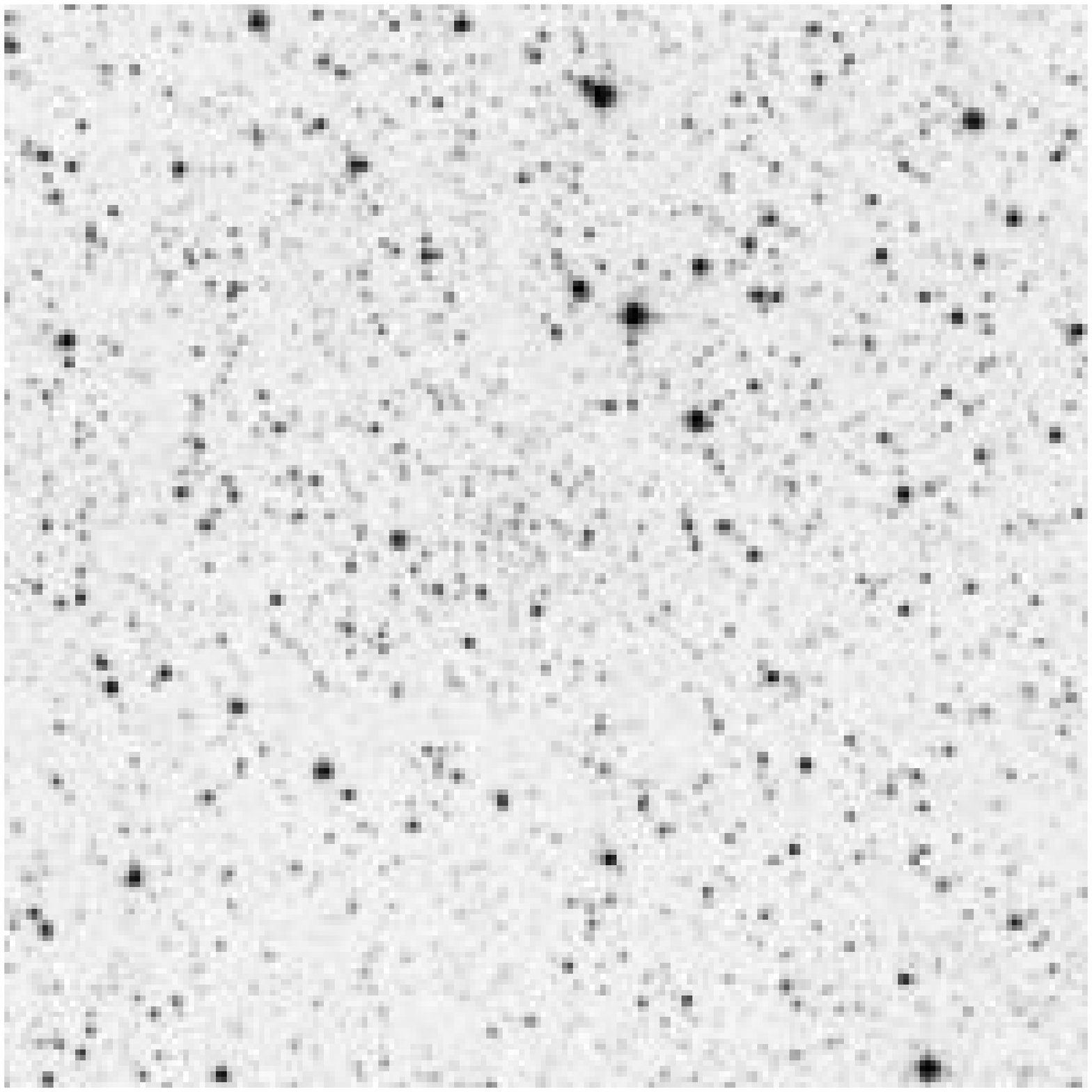}
}

\subfigure[][\object{Alessi~57}]{
\includegraphics[width=8.75cm]{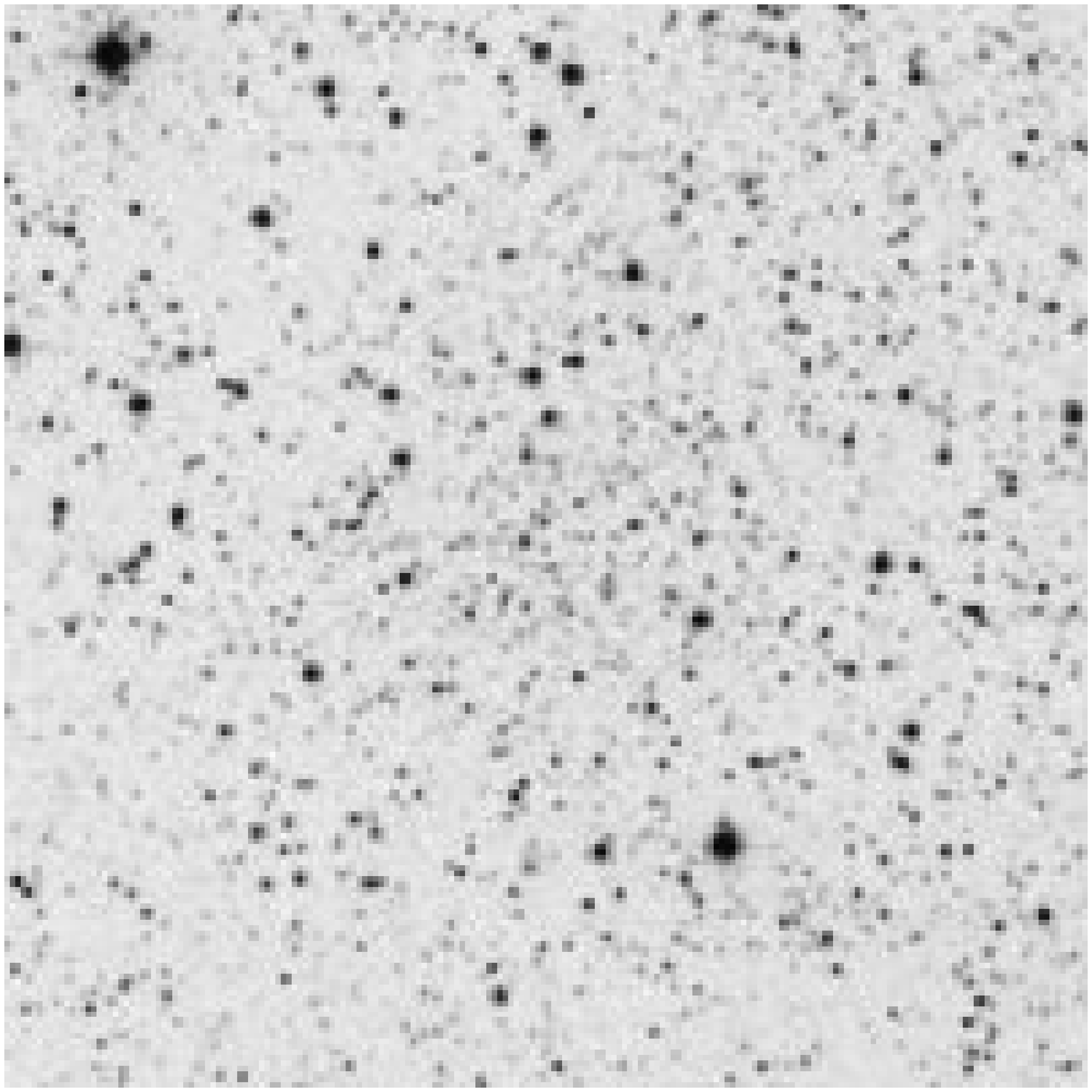}
}
\enskip
\subfigure[][\object{Kronberger~31}]{
\includegraphics[width=8.75cm]{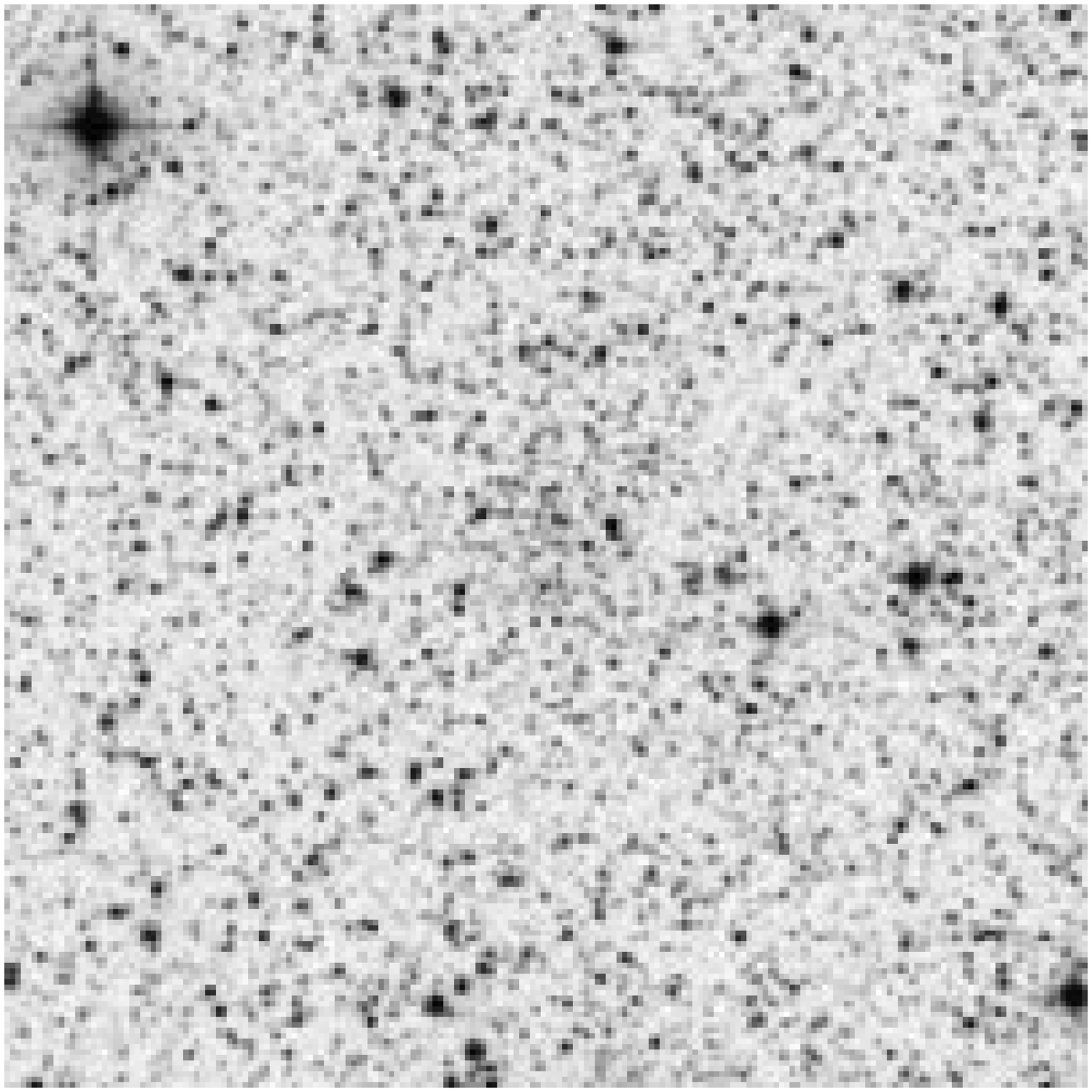}
}

\end{figure*}


\setcounter{figure}{1}
\setcounter{subfigure}{0}
\begin{figure*}
\caption[]{(cont.)}

\subfigure[][\object{Teutsch~43}]{
\includegraphics[width=8.75cm]{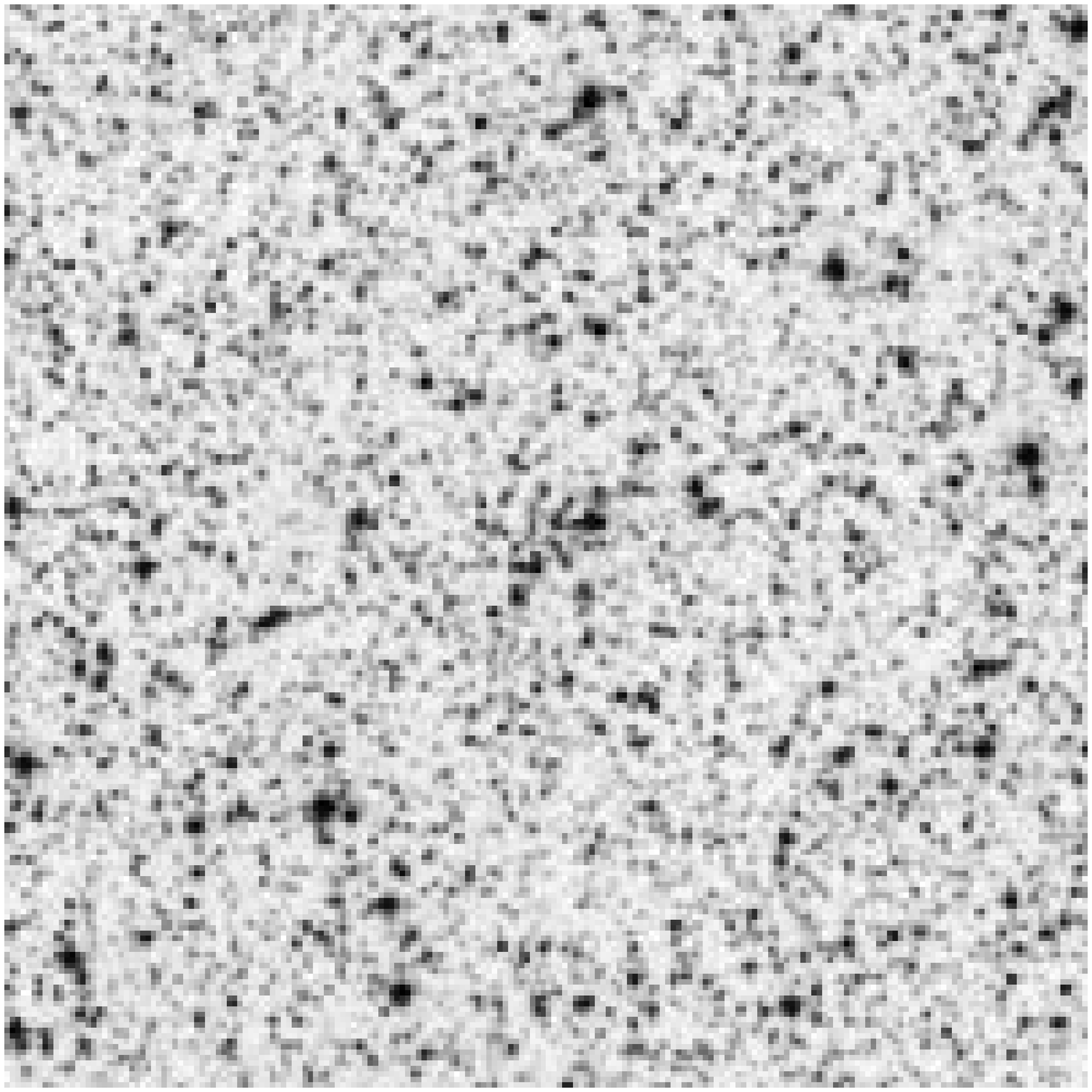}
}
\enskip
\subfigure[][\object{Kronberger~4}]{
\includegraphics[width=8.75cm]{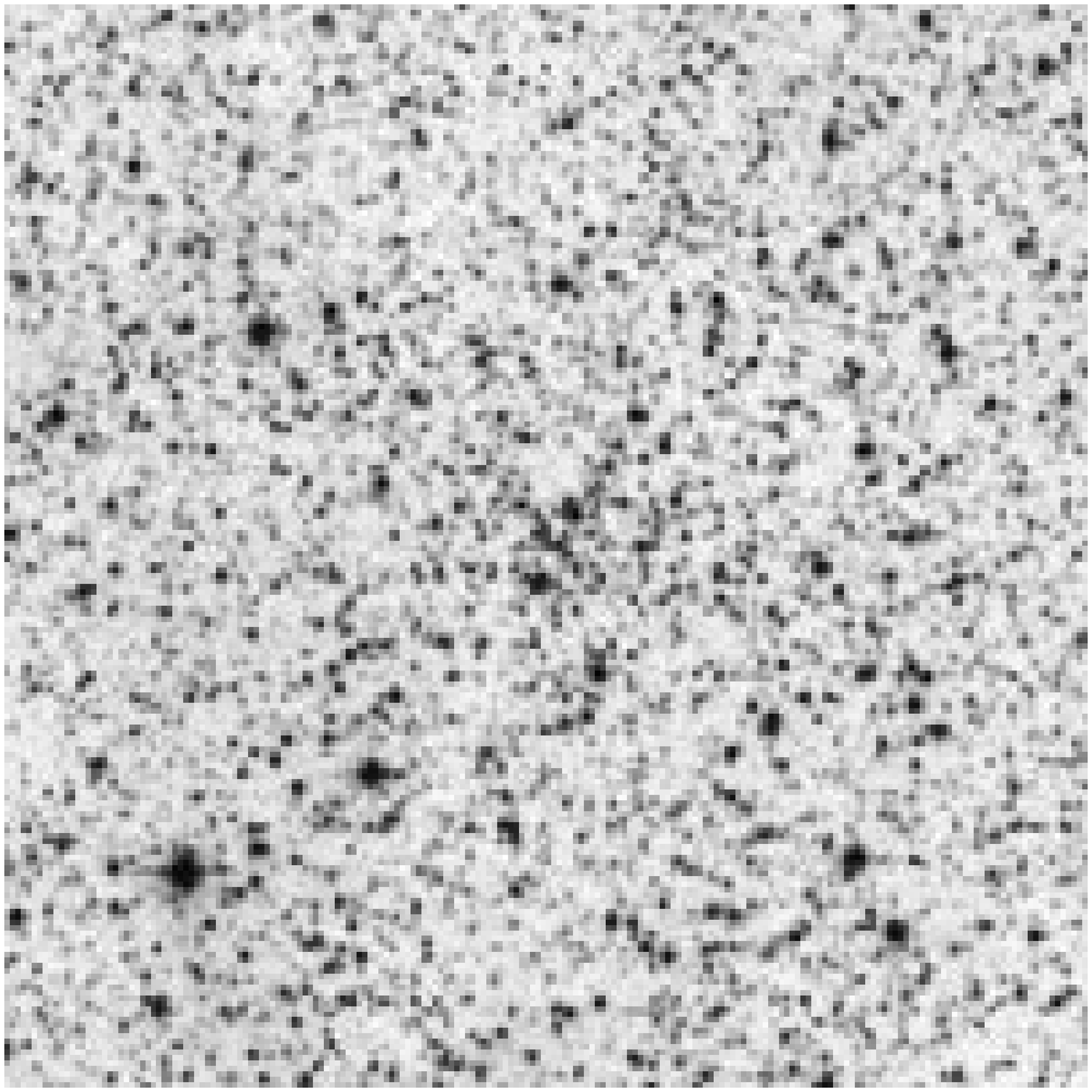}
}

\end{figure*}



\setcounter{figure}{2}
\setcounter{subfigure}{0}
\begin{figure*}
\caption[]{Second generation DSS red images of the cluster candidates presented in
 			Table~\subref{TabNo}. The field of view is in each case $8\arcmin$ $\times$
			$8\arcmin$. North is up and East is left.}
\label{fig3}

\subfigure[][\object{Teutsch~49}]{
\includegraphics[width=8.75cm]{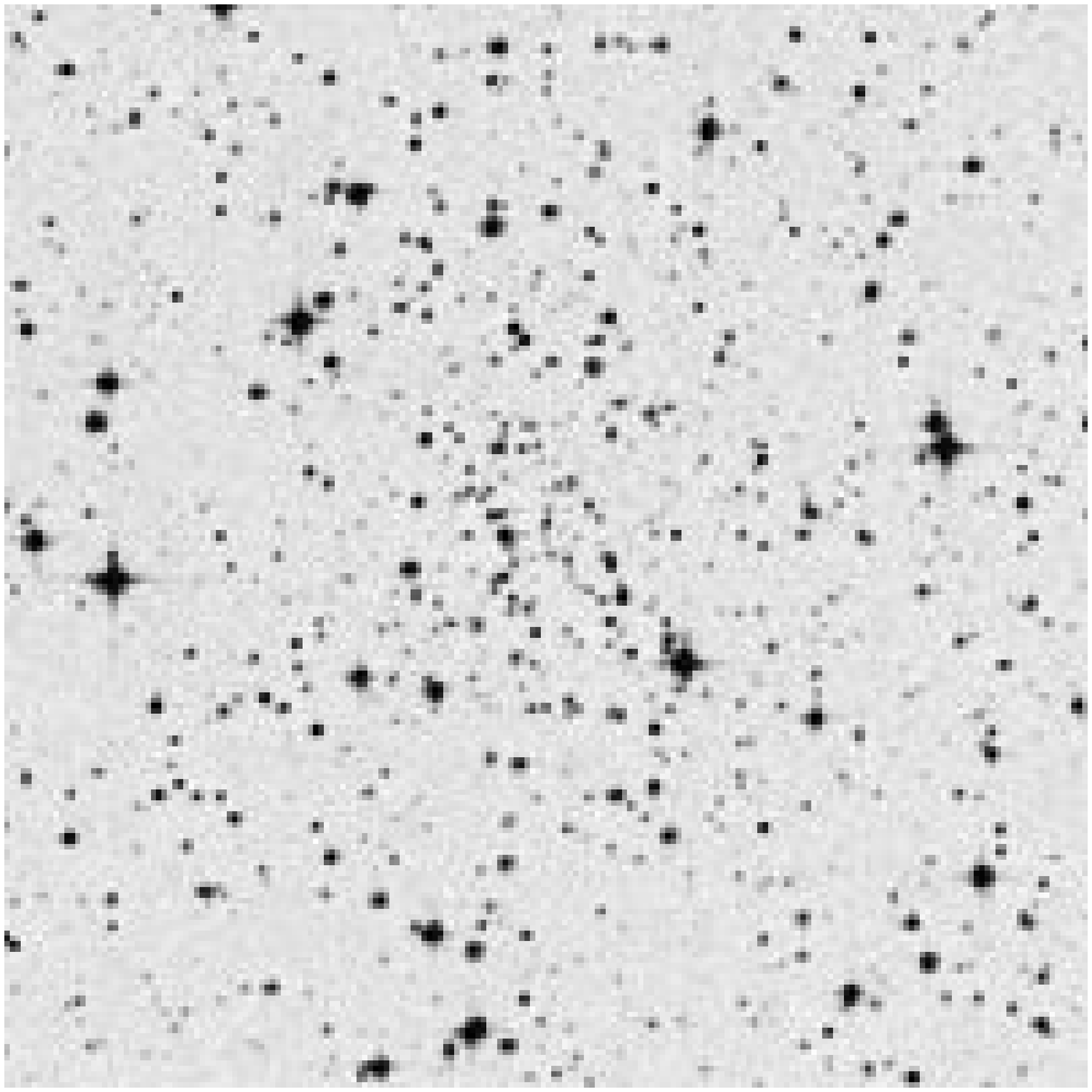}
}
\enskip
\subfigure[][\object{Teutsch~31}]{
\includegraphics[width=8.75cm]{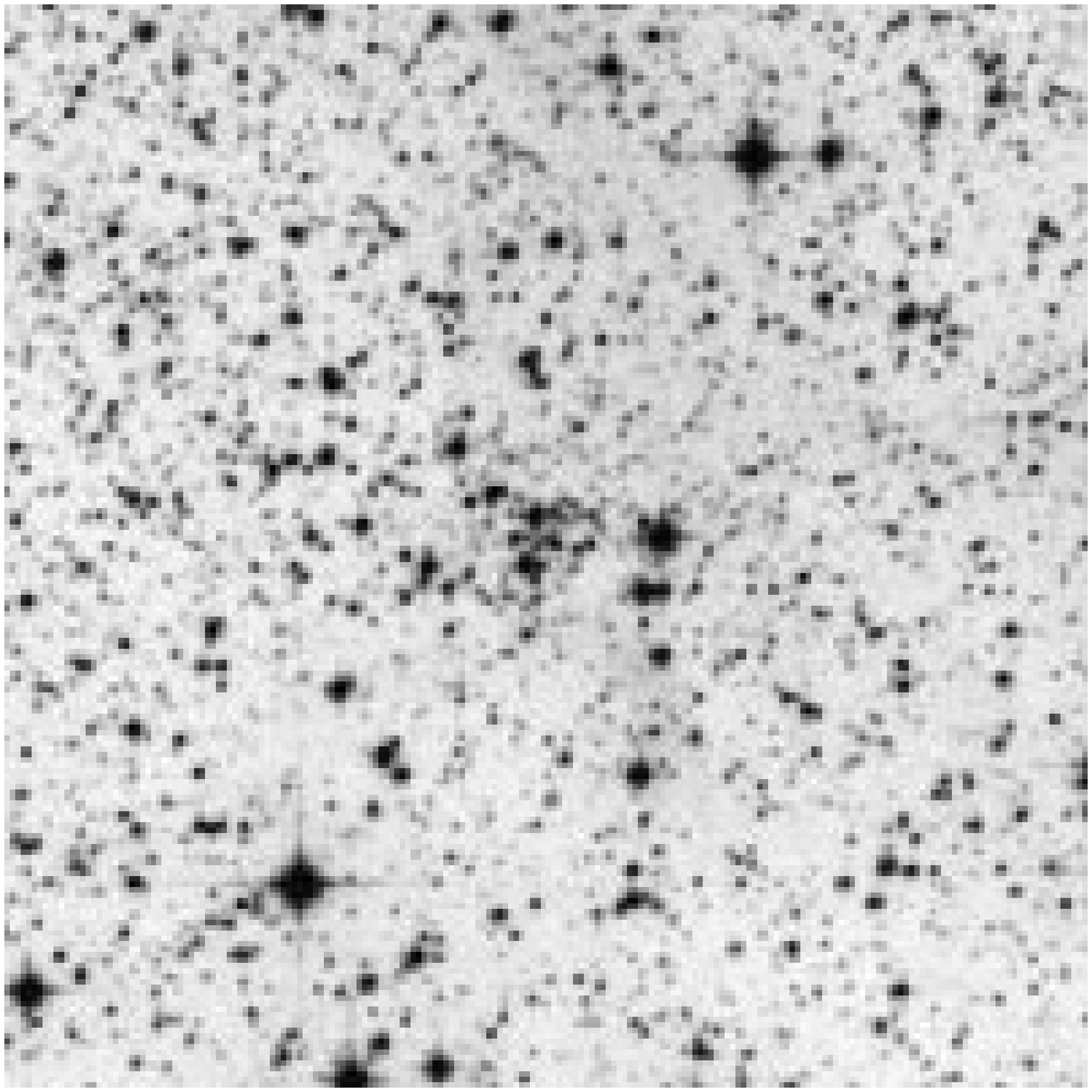}
}

\subfigure[][\object{Kronberger~25}]{
\includegraphics[width=8.75cm]{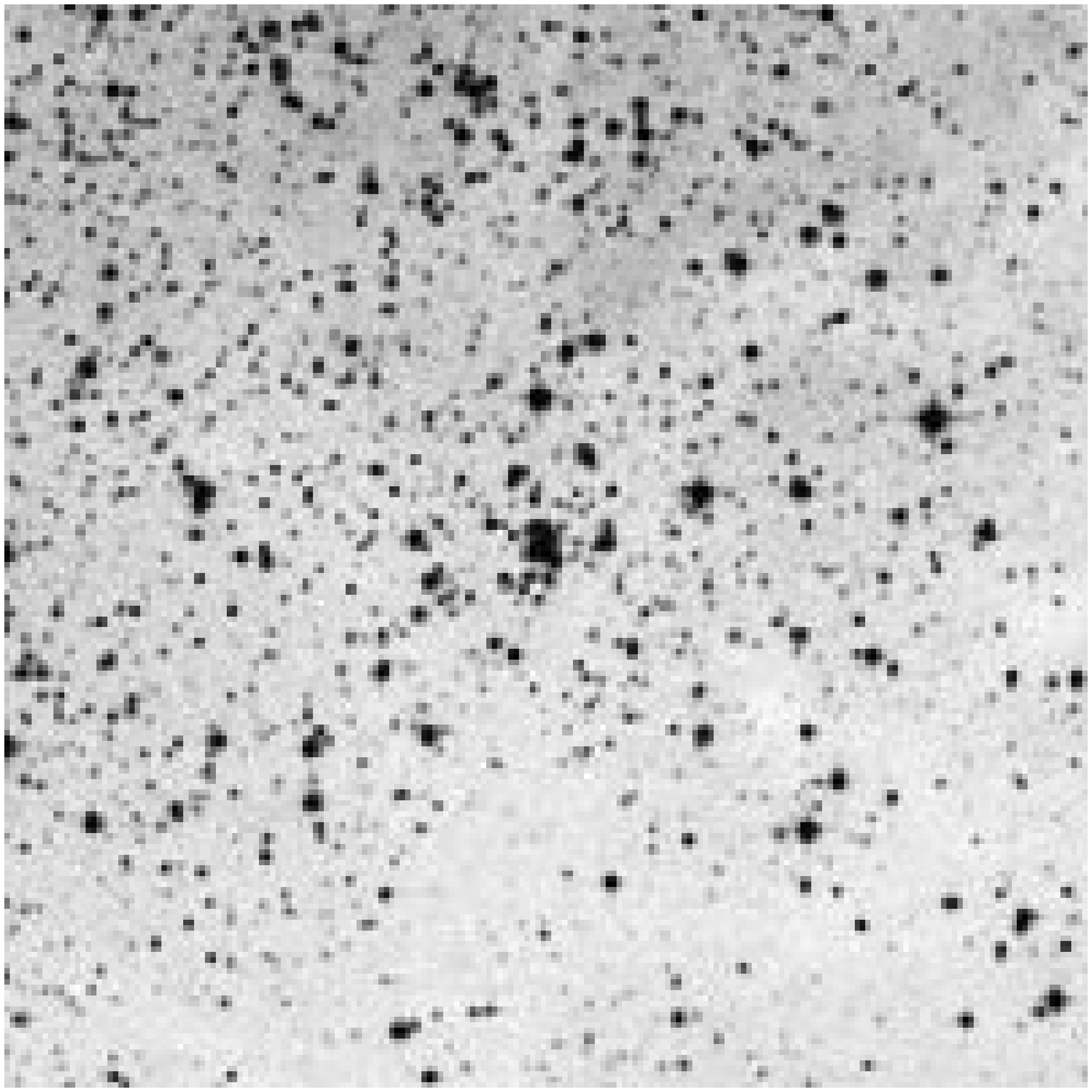}
}
\enskip
\subfigure[][\object{Kronberger~54}]{
\includegraphics[width=8.75cm]{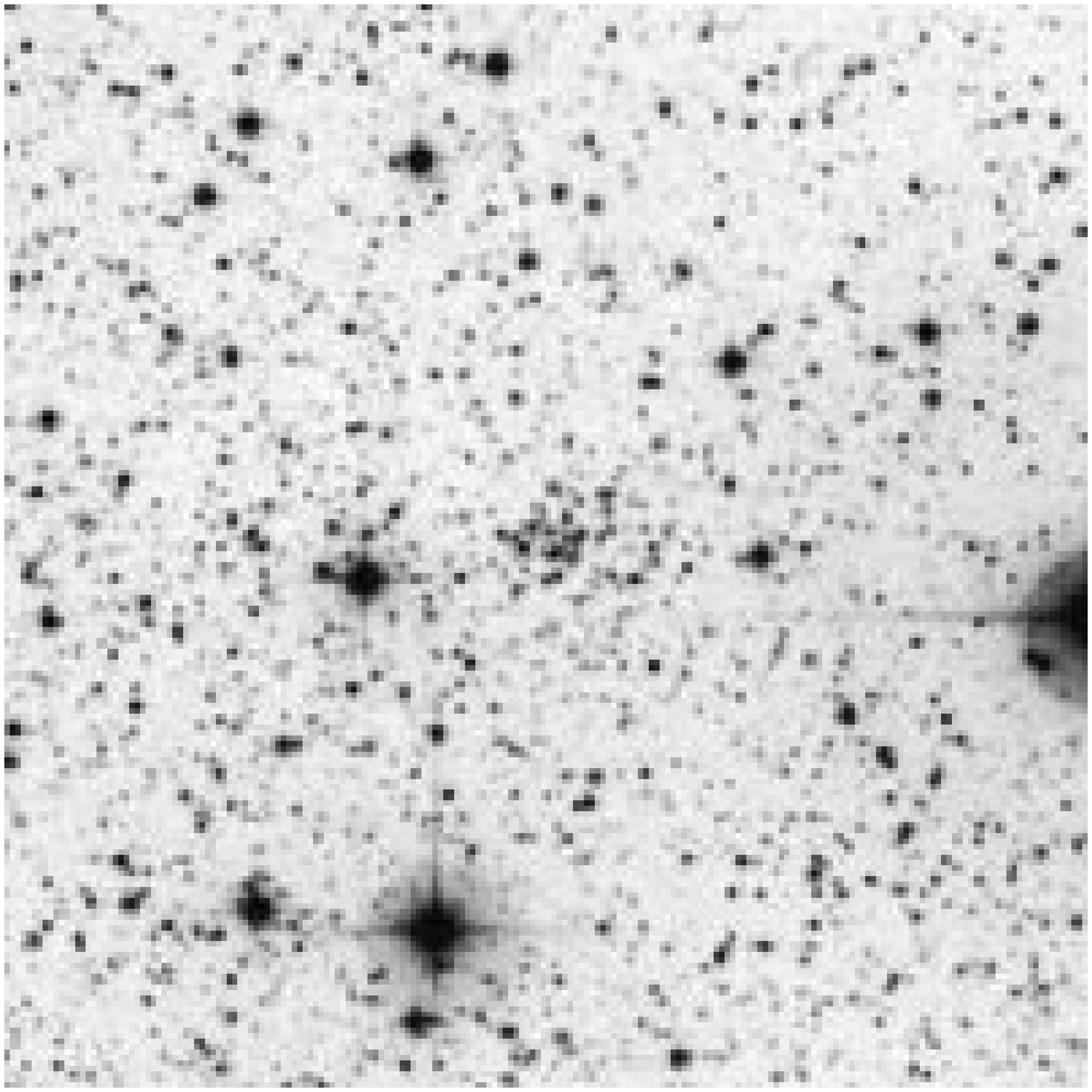}
}

\end{figure*}


\setcounter{figure}{2}
\setcounter{subfigure}{0}
\begin{figure*}
\caption[]{(cont.)}

\subfigure[][\object{Kronberger~80}]{
\includegraphics[width=8.75cm]{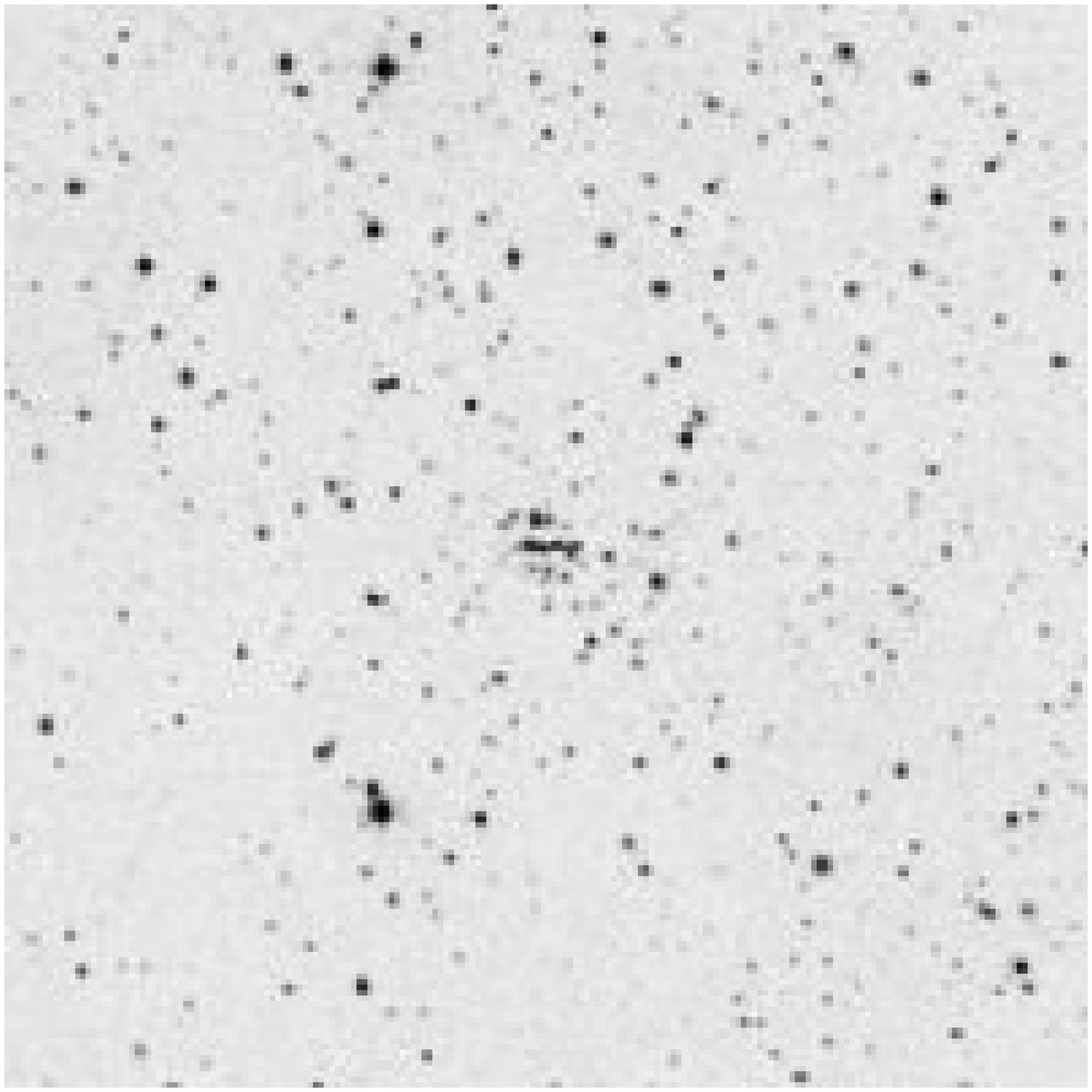}
}

\end{figure*}

\clearpage




\setcounter{figure}{3}
\begin{figure*}
\caption[]{RDPs of the cluster candidates presented in Table~\subref{TabIso}, with the vertical scale normalized to the maximum stellar density in the cluster field. In each diagram, the maximum stellar density (in $\mathrm{stars/arcmin^2}$) and, in brackets, the actual number of stars observed in this area are given. The mean density of the background is shown in each diagram as a horizontal line, with the $2\sigma$ error ranges of the background indicated as shaded areas. Only 2MASS sources with $H < 15.5$ were taken into account.}
\label{fig4}

\includegraphics[width=8.75cm]{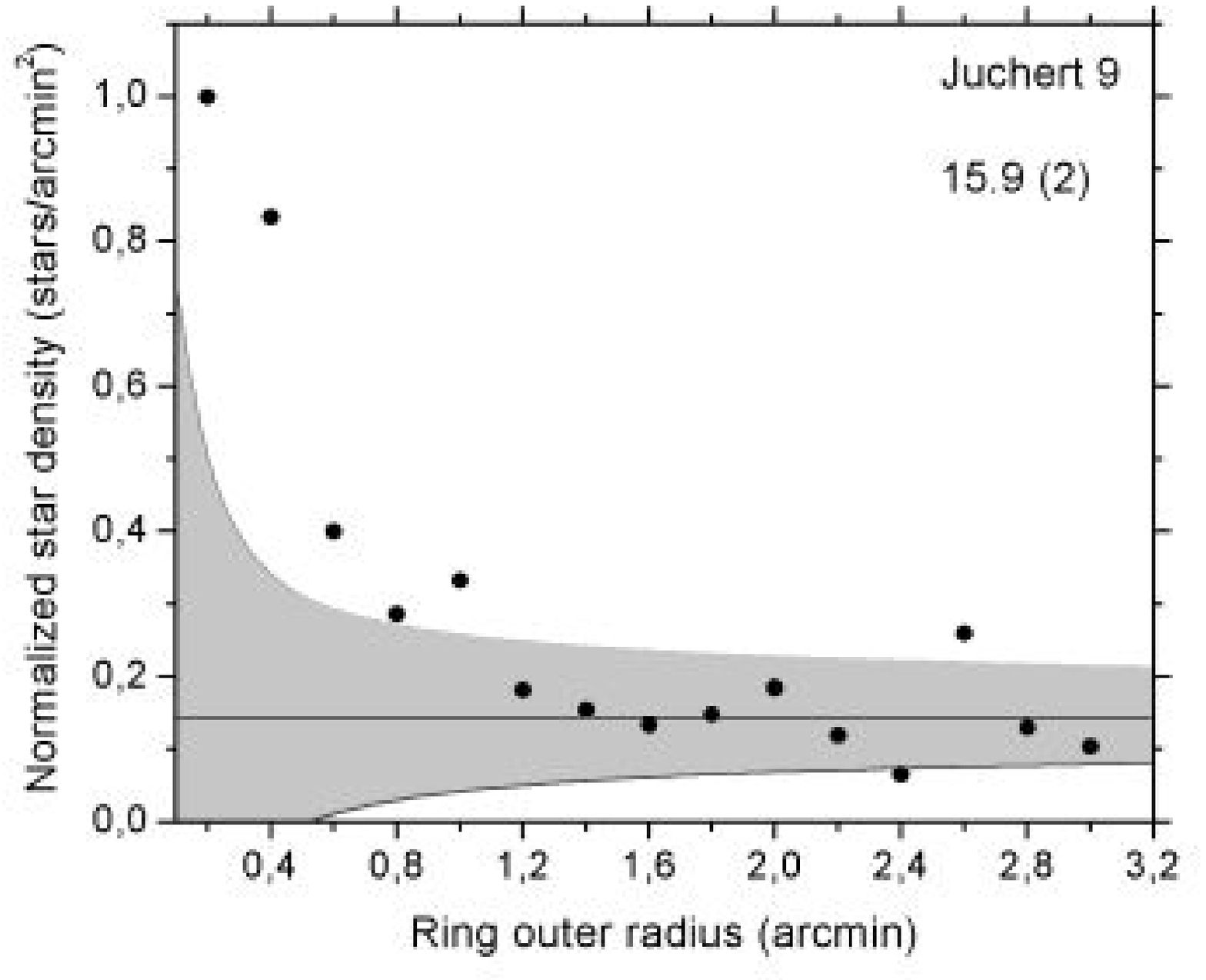}
\enskip
\includegraphics[width=8.75cm]{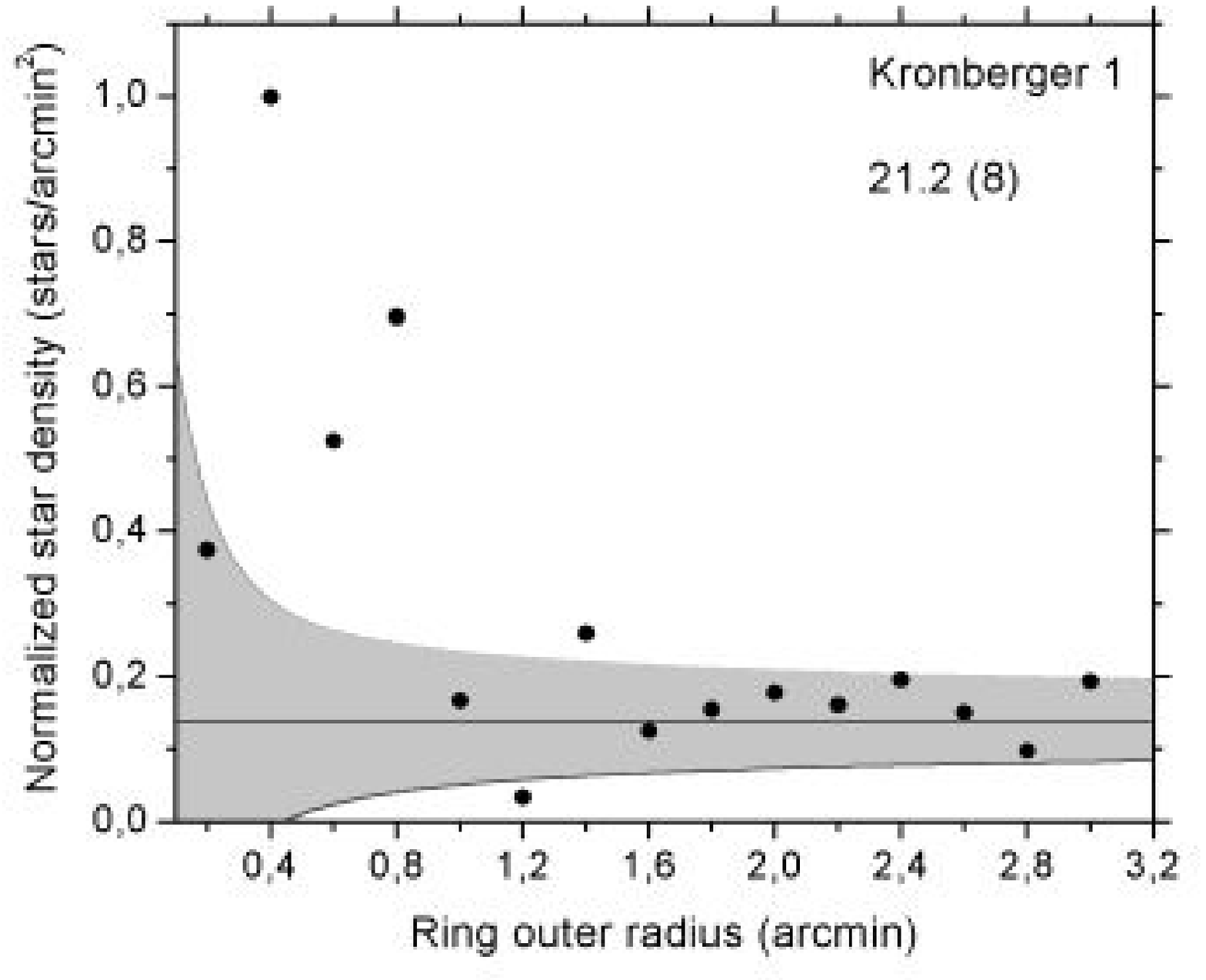}

\includegraphics[width=8.75cm]{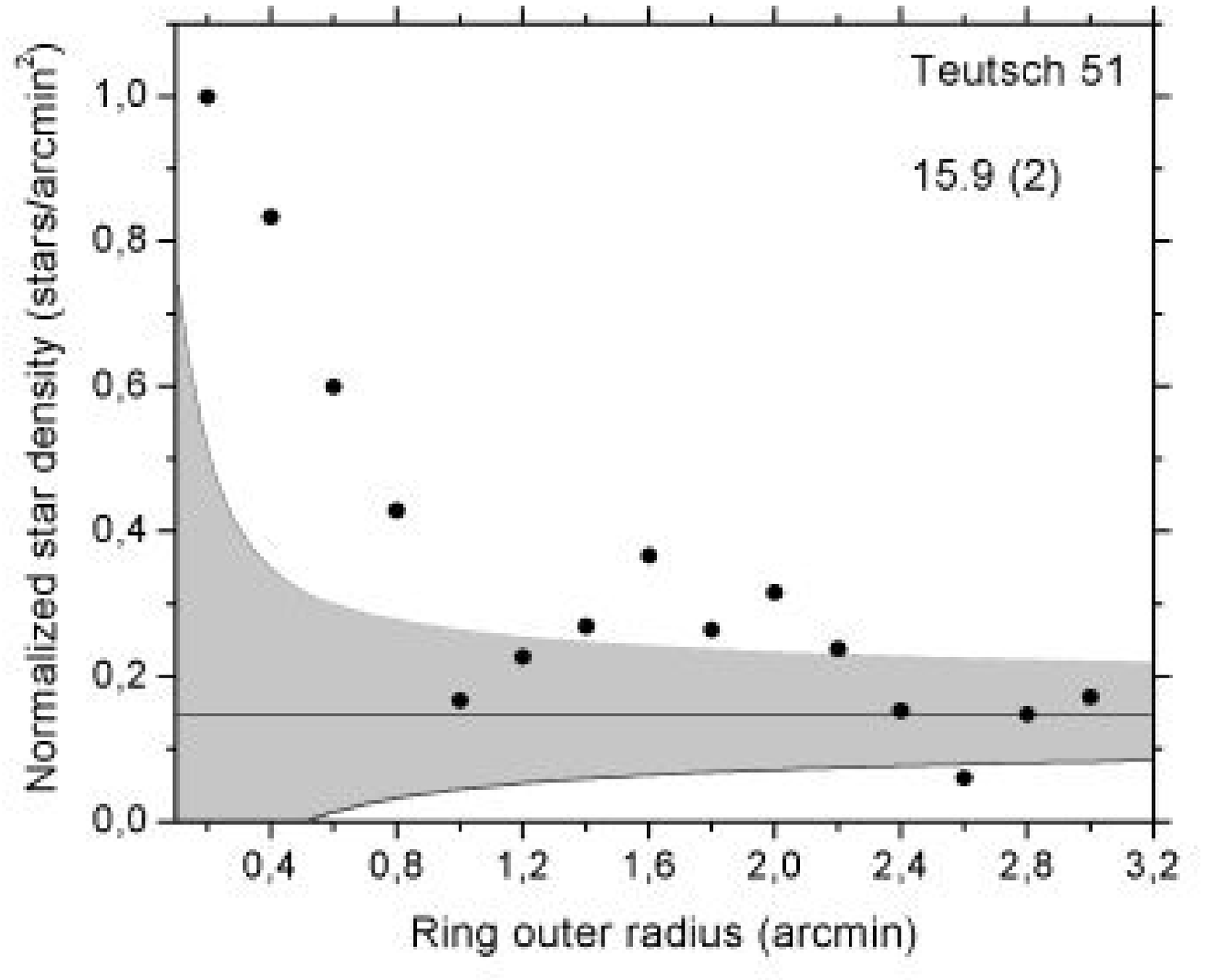}
\enskip
\includegraphics[width=8.75cm]{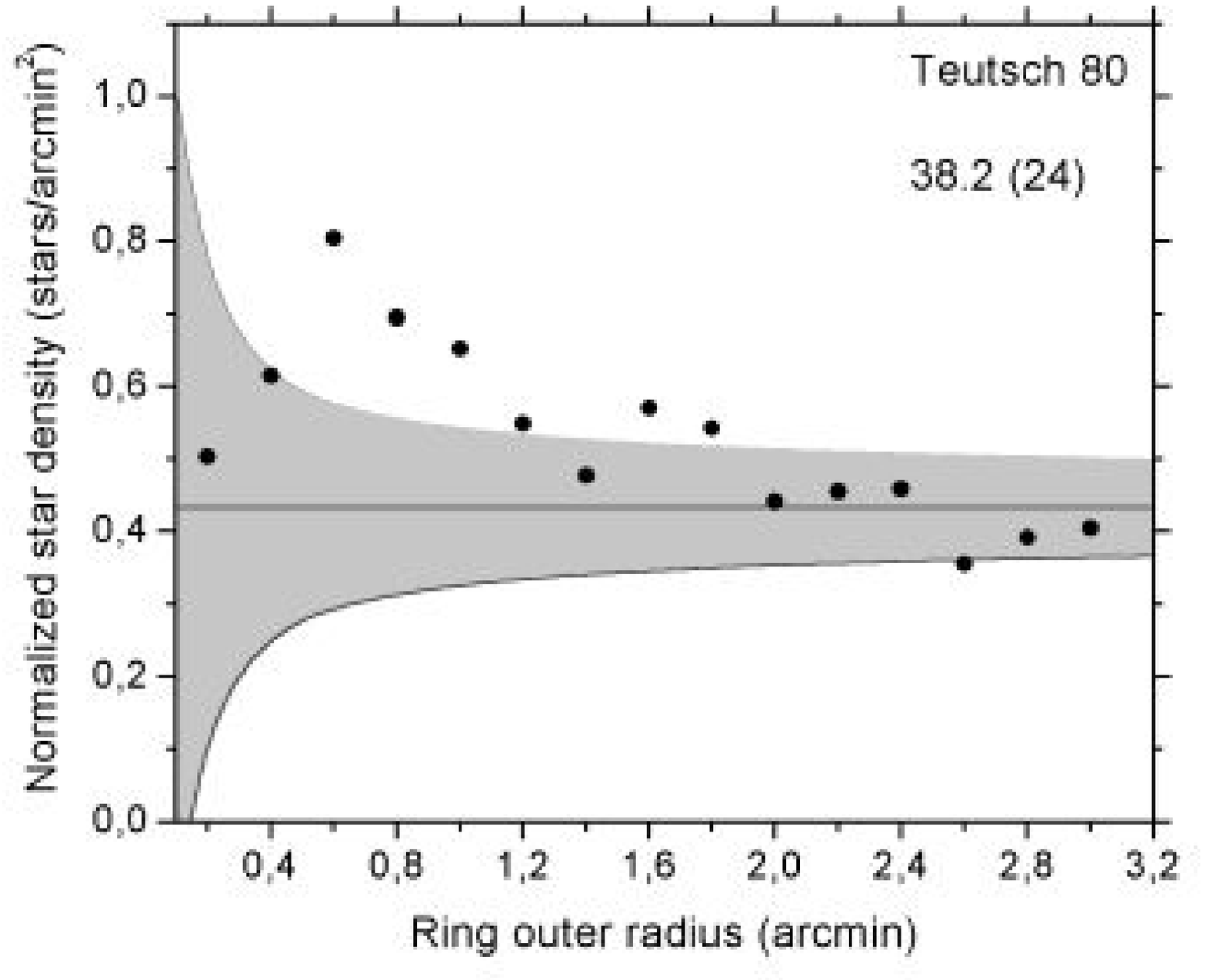}

\includegraphics[width=8.75cm]{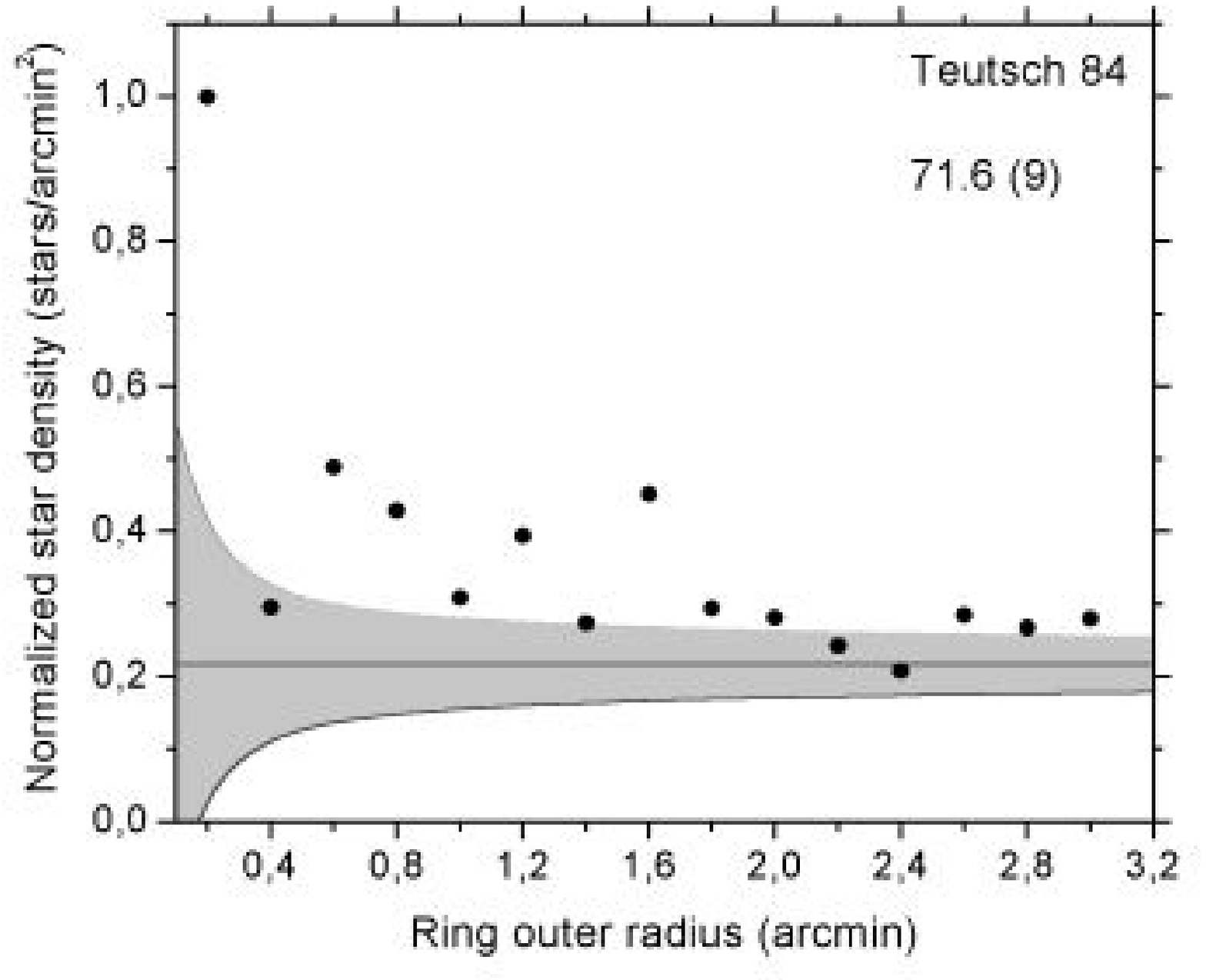}
\enskip
\includegraphics[width=8.75cm]{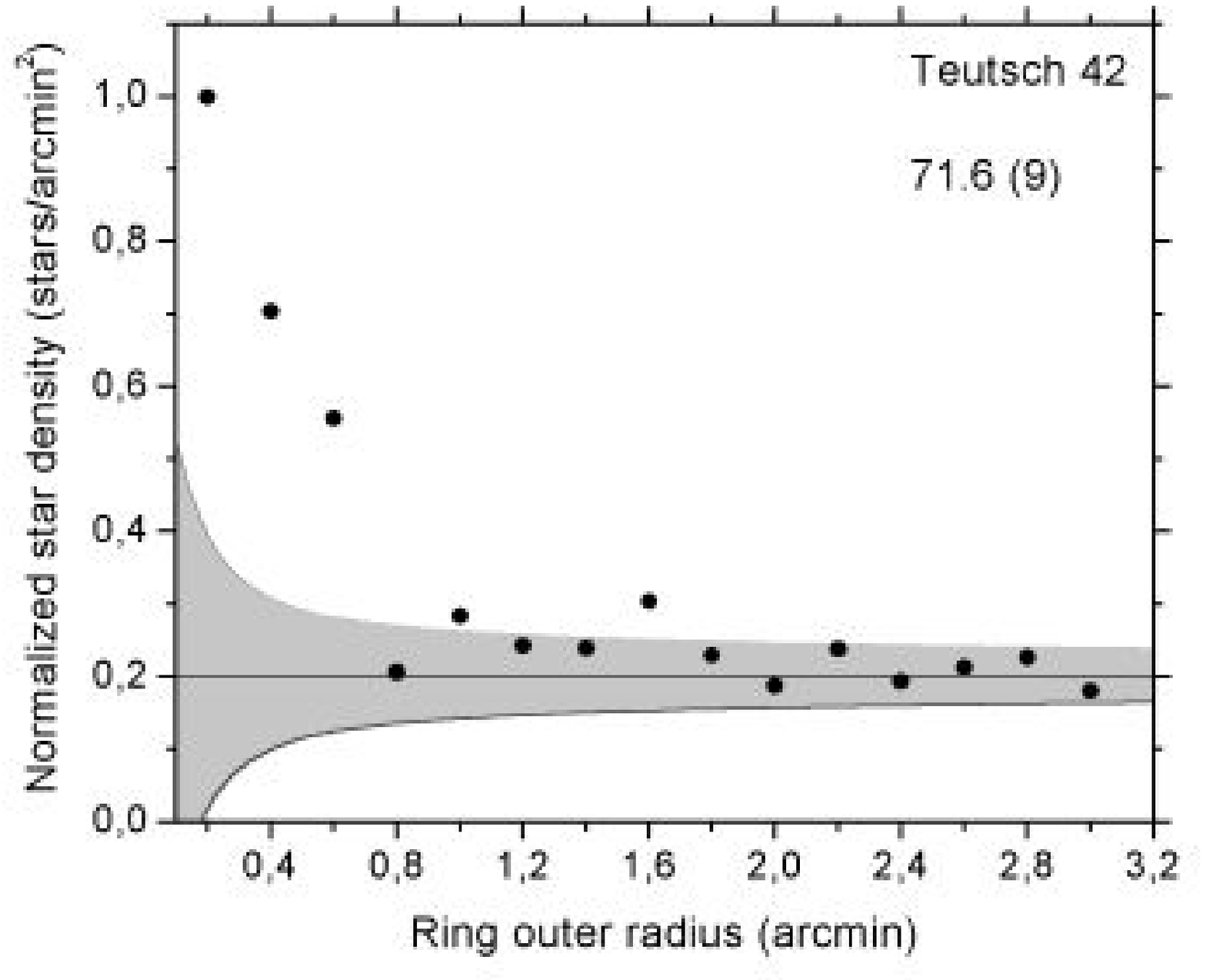}

\end{figure*}


\setcounter{figure}{3}
\begin{figure*}
\caption[]{(cont.)}

\includegraphics[width=8.75cm]{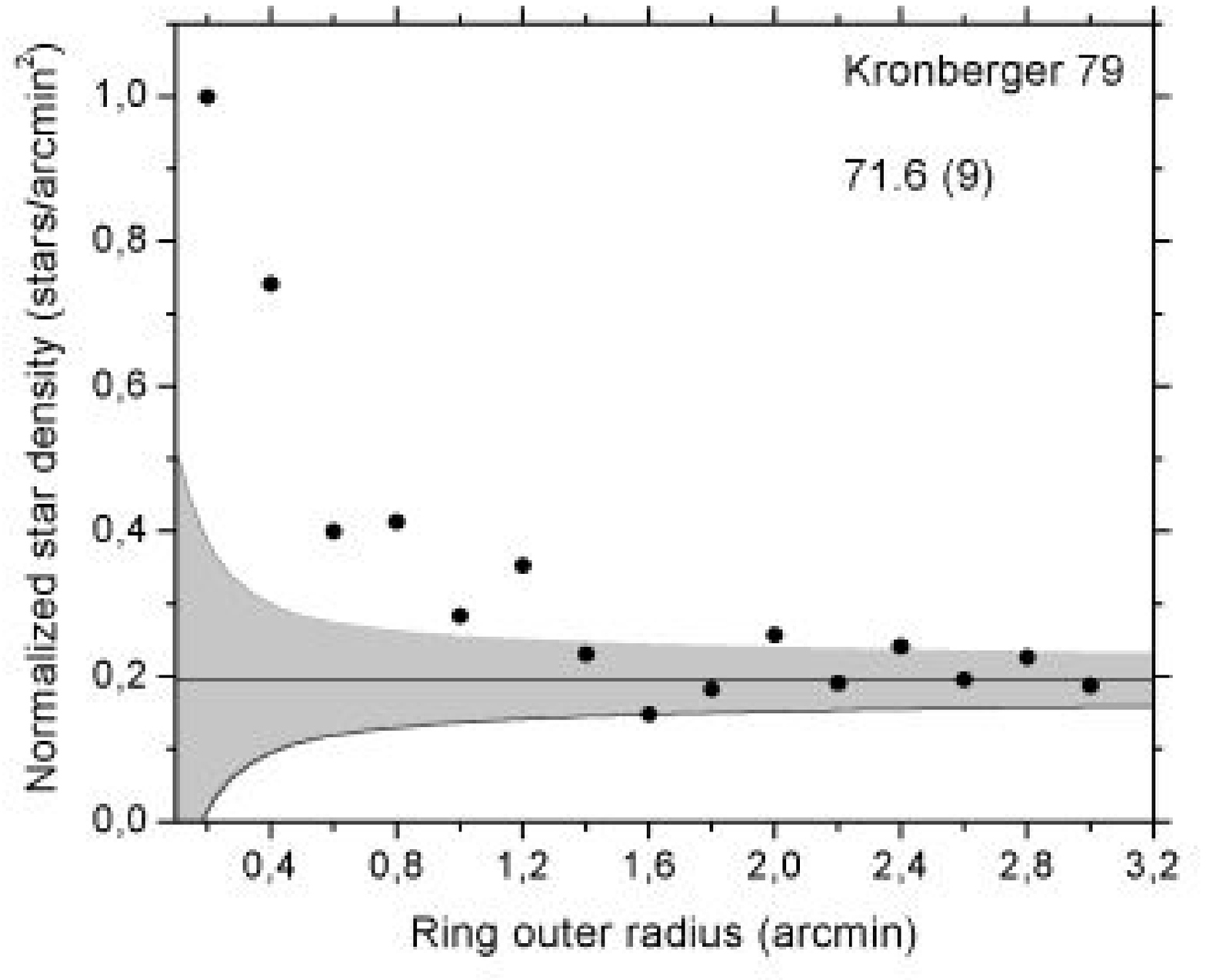}
\enskip
\includegraphics[width=8.75cm]{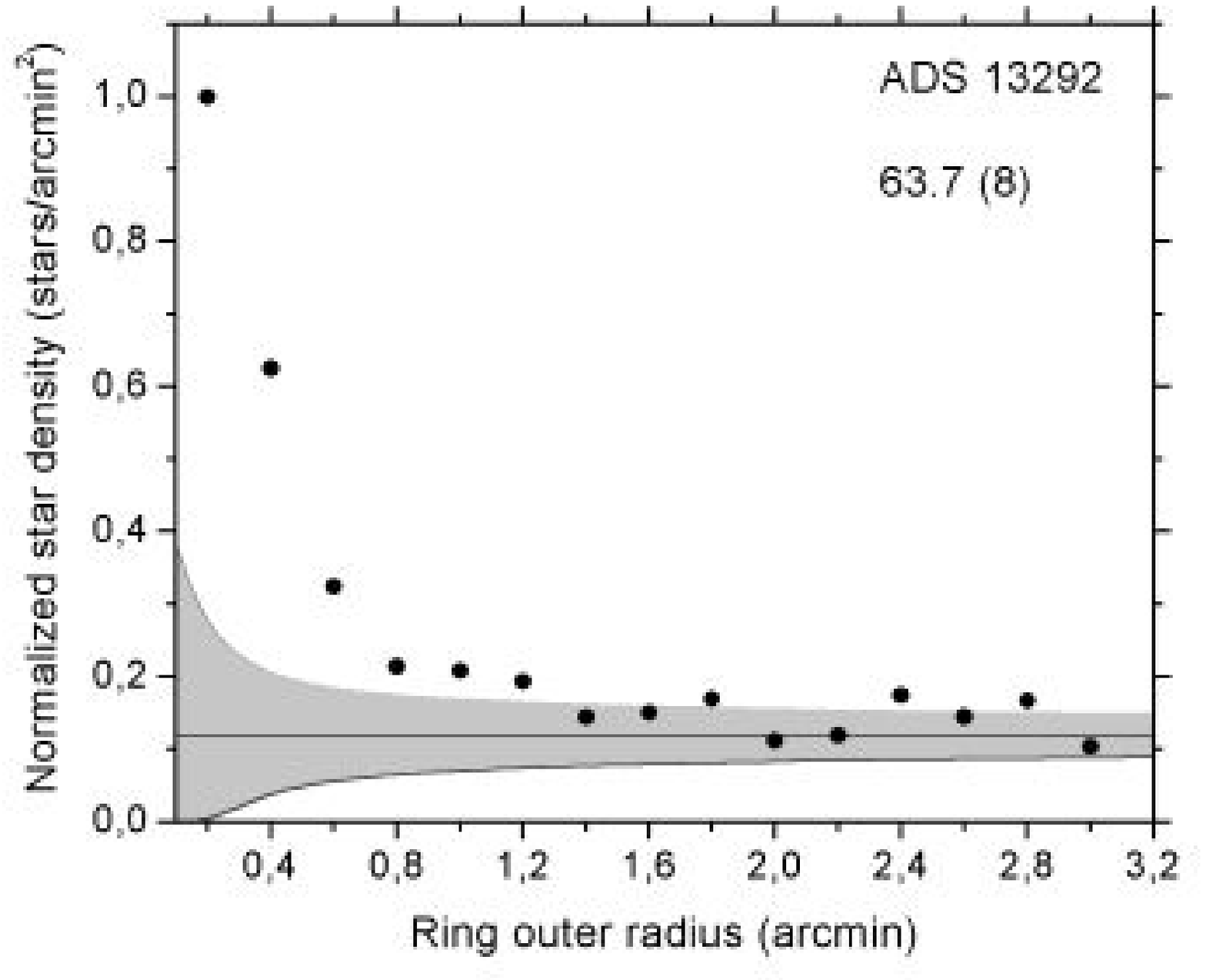}

\includegraphics[width=8.75cm]{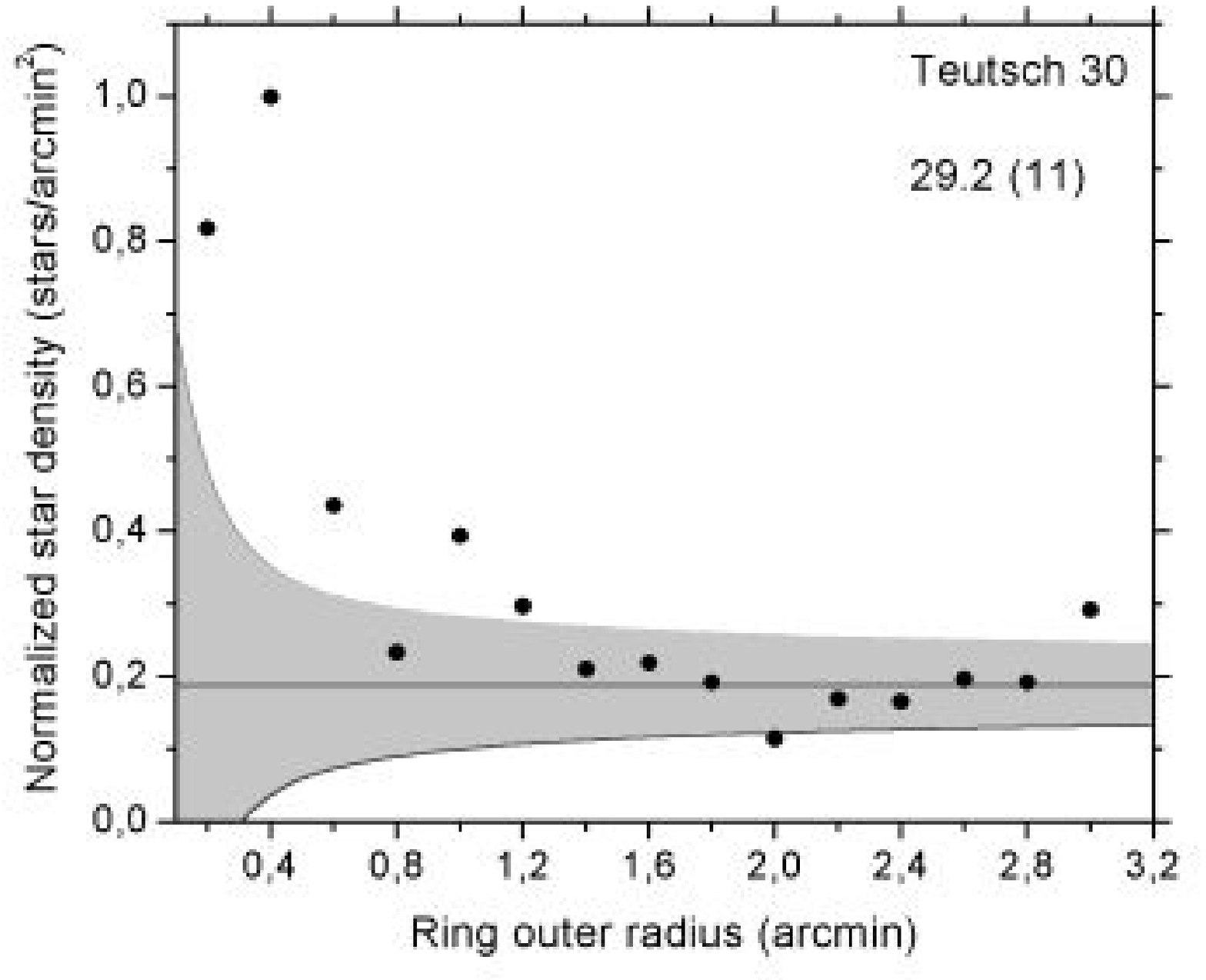}

\end{figure*}



\setcounter{figure}{4}
\begin{figure*}
\caption[]{RDPs of the cluster candidates presented in Table~\subref{TabRC}, with the vertical scale normalized to the maximum stellar density in the cluster field. In each diagram, the maximum stellar density (in $\mathrm{stars/arcmin^2}$) and, in brackets, the actual number of stars observed in this area are given. The mean density of the background is shown in each diagram as a horizontal line, with the $2\sigma$ error ranges of the background indicated as shaded areas. Only 2MASS sources with $H < 15.5$ were taken into account.}
\label{fig5}

\includegraphics[width=8.75cm]{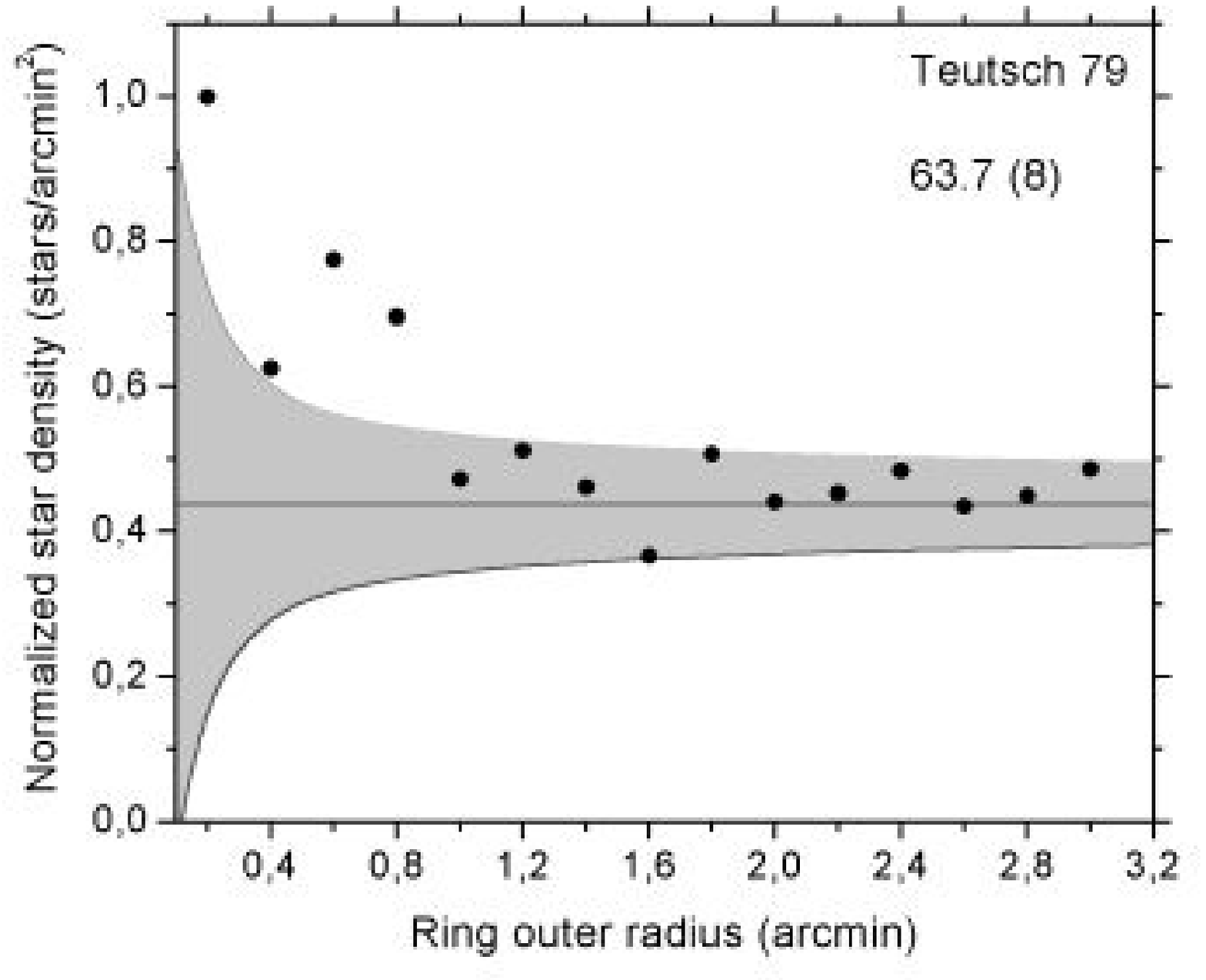}
\enskip
\includegraphics[width=8.75cm]{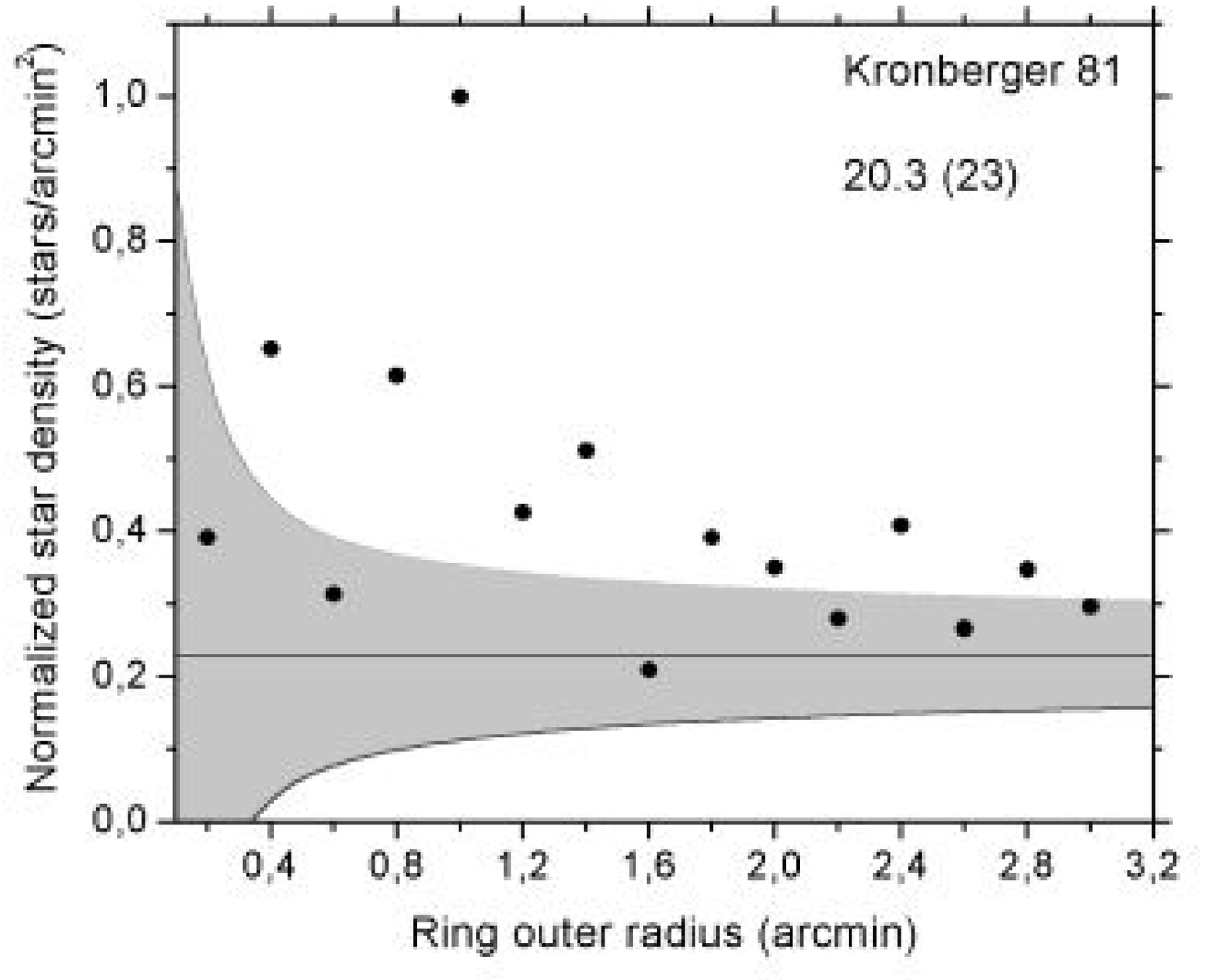}

\includegraphics[width=8.75cm]{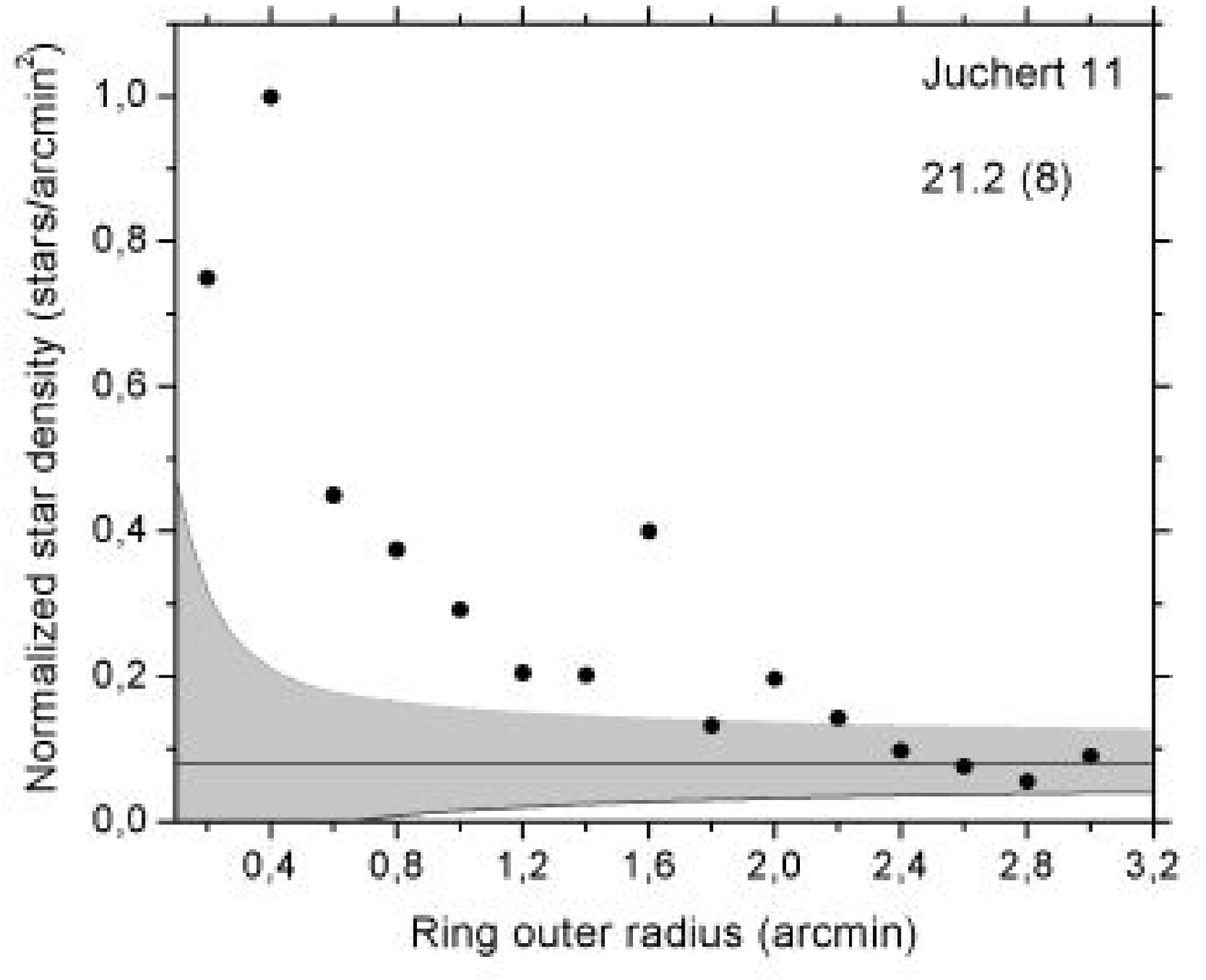}
\enskip
\includegraphics[width=8.75cm]{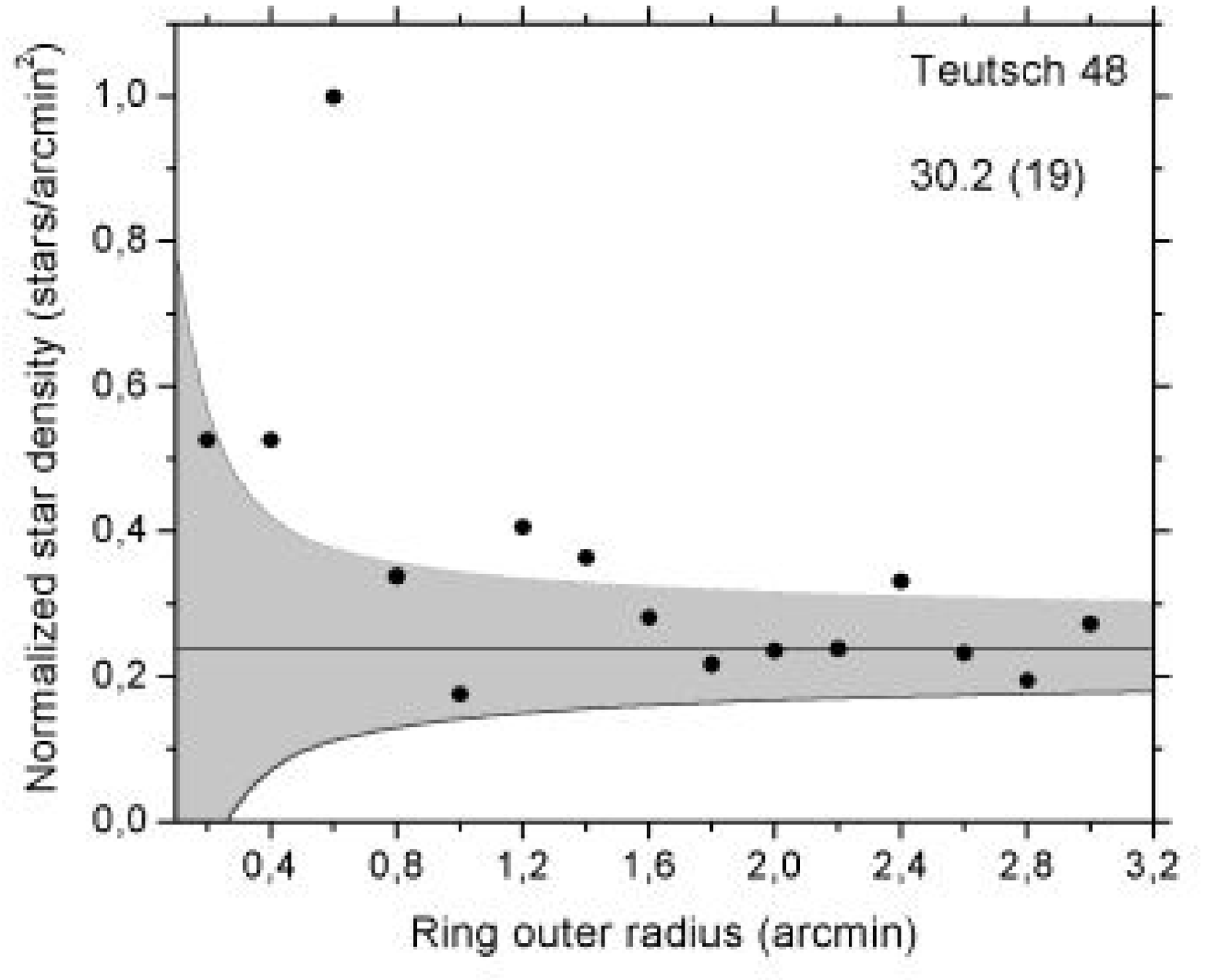}

\includegraphics[width=8.75cm]{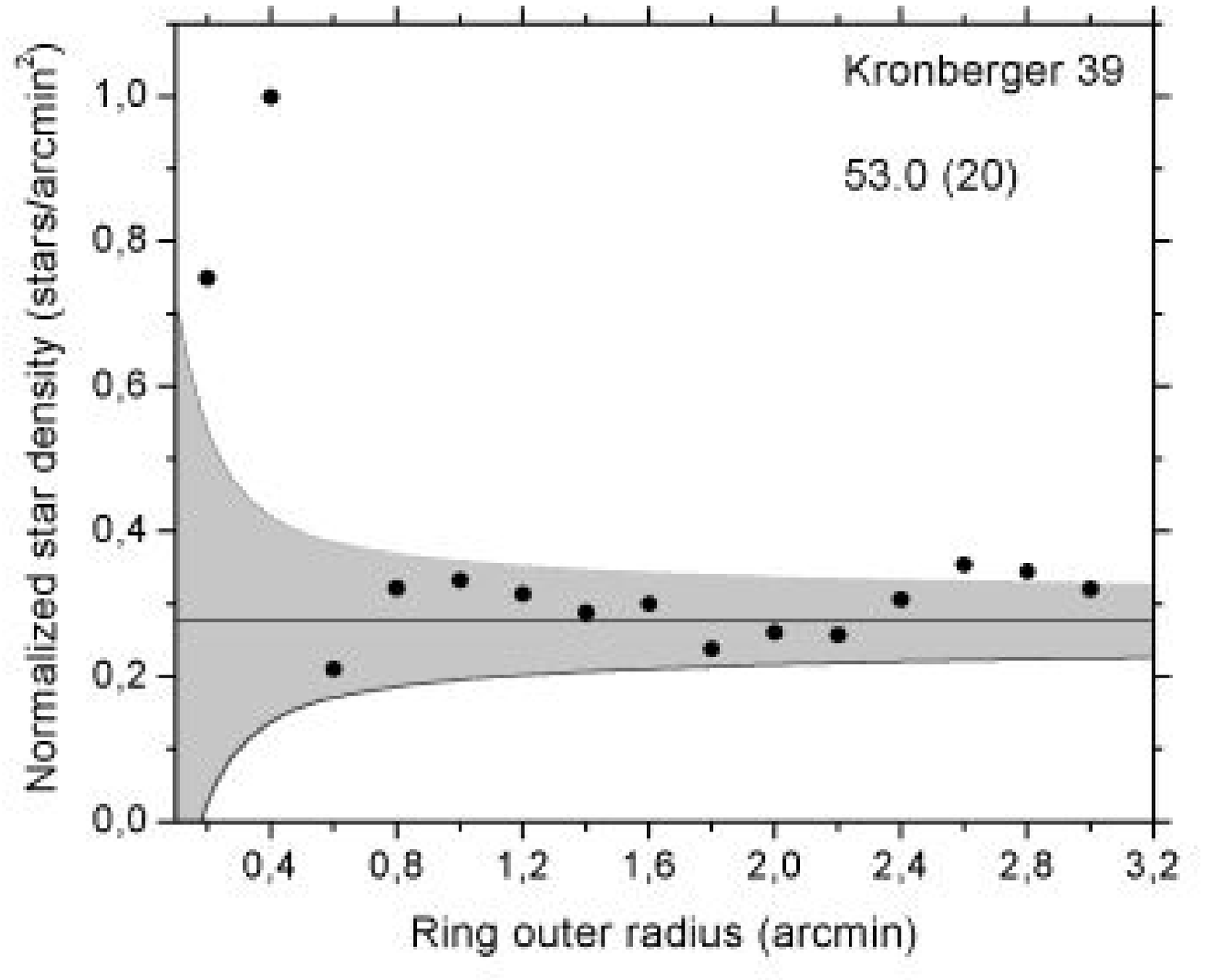}
\enskip
\includegraphics[width=8.75cm]{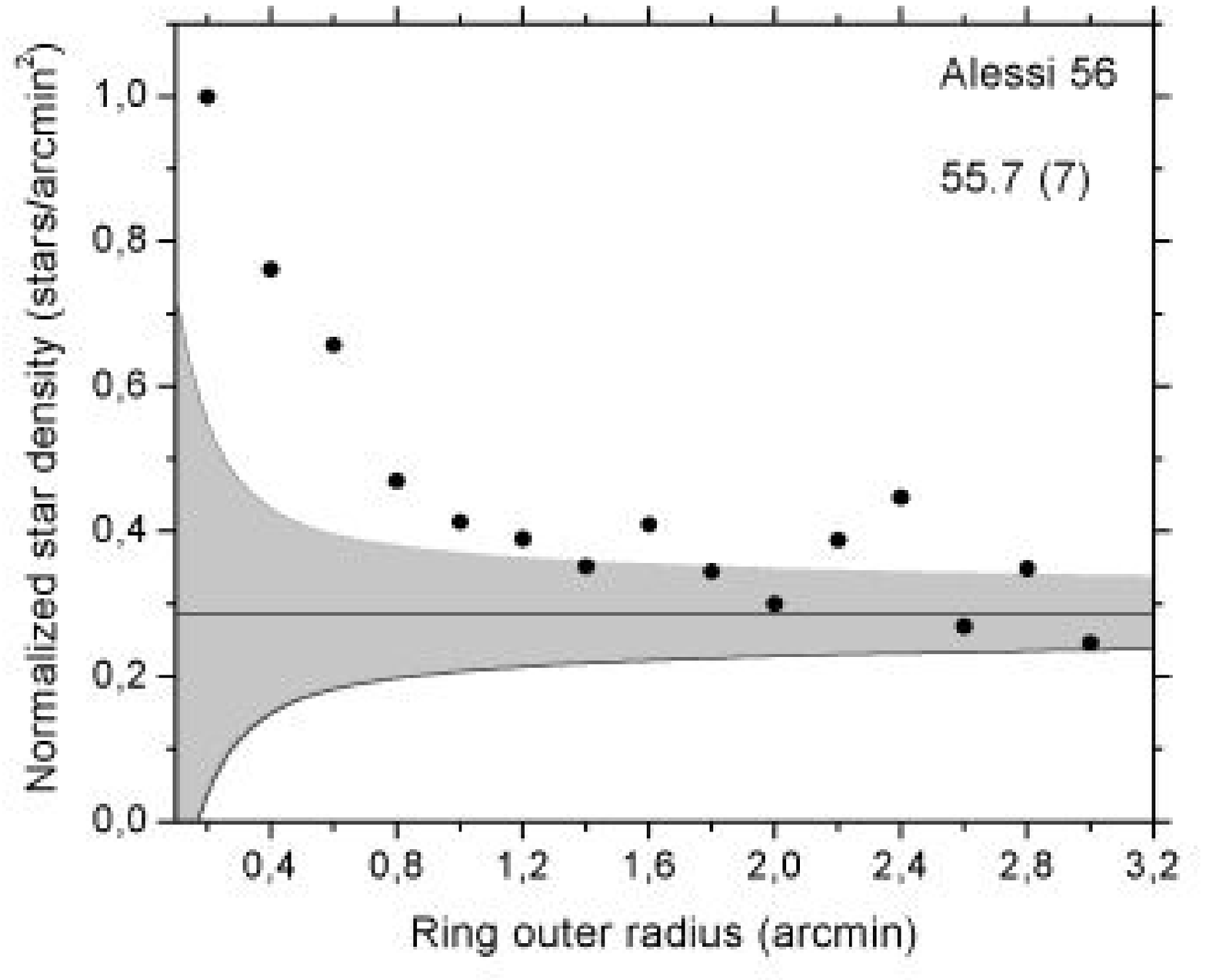}

\end{figure*}


\setcounter{figure}{4}
\begin{figure*}
\caption[]{(cont.)}

\includegraphics[width=8.75cm]{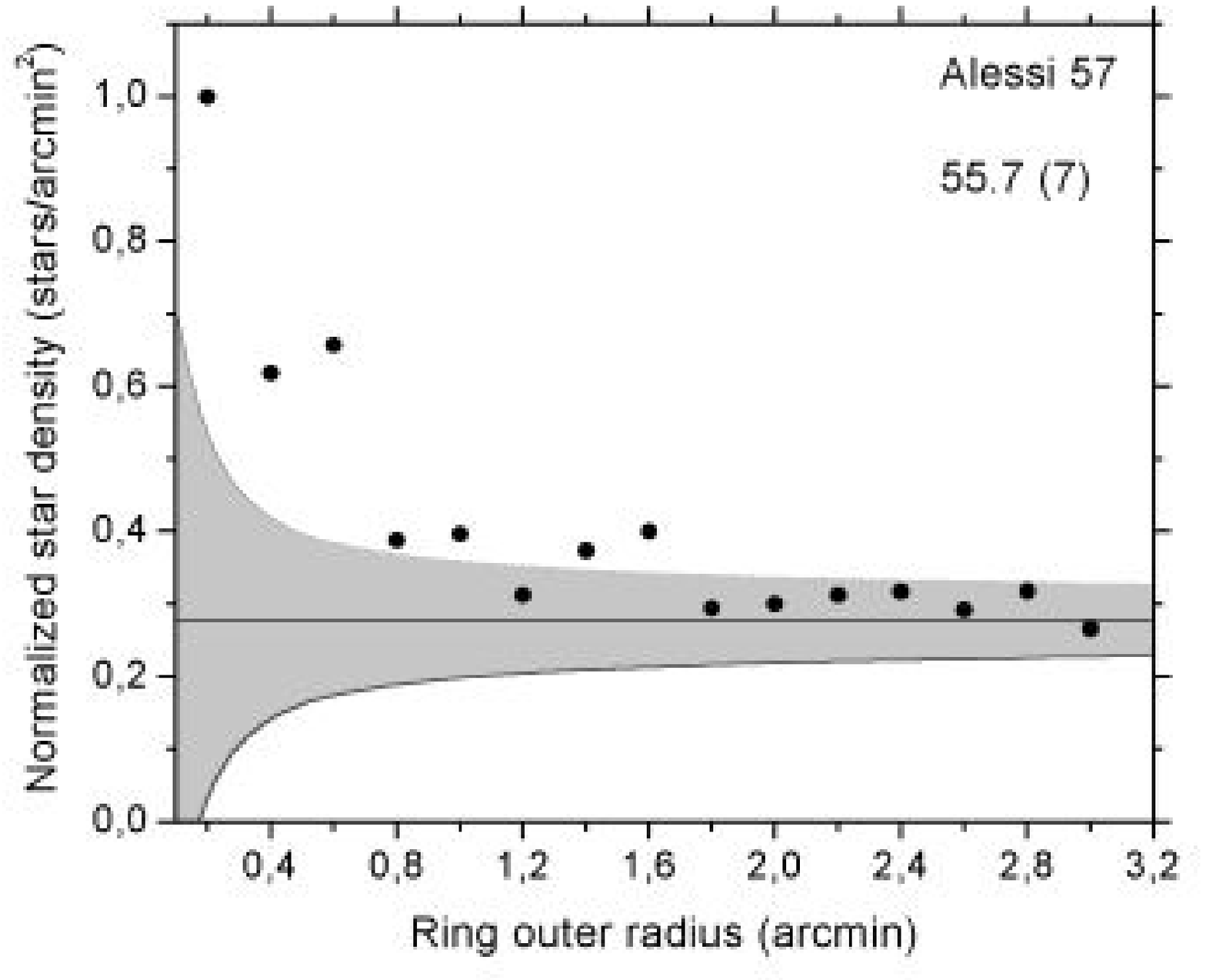}
\enskip
\includegraphics[width=8.75cm]{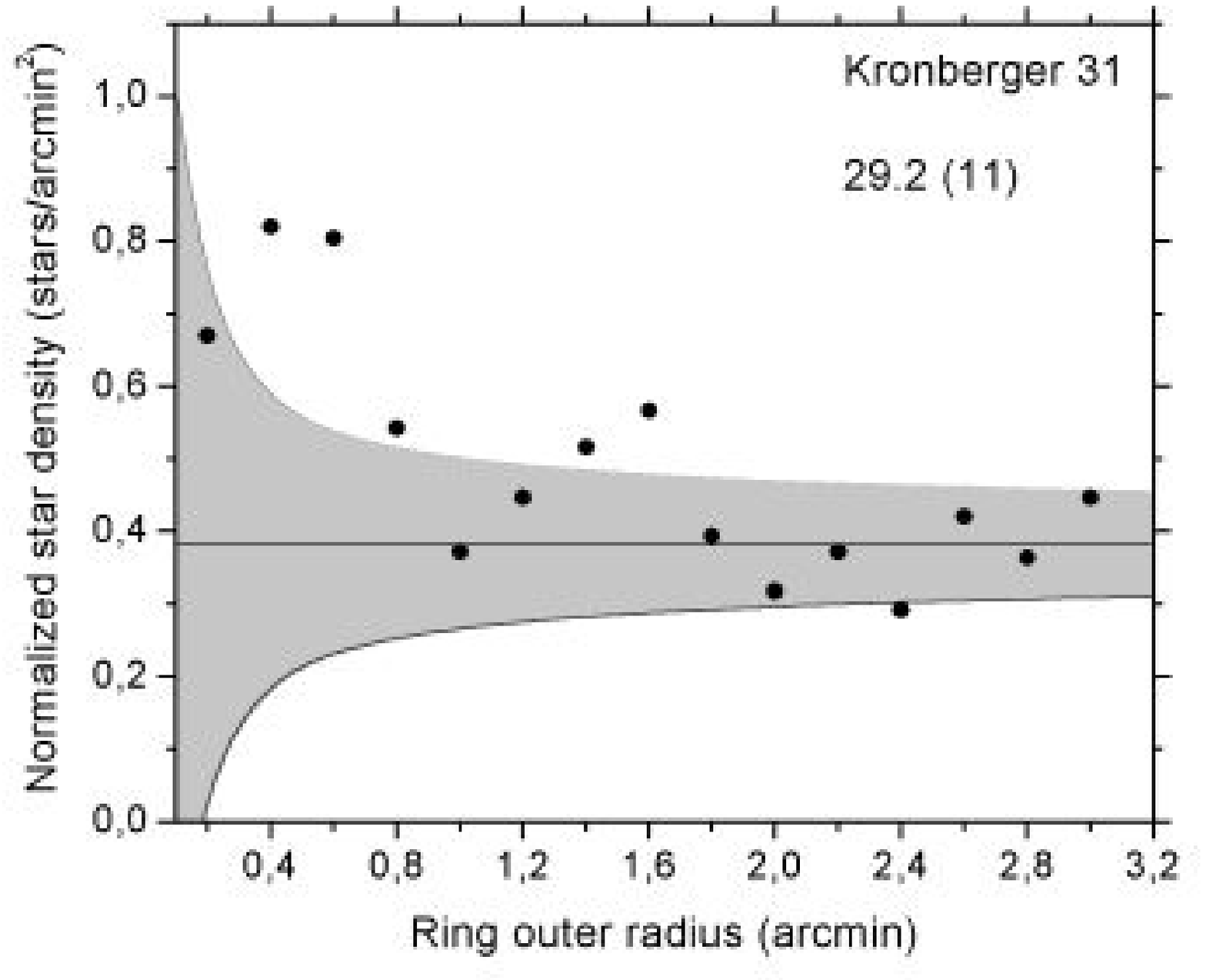}

\includegraphics[width=8.75cm]{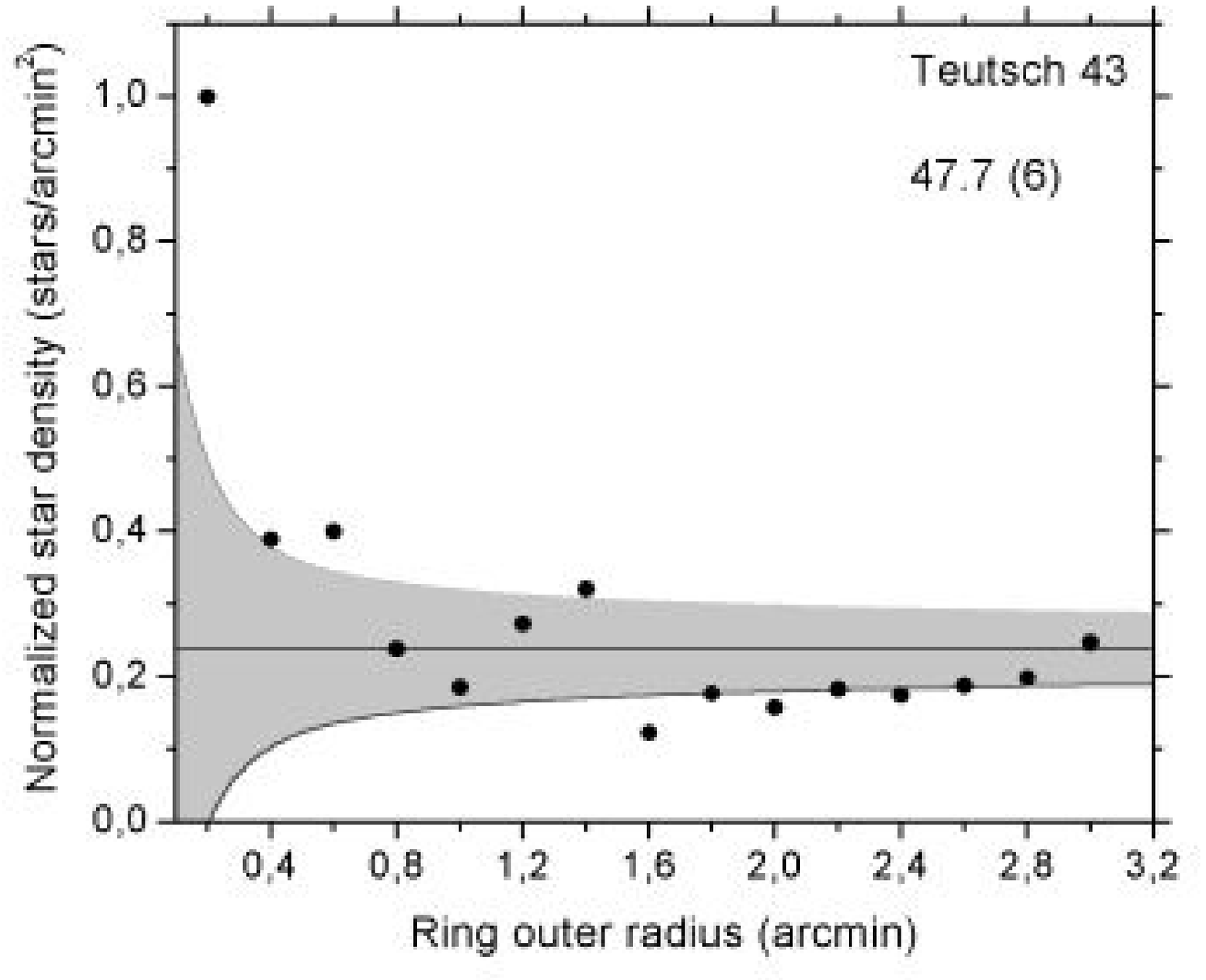}
\enskip
\includegraphics[width=8.75cm]{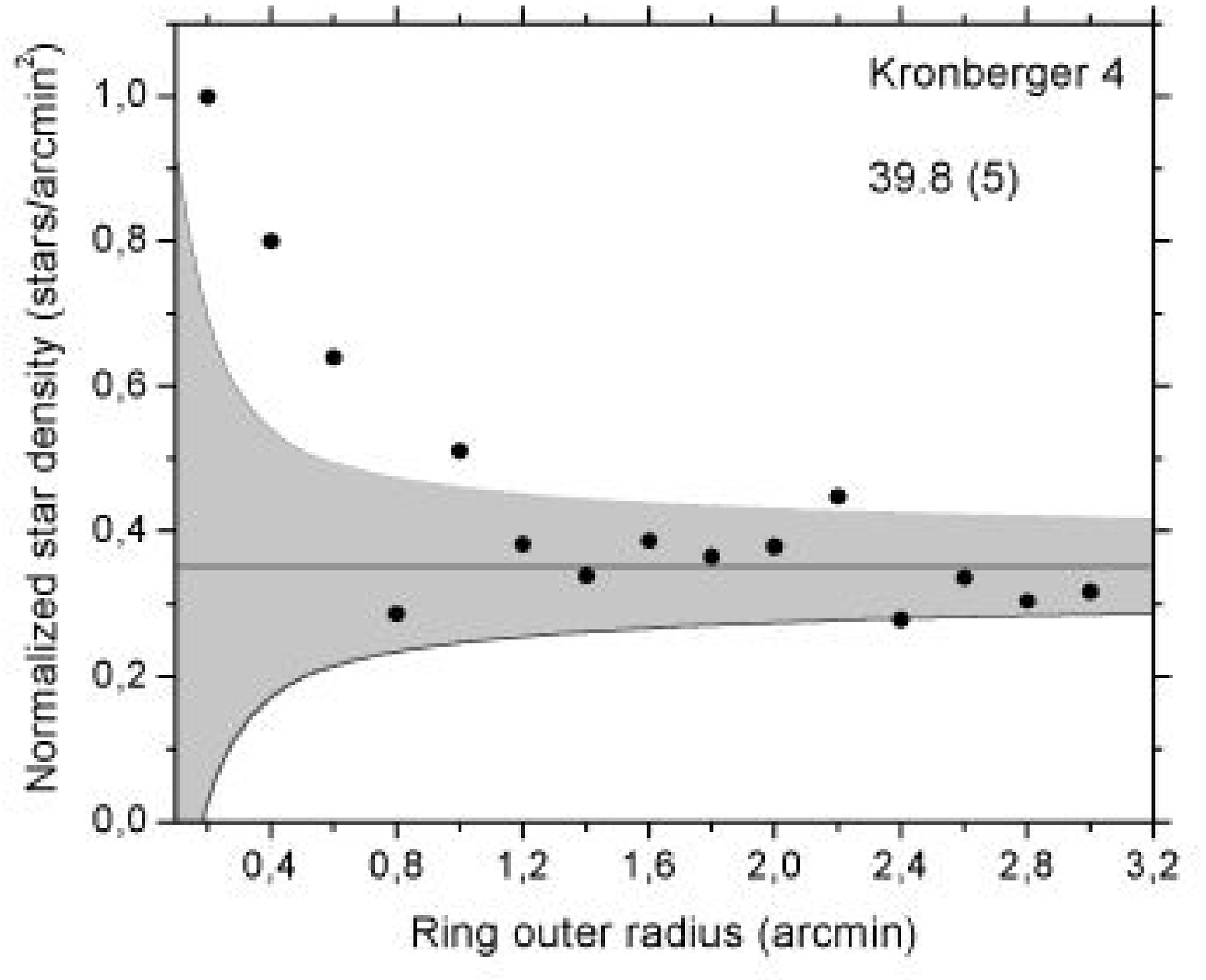}

\end{figure*}



\setcounter{figure}{5}
\begin{figure*}
\caption[]{RDPs of the cluster candidates presented in Table~\subref{TabNo}, with the vertical scale normalized to the maximum stellar density in the cluster field. In each diagram, the maximum stellar density (in $\mathrm{stars/arcmin^2}$) and, in brackets, the actual number of stars observed in this area are given. The mean density of the background is shown in each diagram as a horizontal line, with the $2\sigma$ error ranges of the background indicated as shaded areas. Only 2MASS sources with $H < 15.5$ were taken into account.}
\label{fig6}

\includegraphics[width=8.75cm]{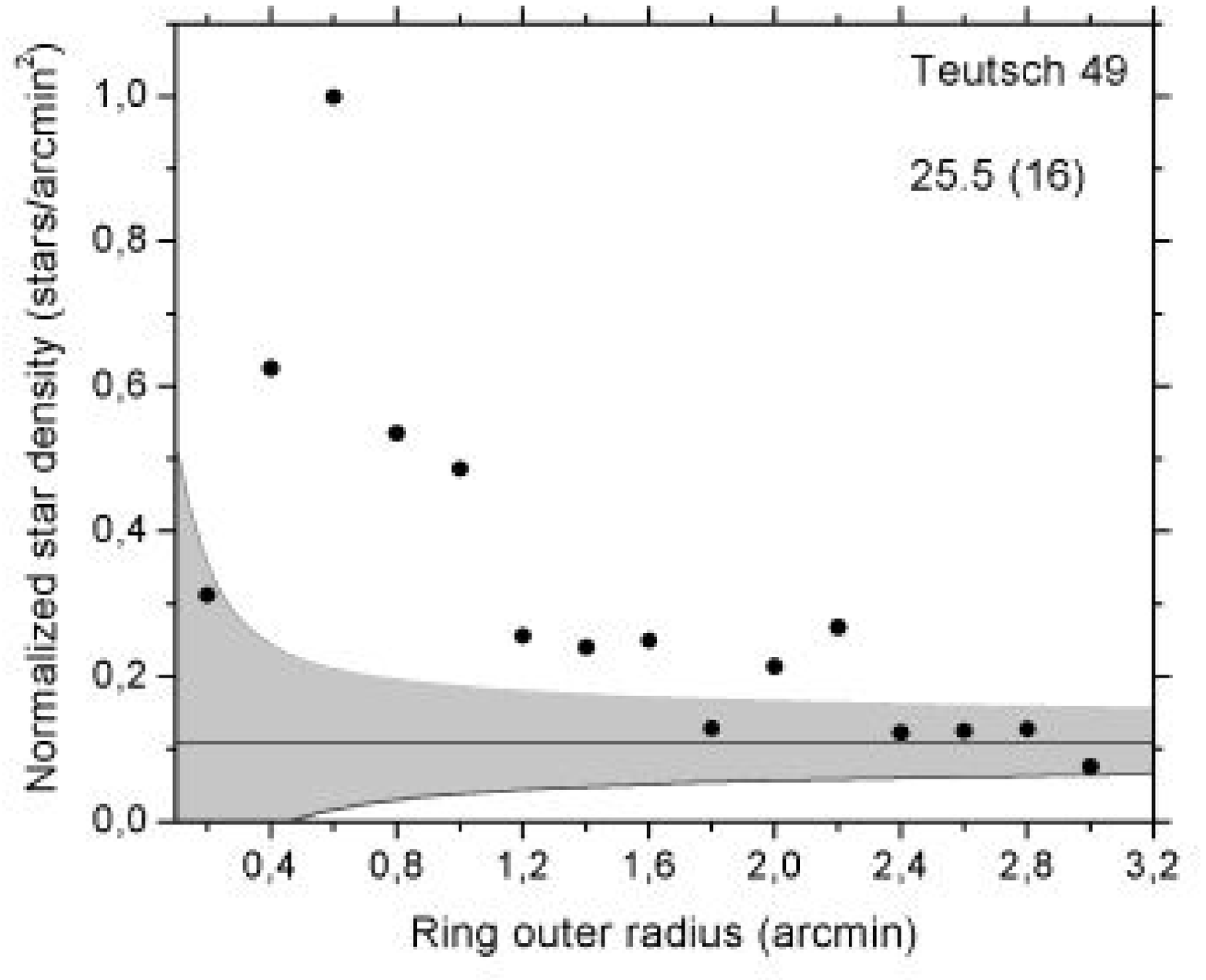}
\enskip
\includegraphics[width=8.75cm]{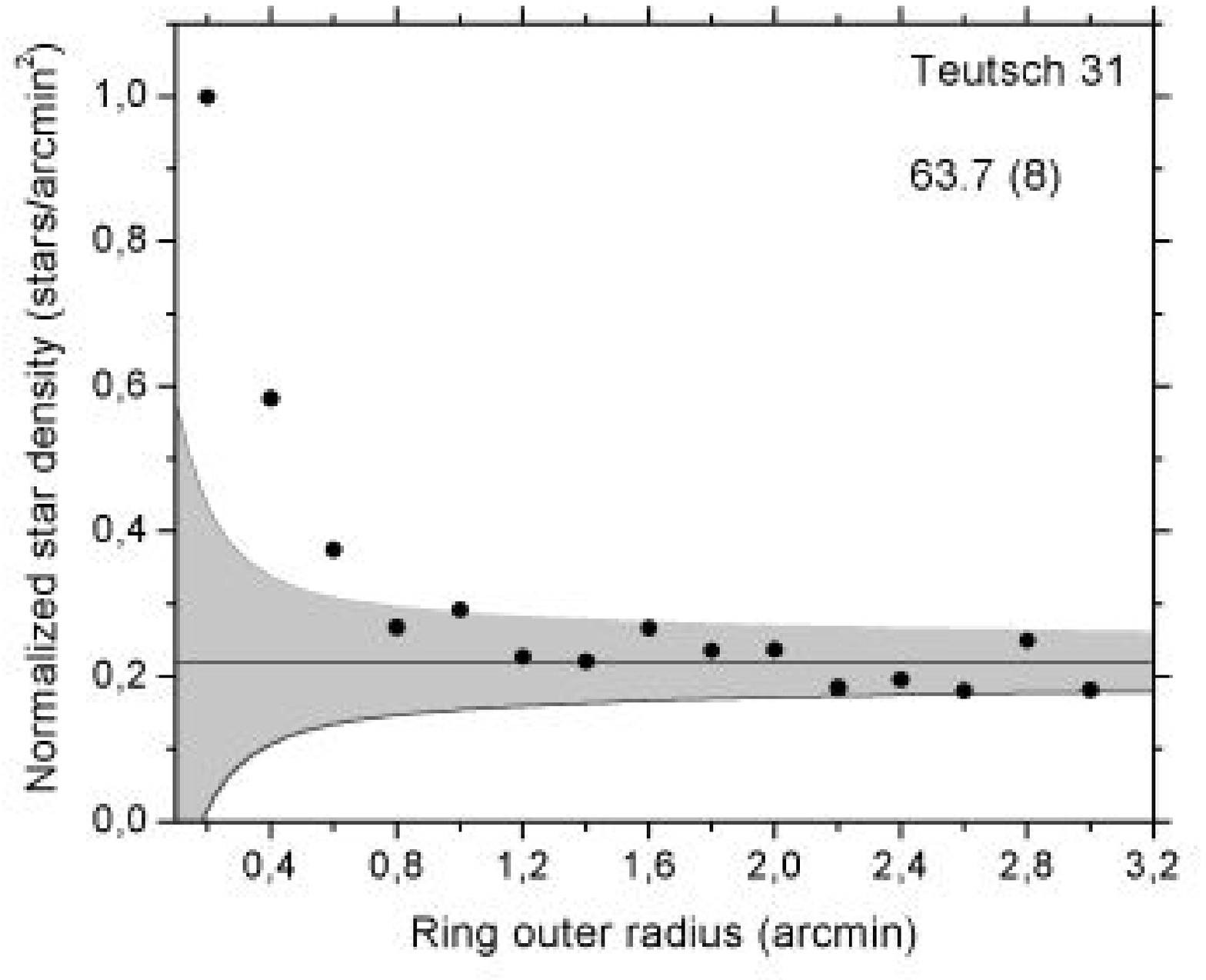}

\includegraphics[width=8.75cm]{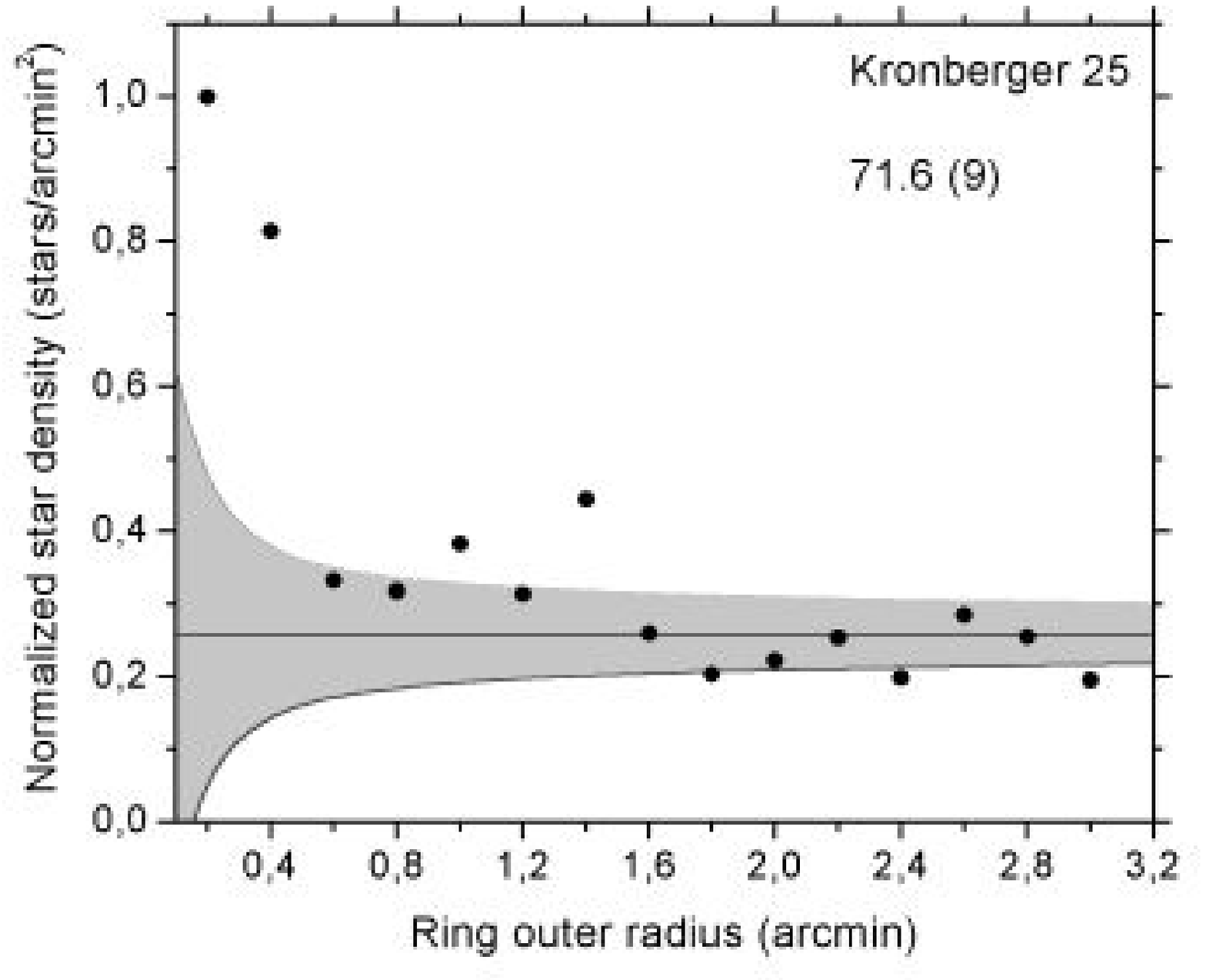}
\enskip
\includegraphics[width=8.75cm]{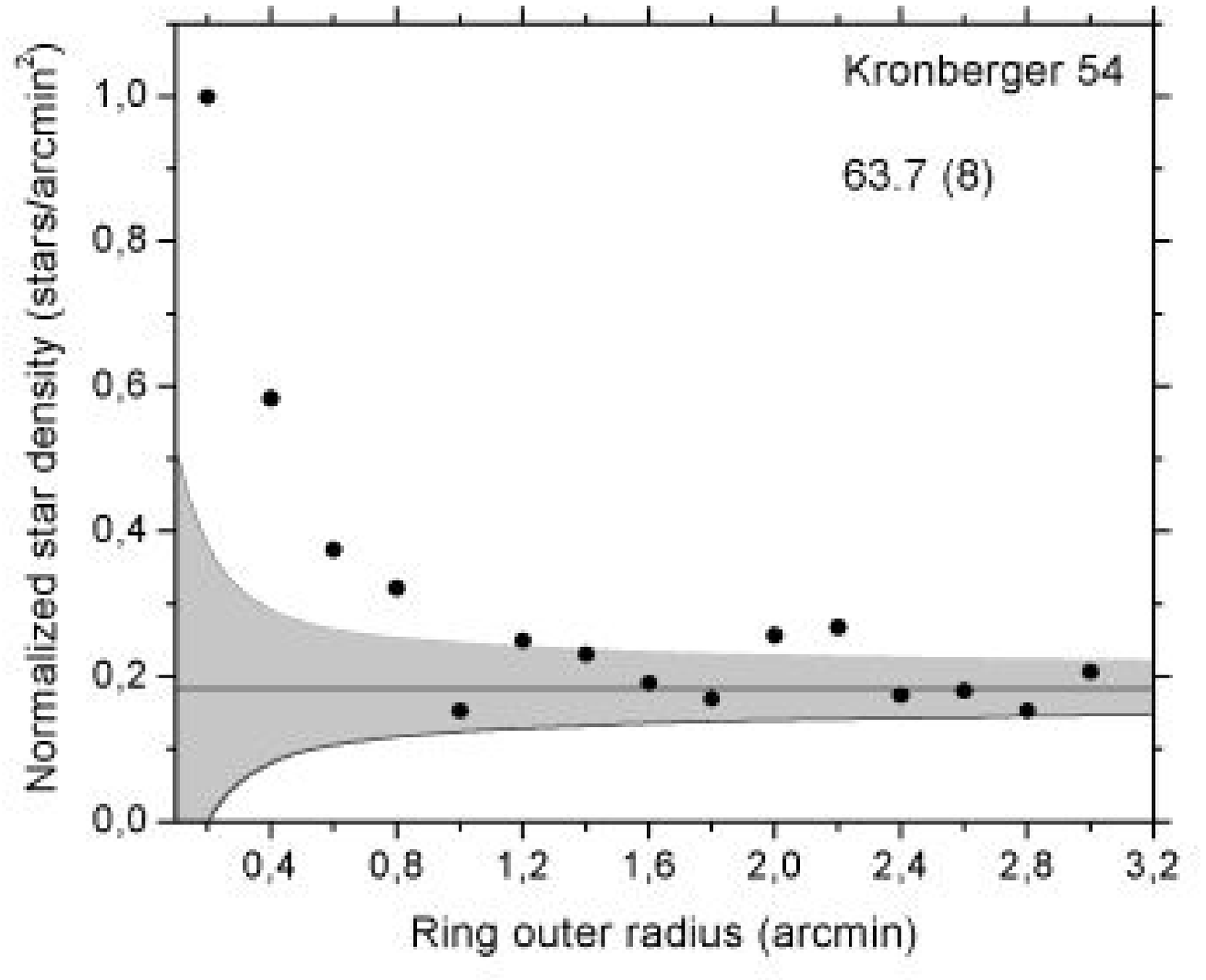}

\includegraphics[width=8.75cm]{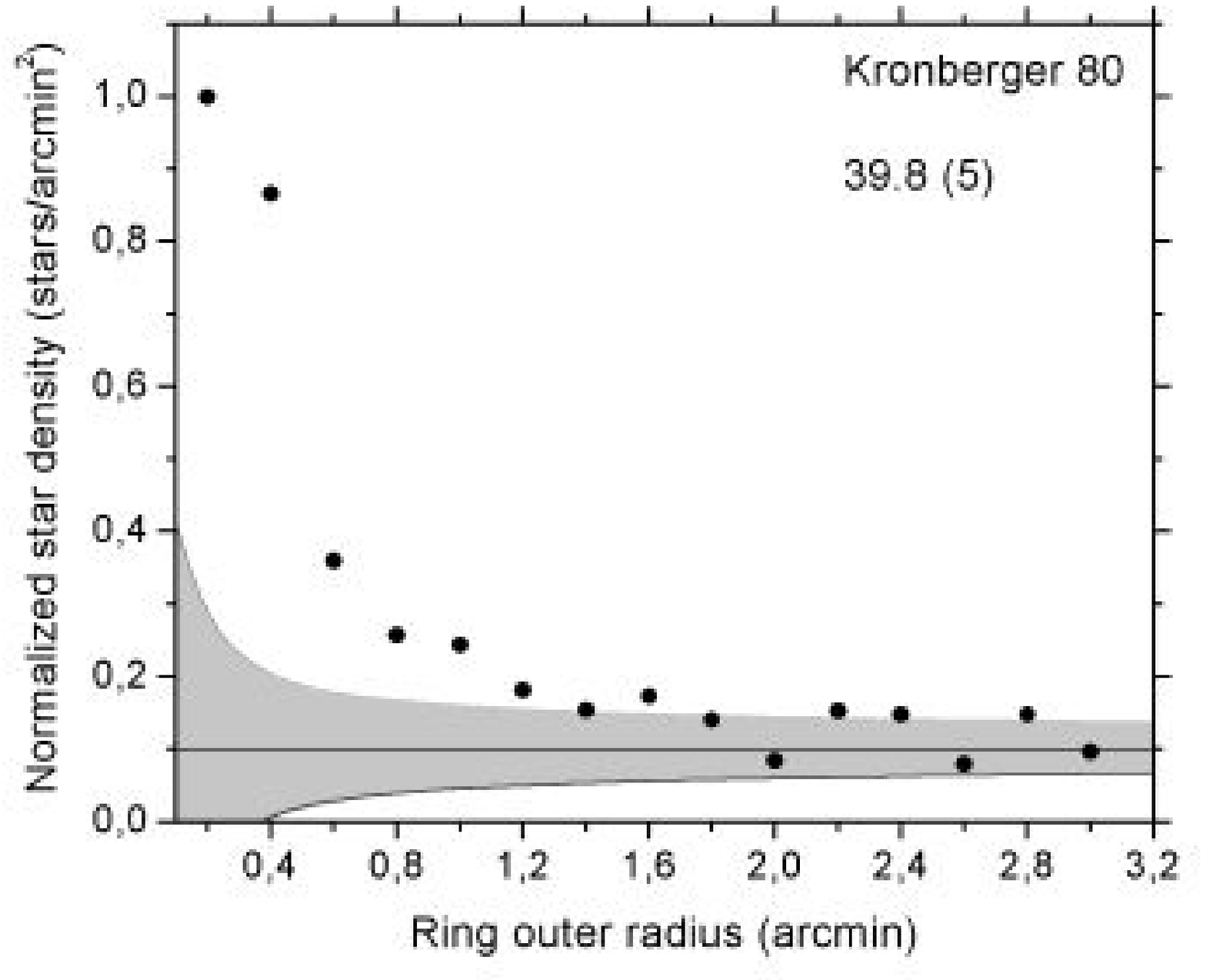}

\end{figure*}

\clearpage




\setcounter{figure}{6}
\begin{figure*}
\caption[]{CMDs of the cluster candidates presented in Table~\subref{TabIso}. Each diagram contains only those stars
 			with a distance from the cluster center that is less than the visual cluster radius $R$ and with $J$ and
			$H$ magnitudes derived either via aperture photometry ($rd\_flg = 1$) or via point-spread function fitting 
			($rd\_flg = 2$). Also included are the best-fit solar metallicity isochrones of the CMD.}
\label{fig7}
\centering

\includegraphics[width=11cm]{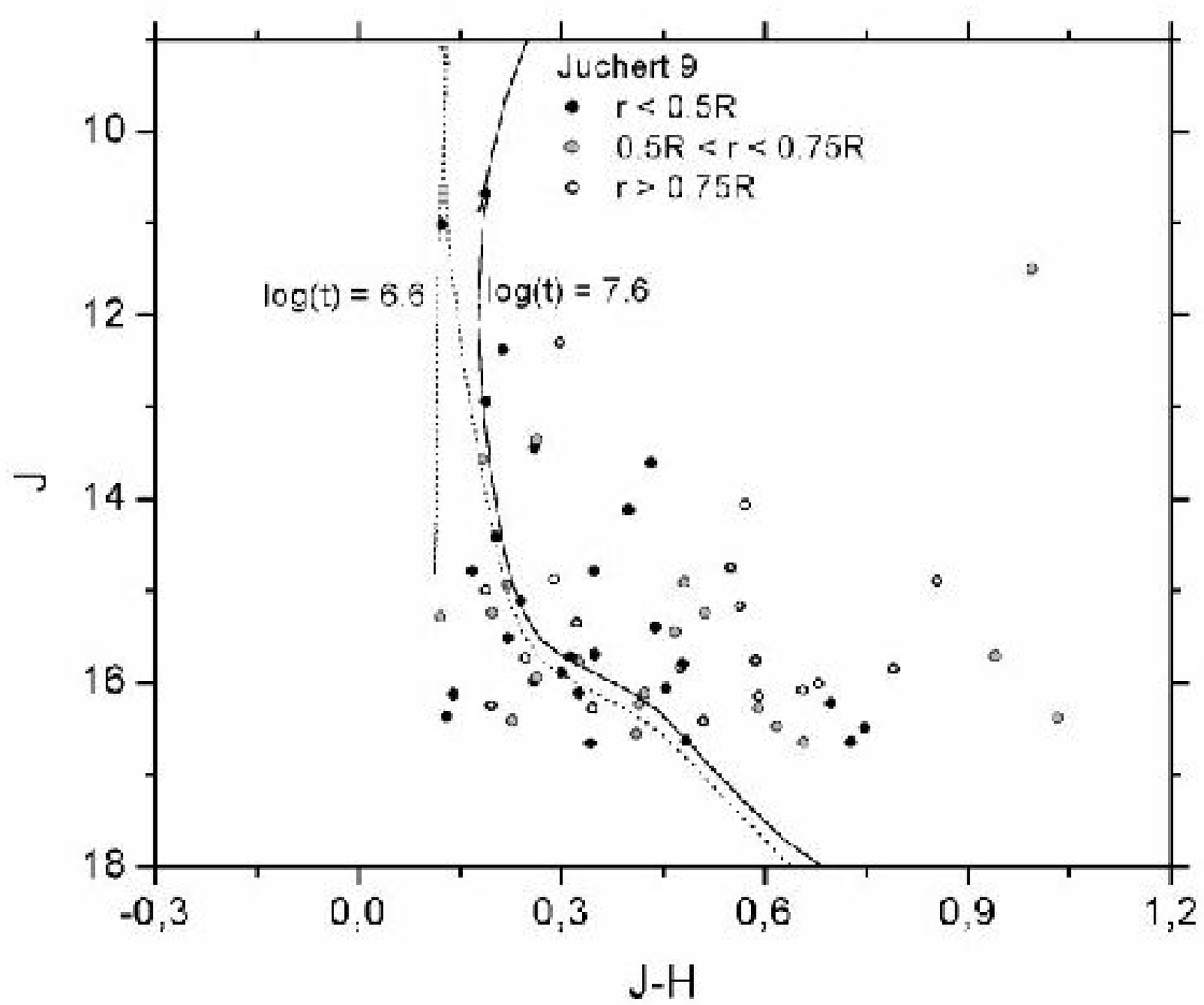}

\includegraphics[width=11cm]{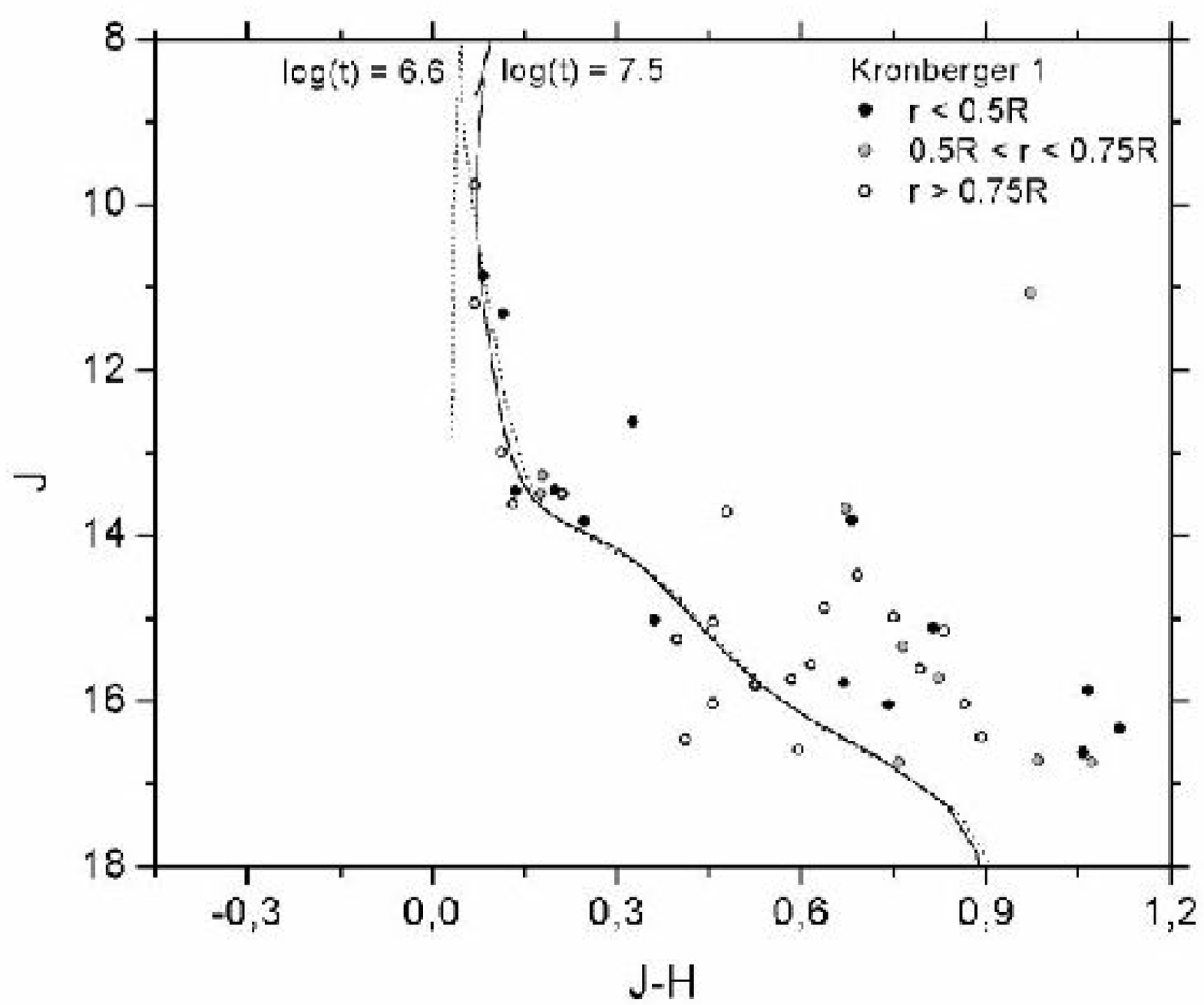}

\includegraphics[width=11cm]{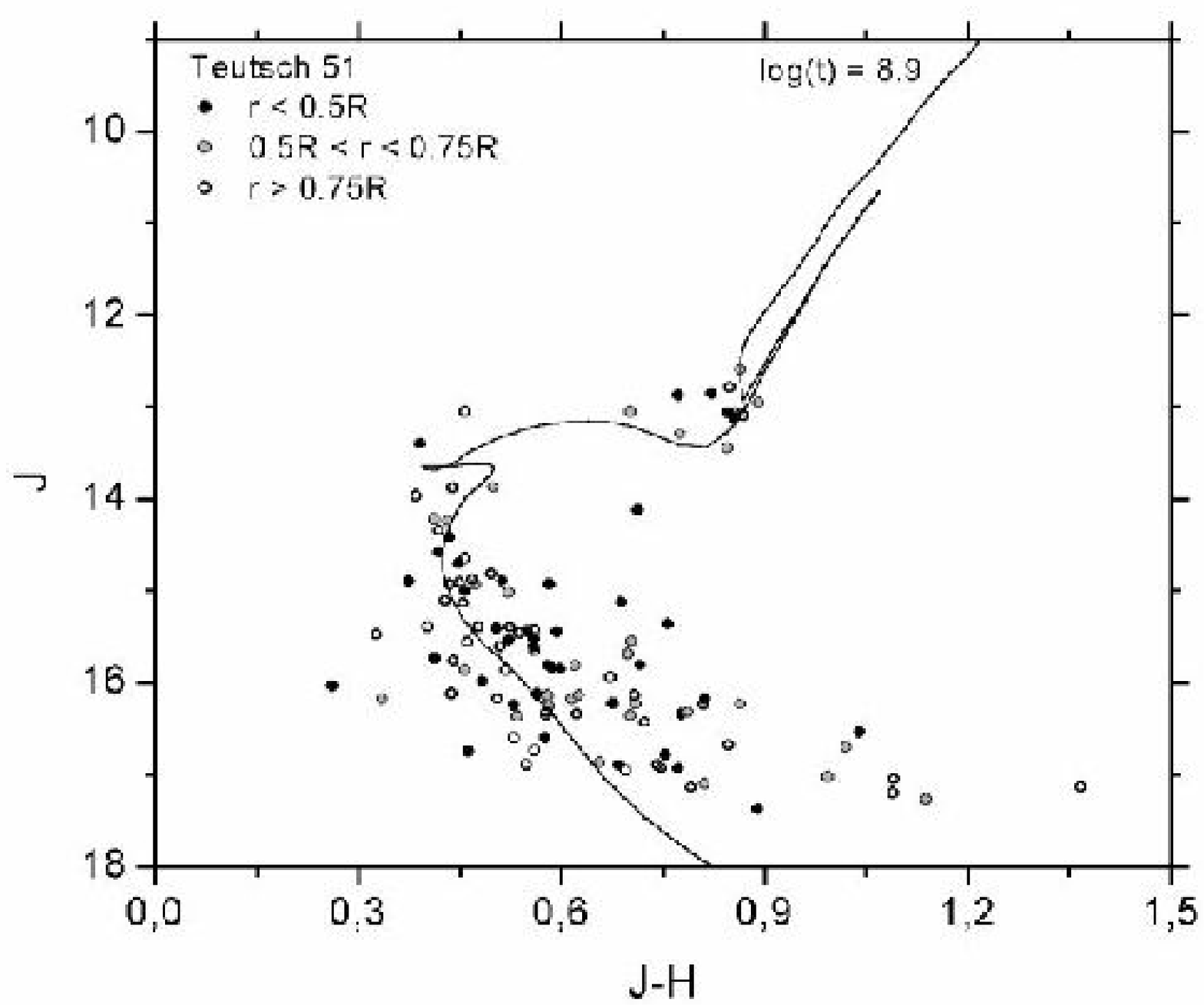}

\end{figure*}


\setcounter{figure}{6}
\begin{figure*}
\caption[]{(cont.)}
\centering

\includegraphics[width=11cm]{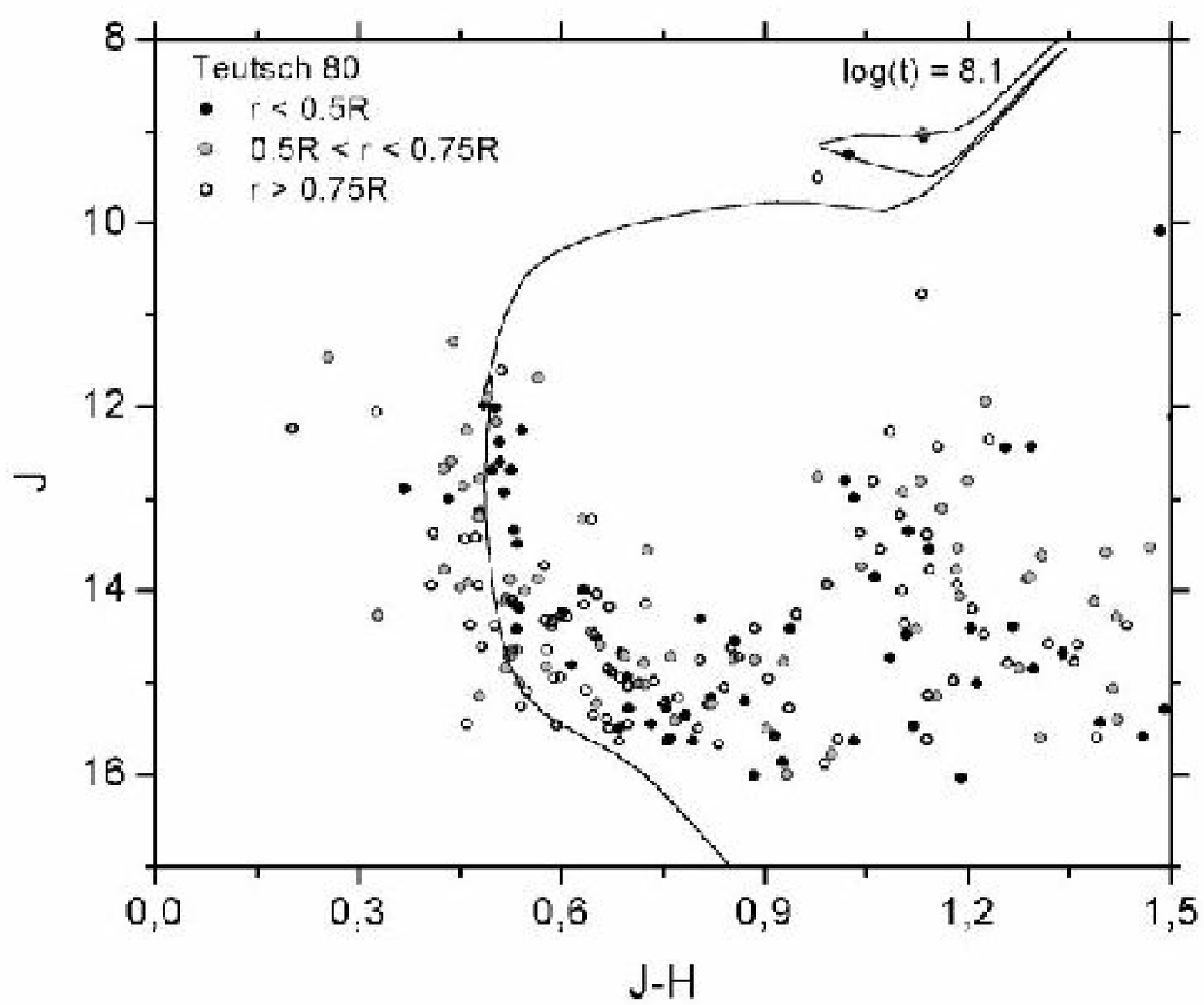}

\includegraphics[width=11cm]{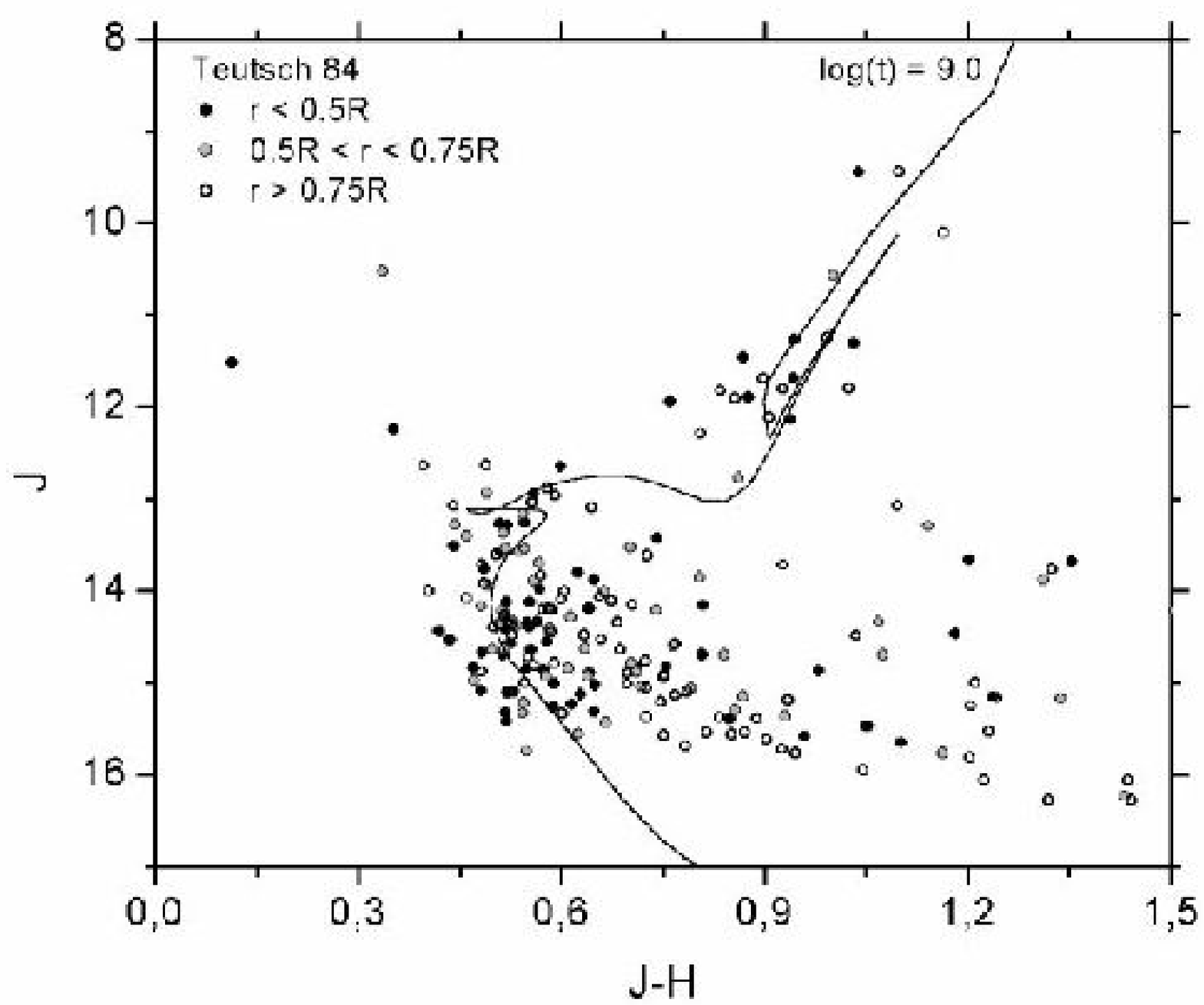}

\includegraphics[width=11cm]{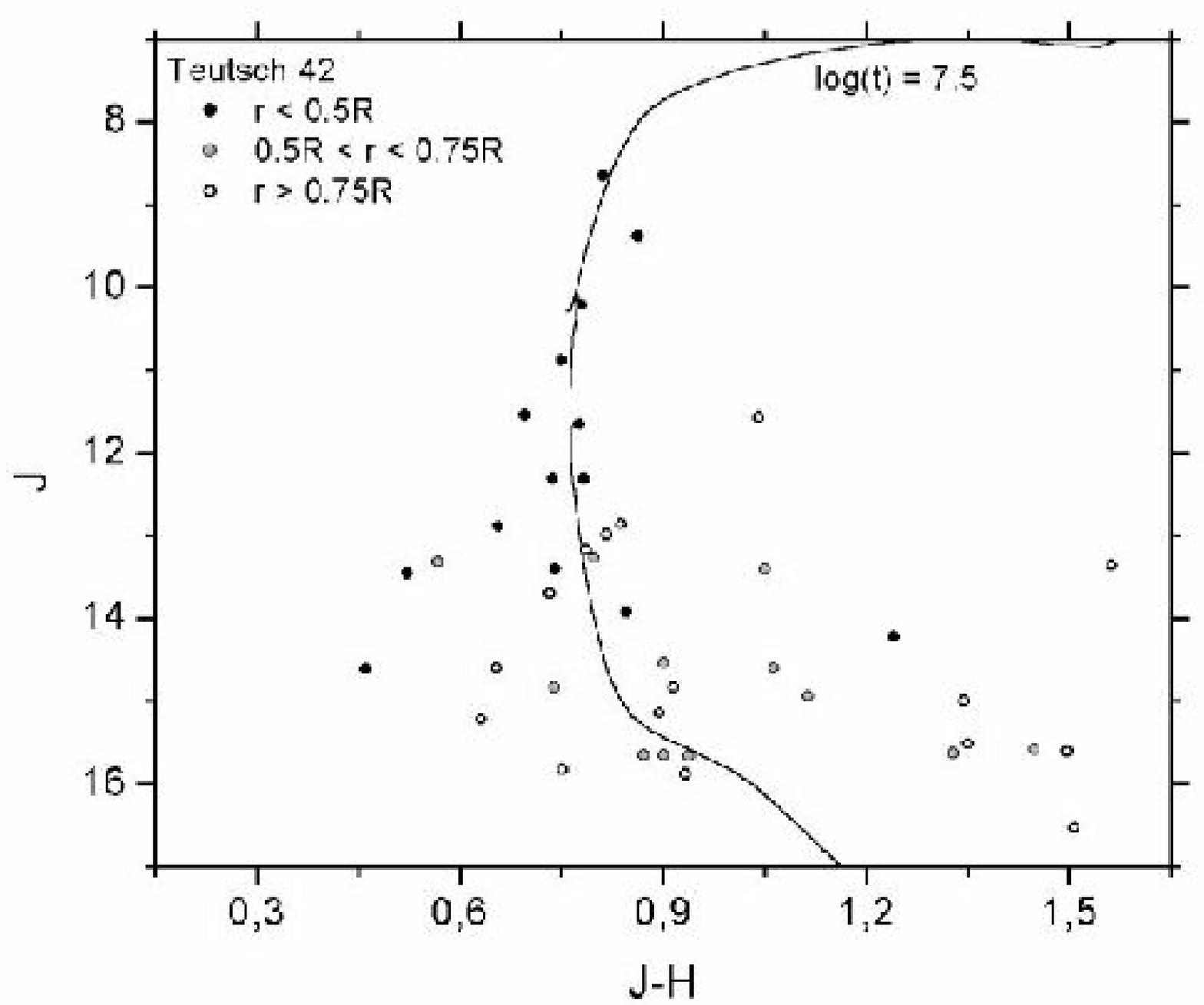}

\end{figure*}


\setcounter{figure}{6}
\begin{figure*}
\caption[]{(cont.)}
\centering

\includegraphics[width=11cm]{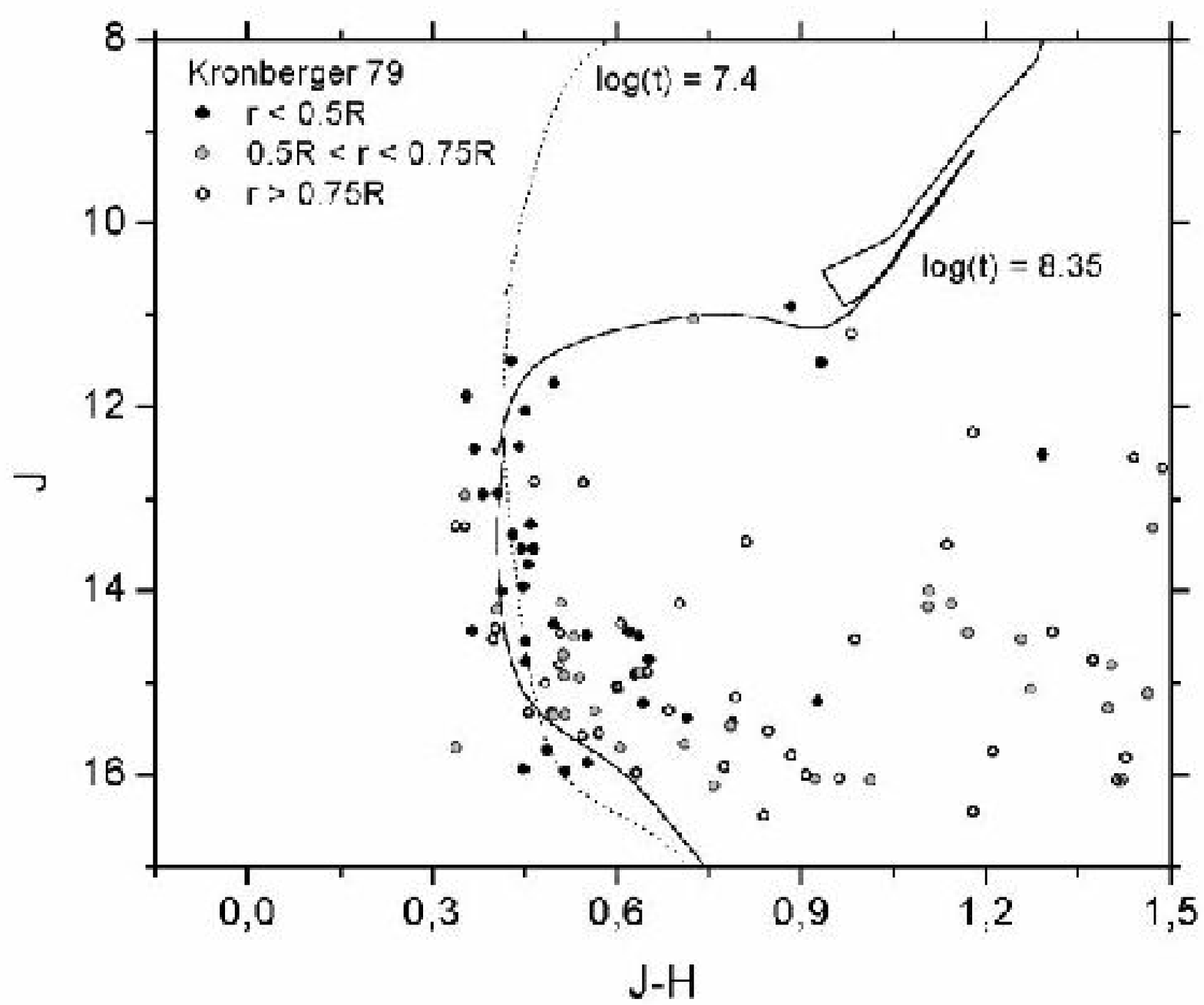}

\includegraphics[width=11cm]{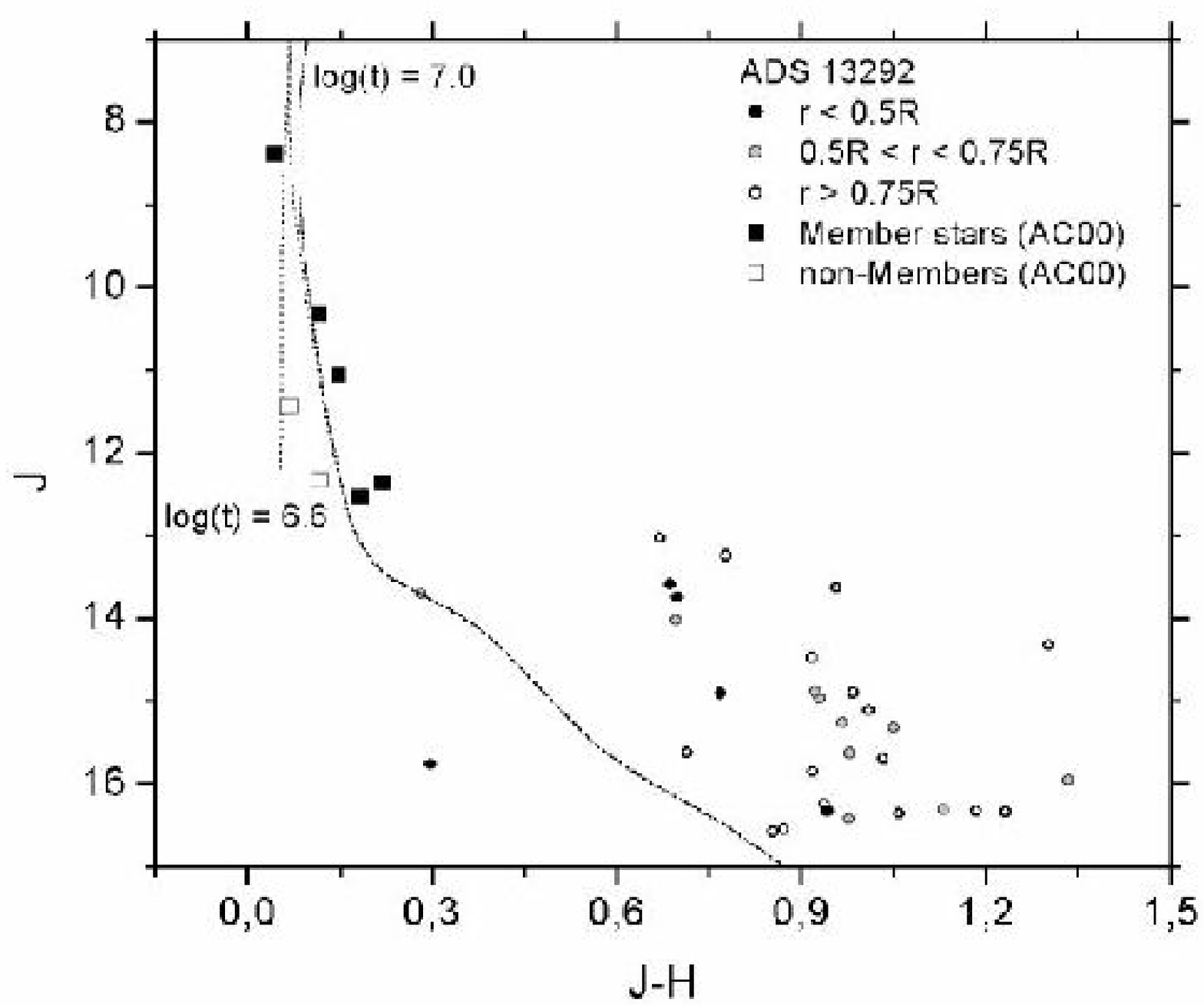}

\includegraphics[width=11cm]{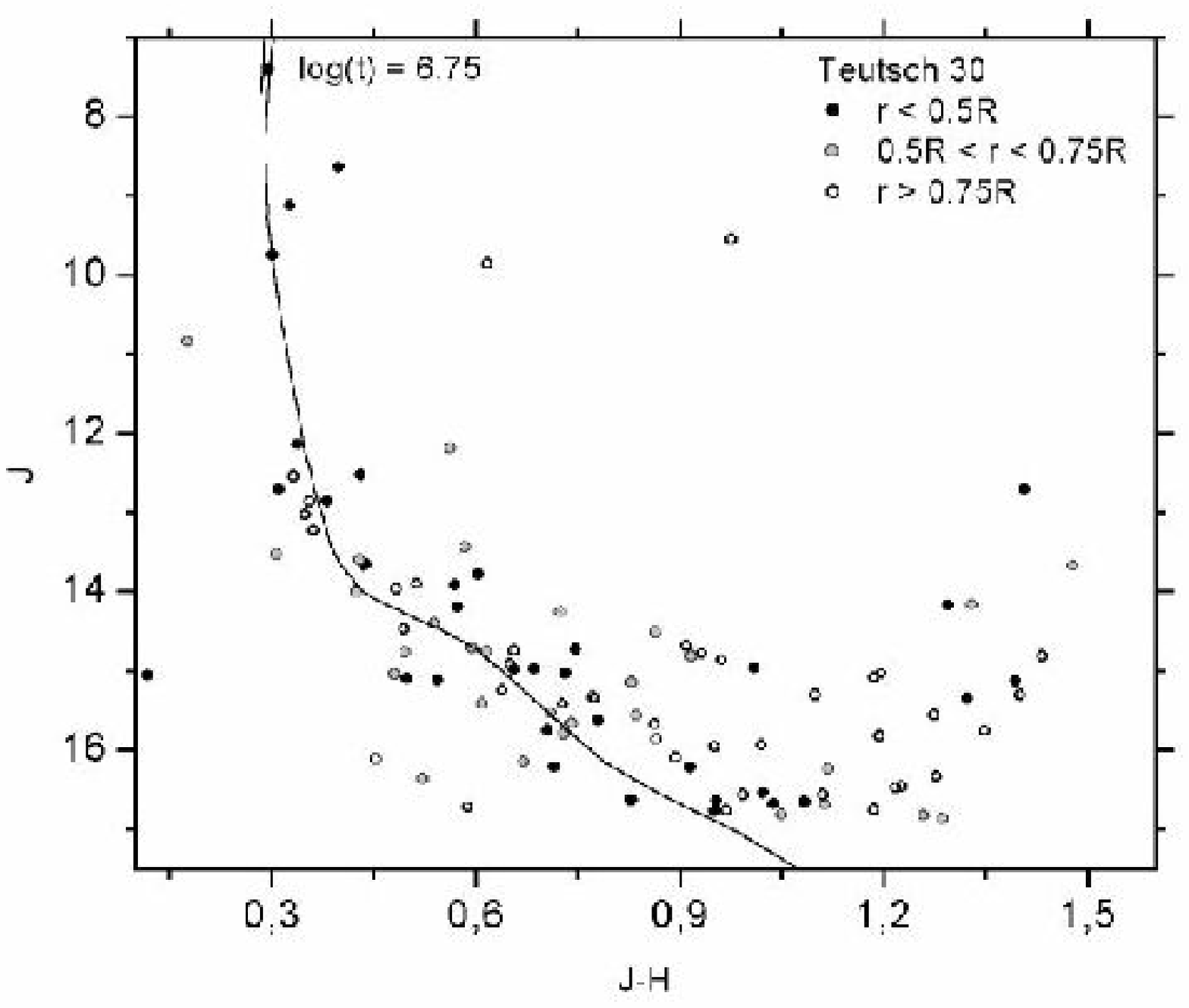}

\end{figure*}



\setcounter{figure}{7}
\begin{figure*}
\caption[]{CMDs of the cluster candidates presented in Table~\subref{TabRC}. Each diagram contains only those stars
			with a
		 	distance from the cluster center that is less than the visual cluster radius $R$ and with $J$ and $H$ 
			magnitudes derived either via aperture photometry ($rd\_flg = 1$) or via point-spread function fitting 
			($rd\_flg = 2$).}
\label{fig8}
\centering

\includegraphics[width=11cm]{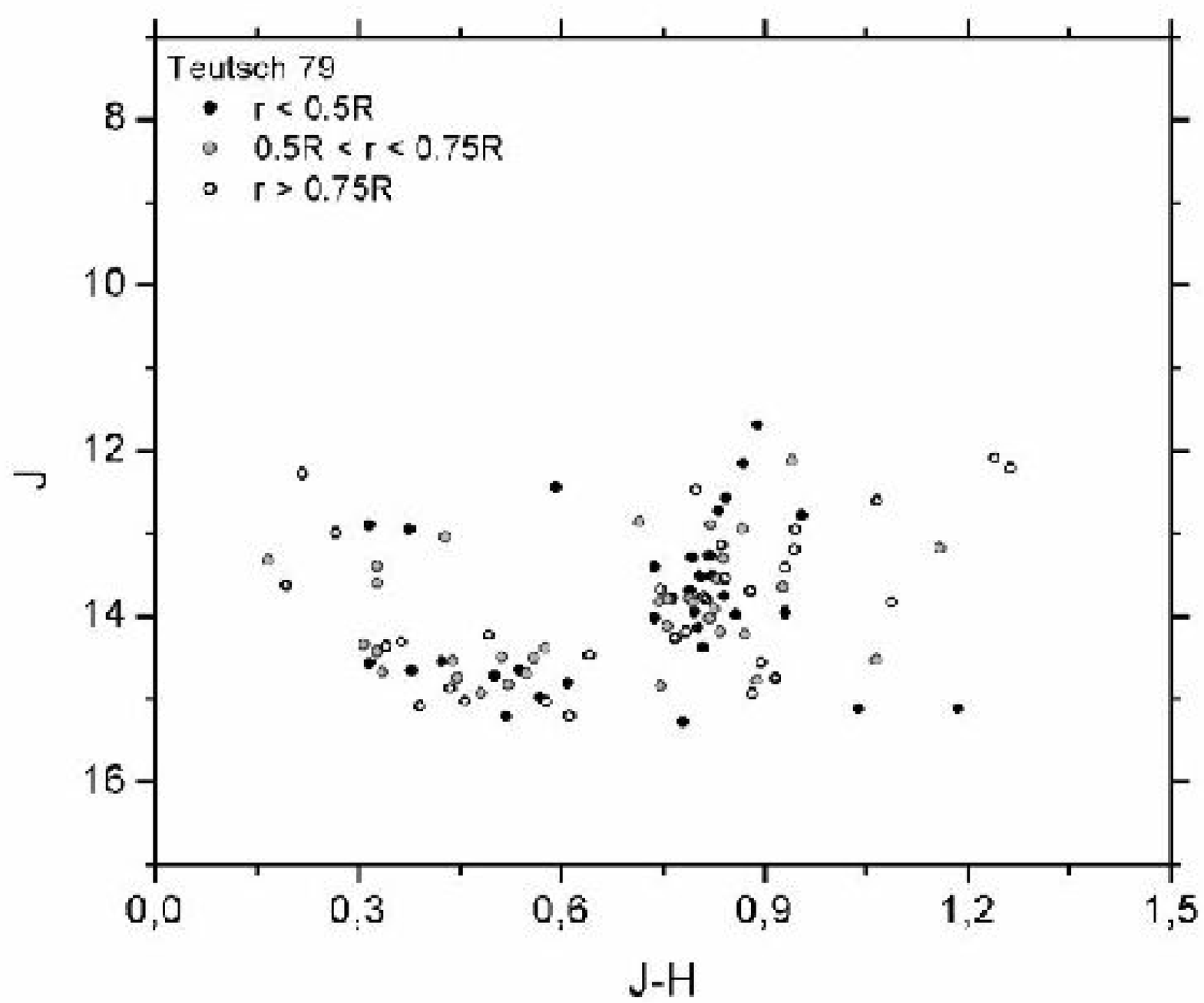}

\includegraphics[width=11cm]{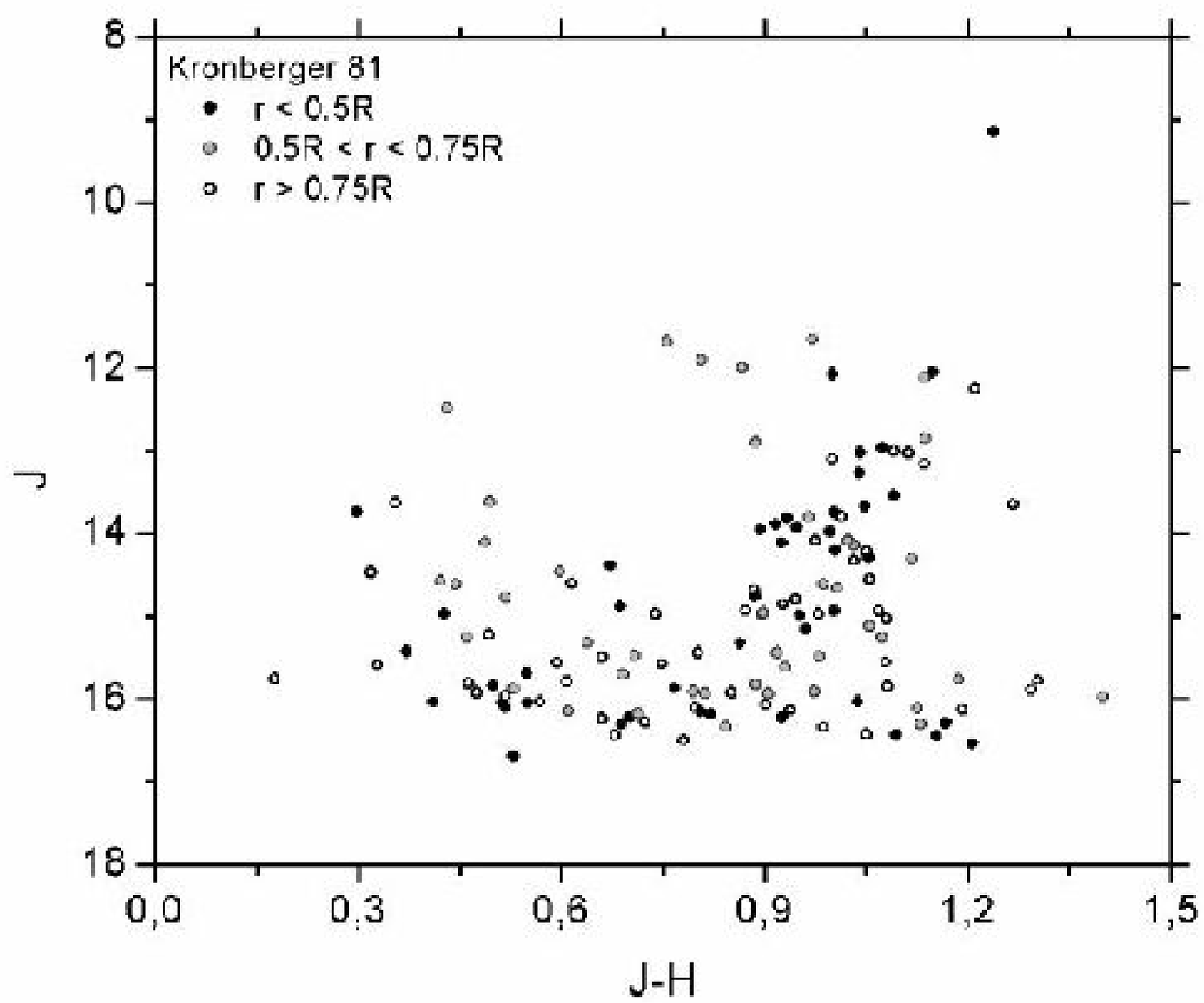}

\includegraphics[width=11cm]{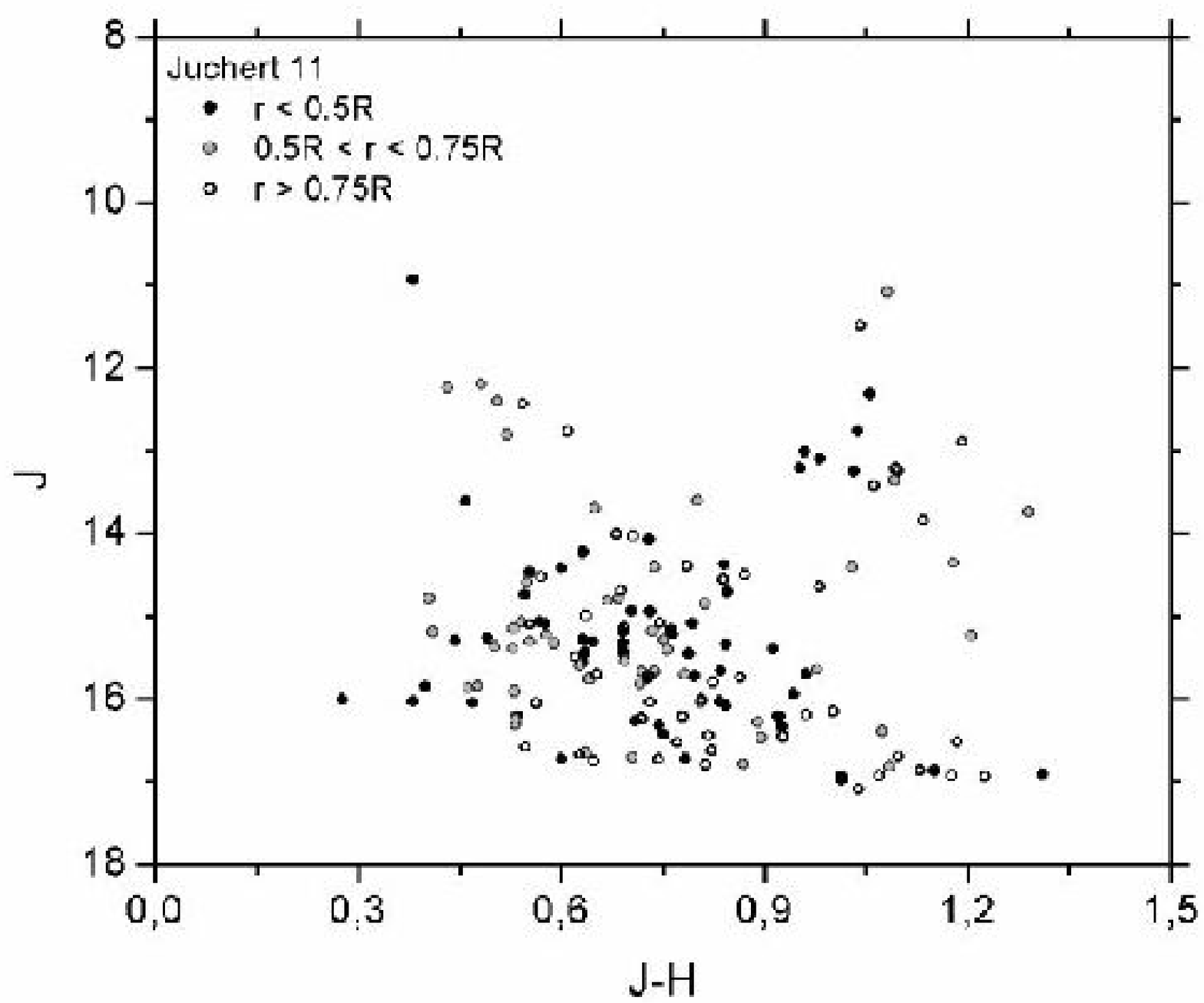}

\end{figure*}


\setcounter{figure}{7}
\begin{figure*}
\caption[]{(cont.)}
\centering

\includegraphics[width=11cm]{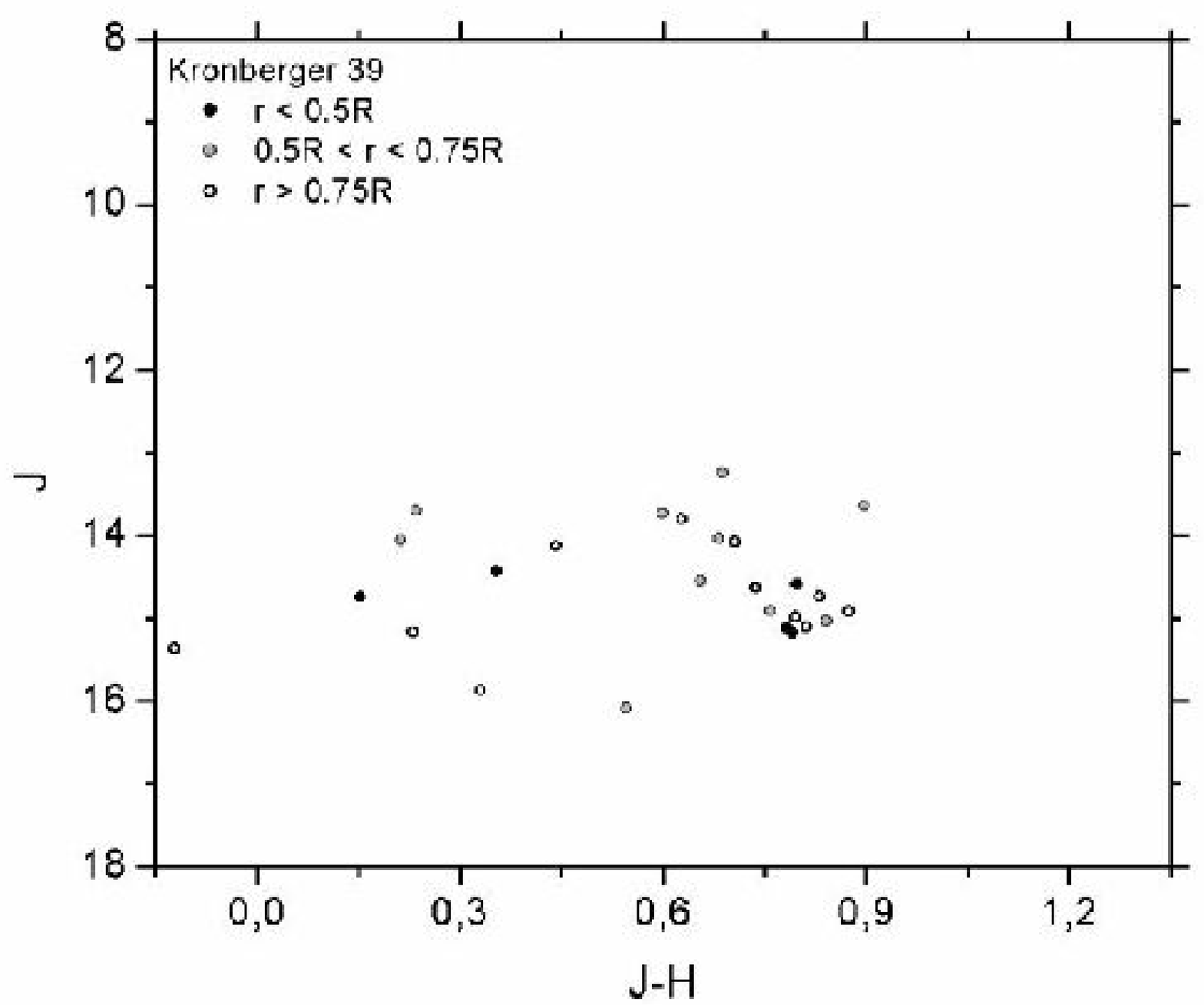}

\includegraphics[width=11cm]{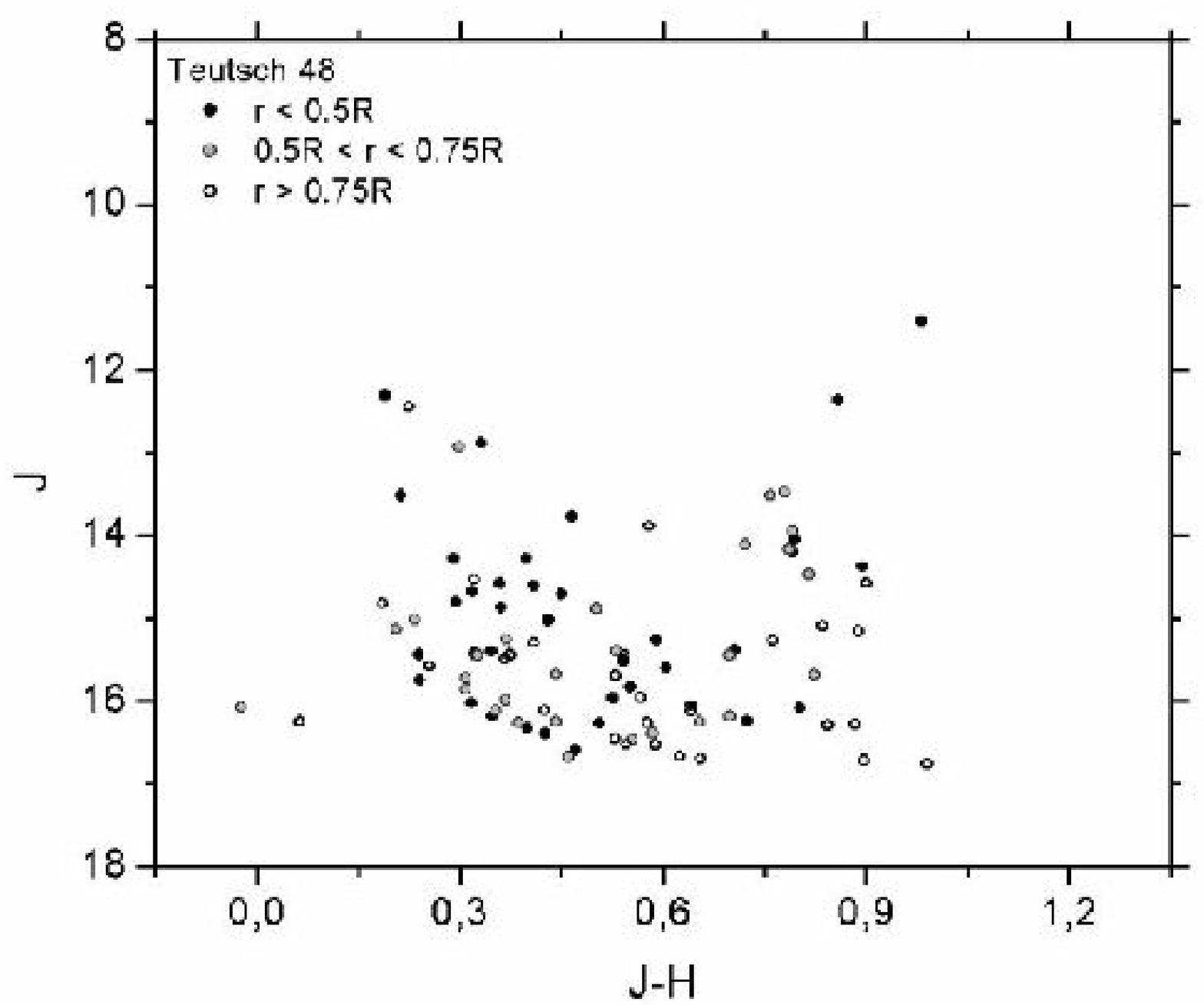}

\includegraphics[width=11cm]{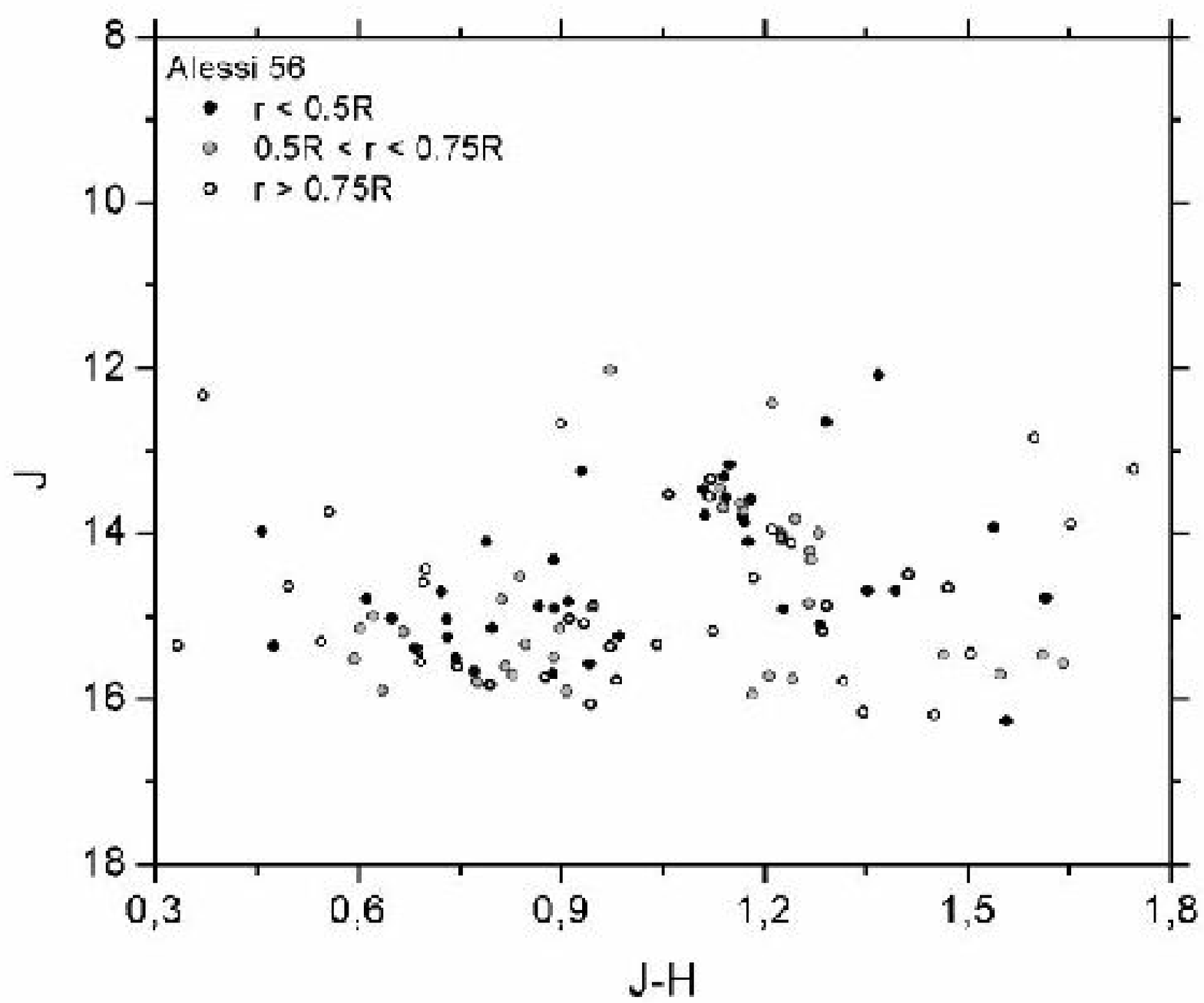}

\end{figure*}


\setcounter{figure}{7}
\begin{figure*}
\caption[]{(cont.)}
\centering

\includegraphics[width=11cm]{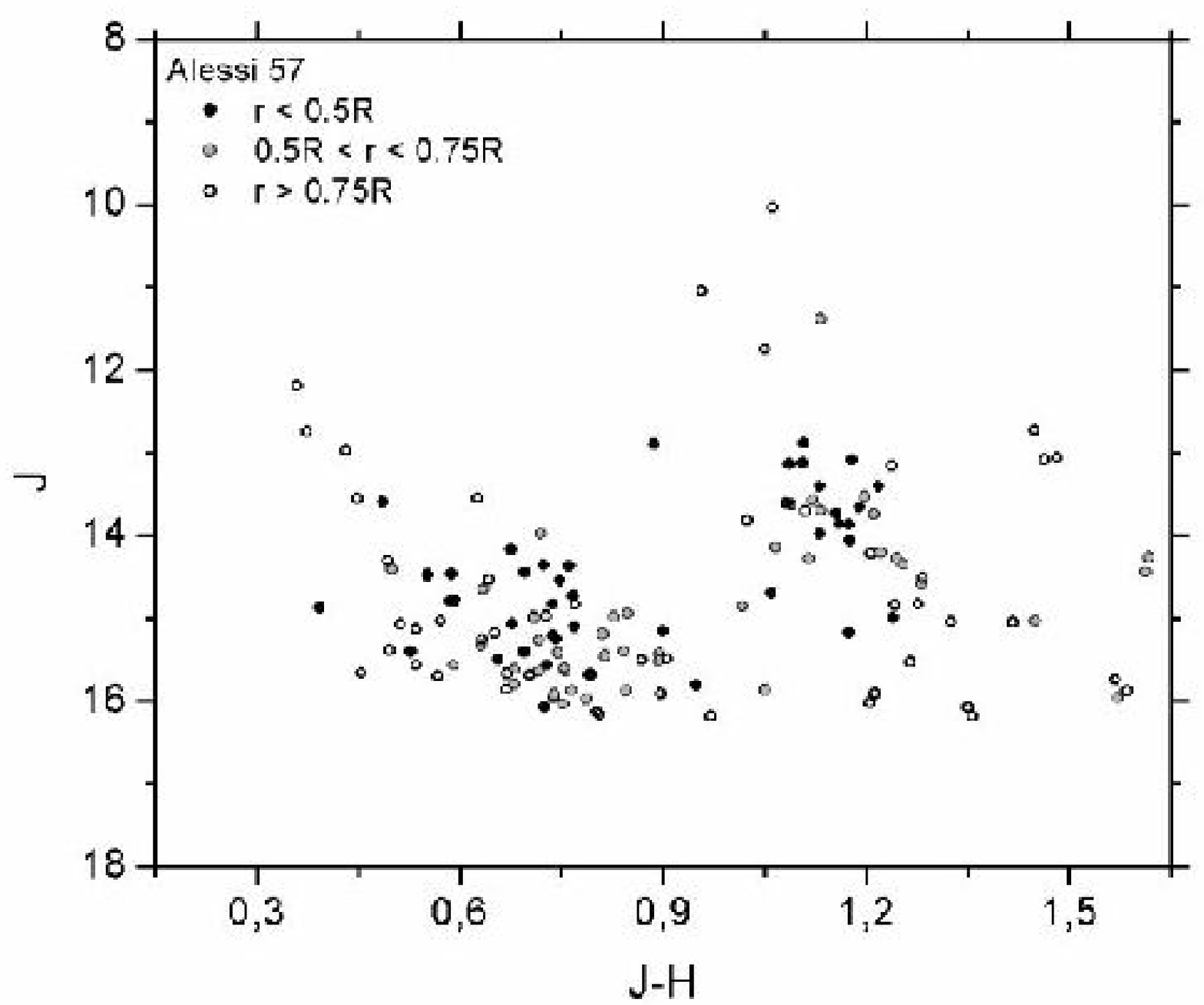}

\includegraphics[width=11cm]{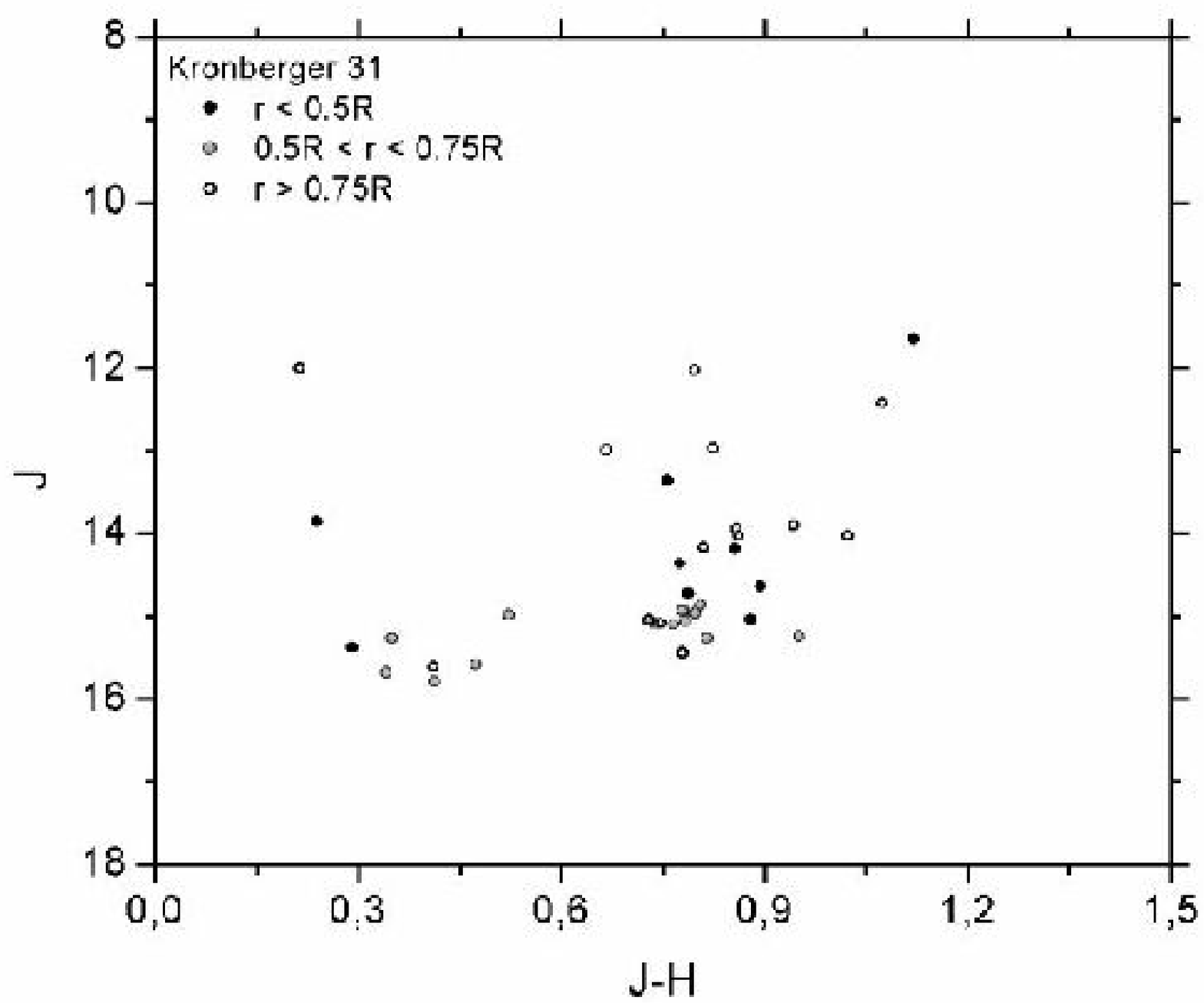}

\includegraphics[width=11cm]{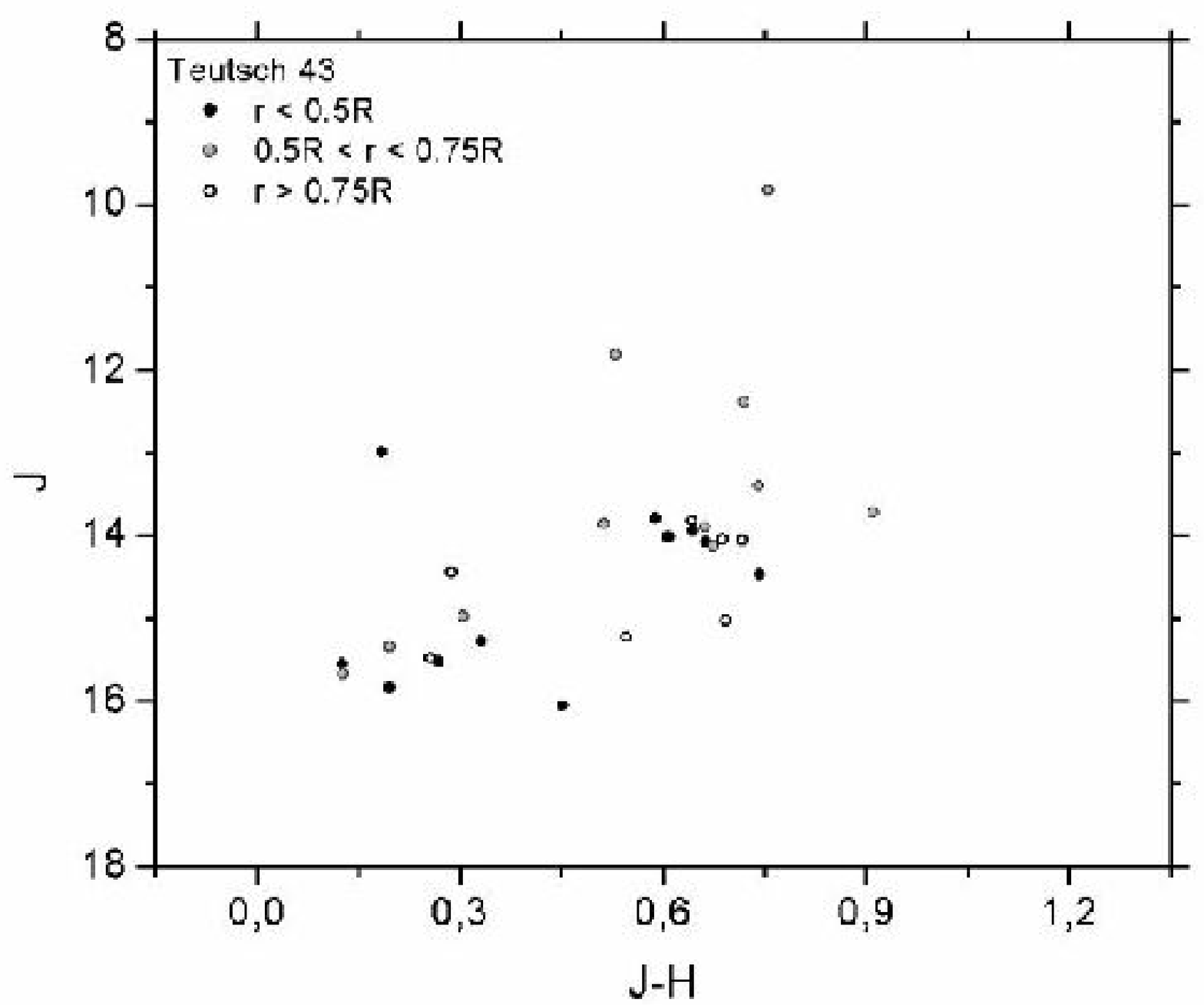}

\end{figure*}


\setcounter{figure}{7}
\begin{figure*}
\caption[]{(cont.)}
\centering

\includegraphics[width=11cm]{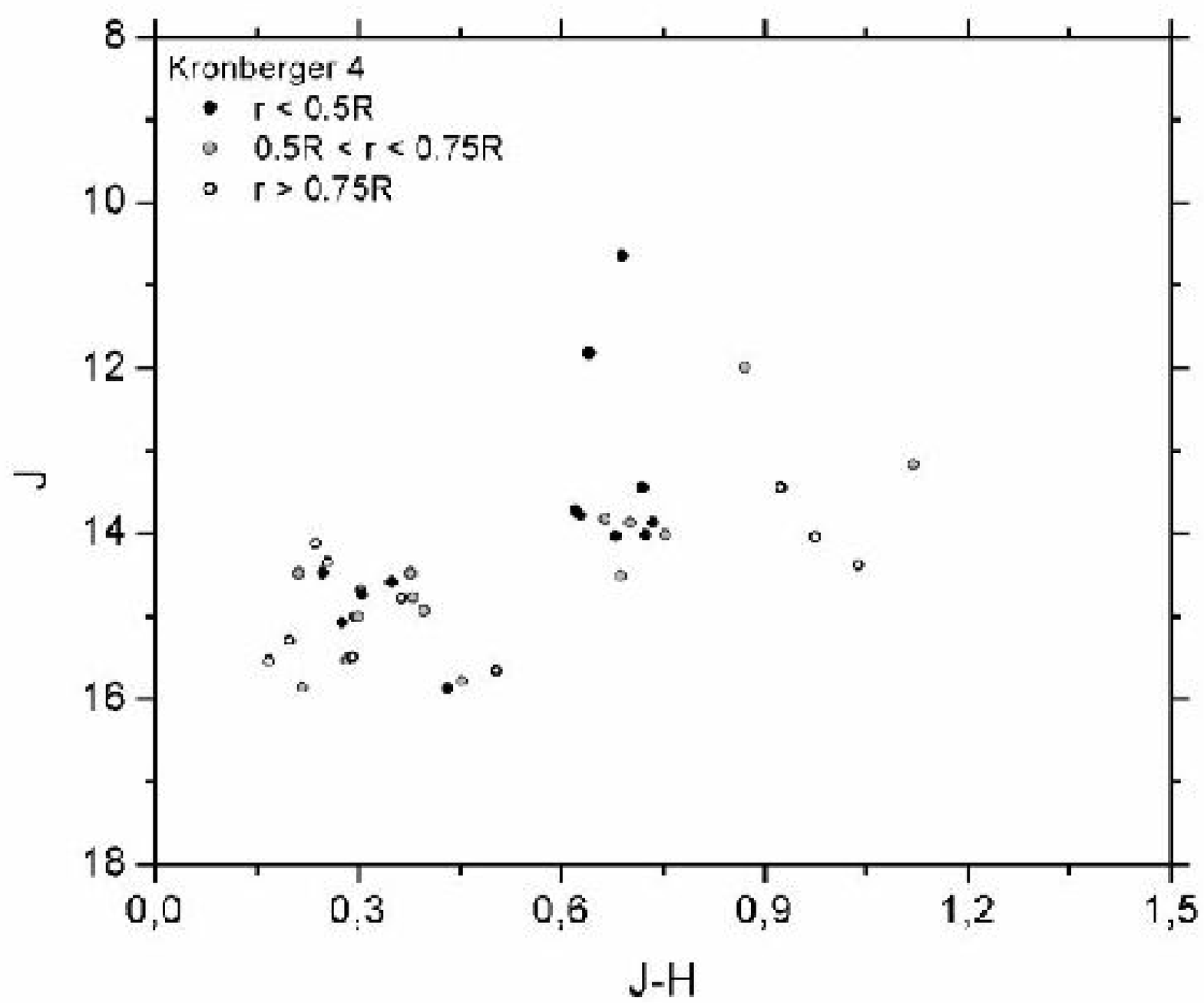}

\end{figure*}



\setcounter{figure}{8}
\begin{figure*}
\caption[]{CMDs of the cluster candidates presented in Table~\subref{TabNo}. Each diagram contains only those stars
 			with a
 			distance from the cluster center that is less than the visual cluster radius $R$ and with $J$ and $H$ 
			magnitudes derived either via aperture photometry ($rd\_flg = 1$) or via point-spread function fitting 
			($rd\_flg = 2$).}
\label{fig9}
\centering

\includegraphics[width=11cm]{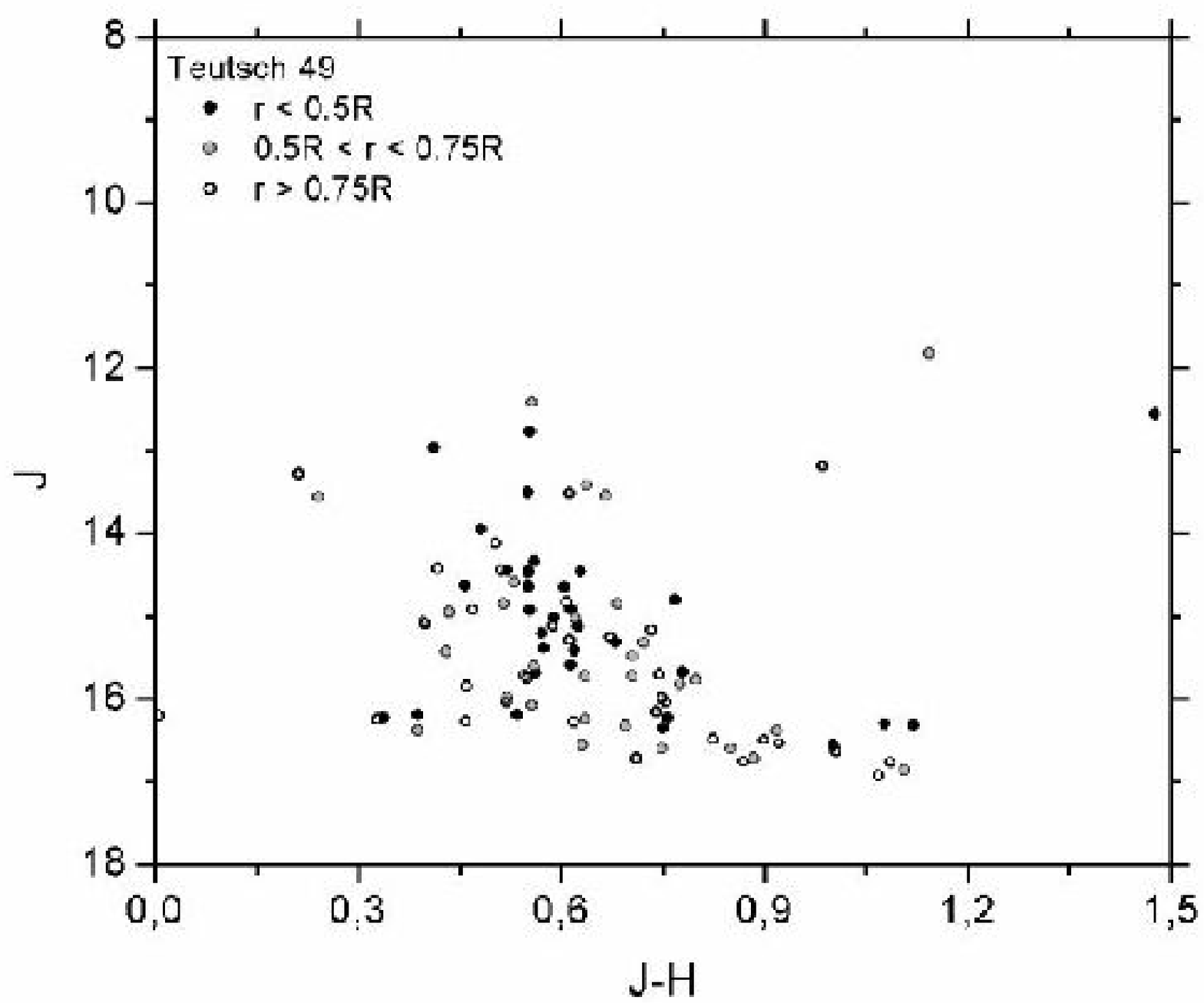}

\includegraphics[width=11cm]{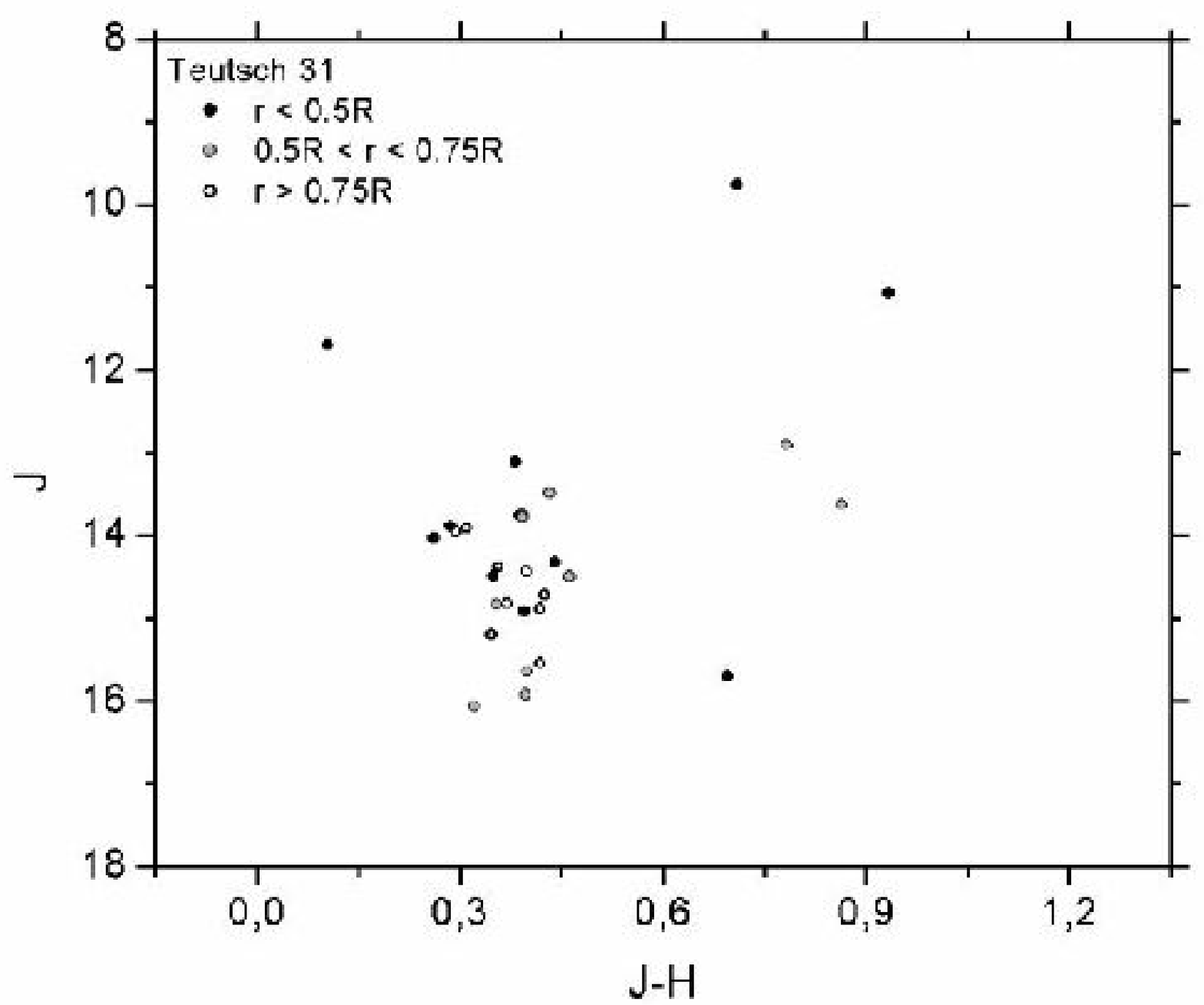}

\includegraphics[width=11cm]{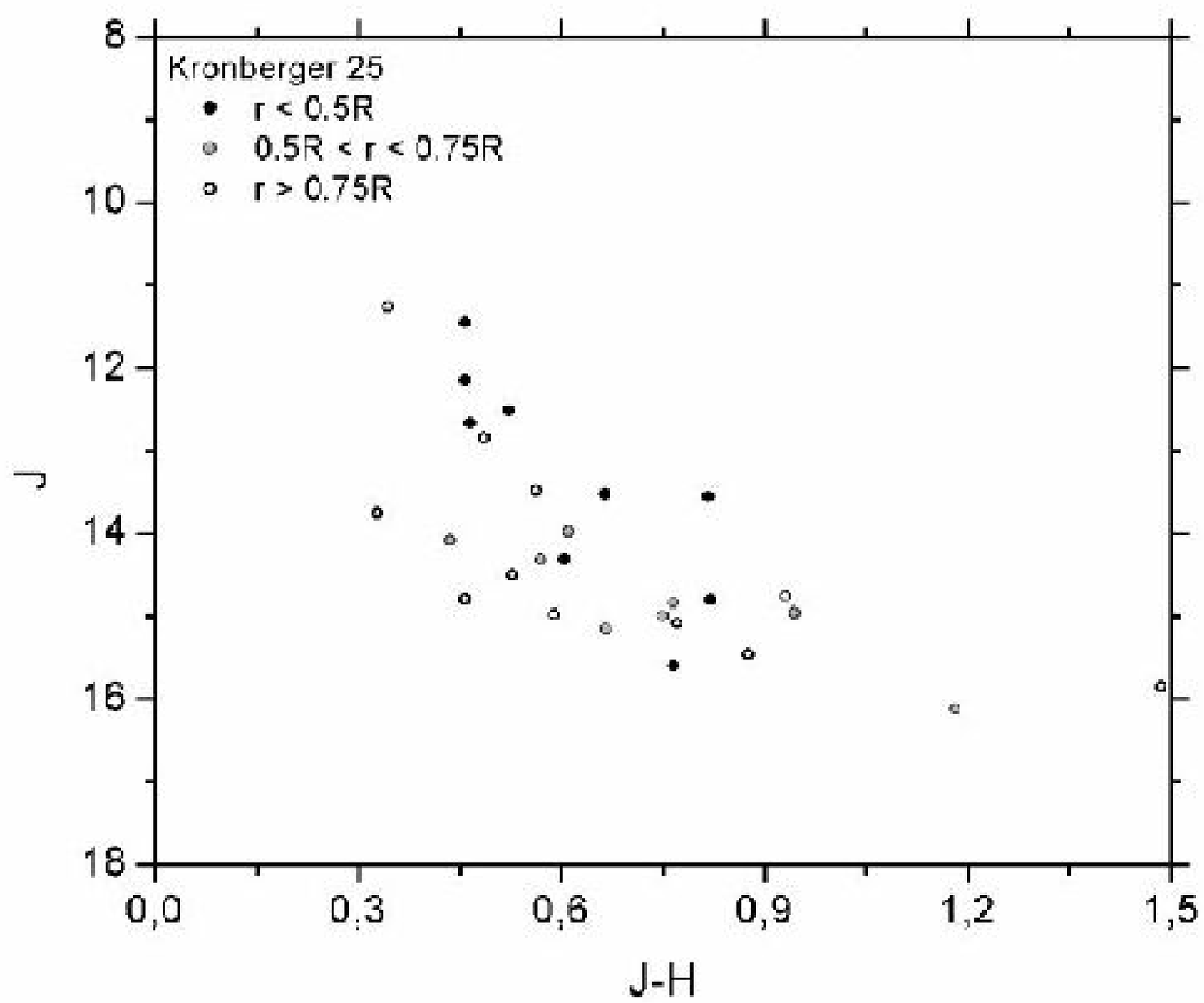}

\end{figure*}


\setcounter{figure}{8}
\begin{figure*}
\caption[]{(cont.)}
\centering

\includegraphics[width=11cm]{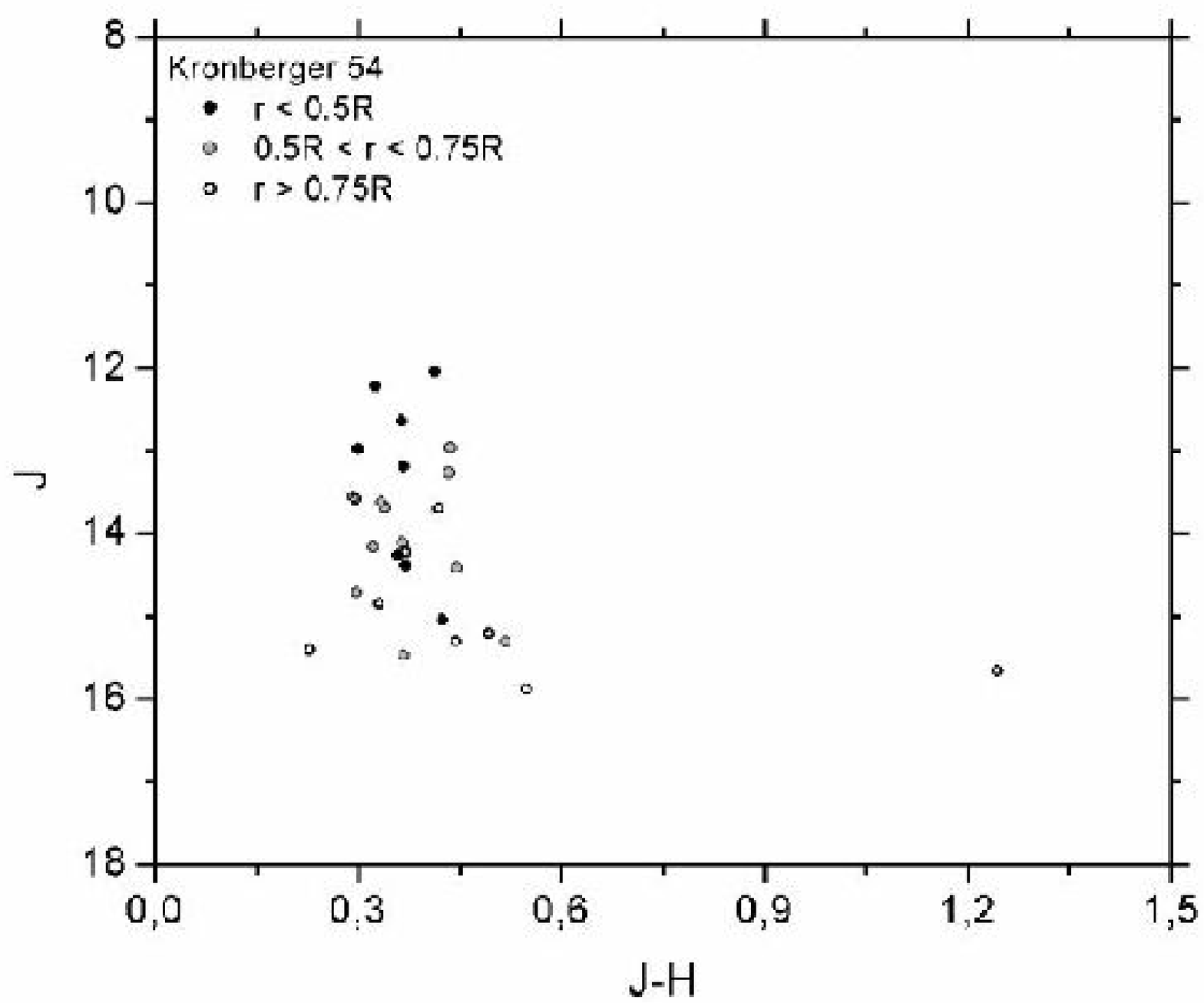}

\includegraphics[width=11cm]{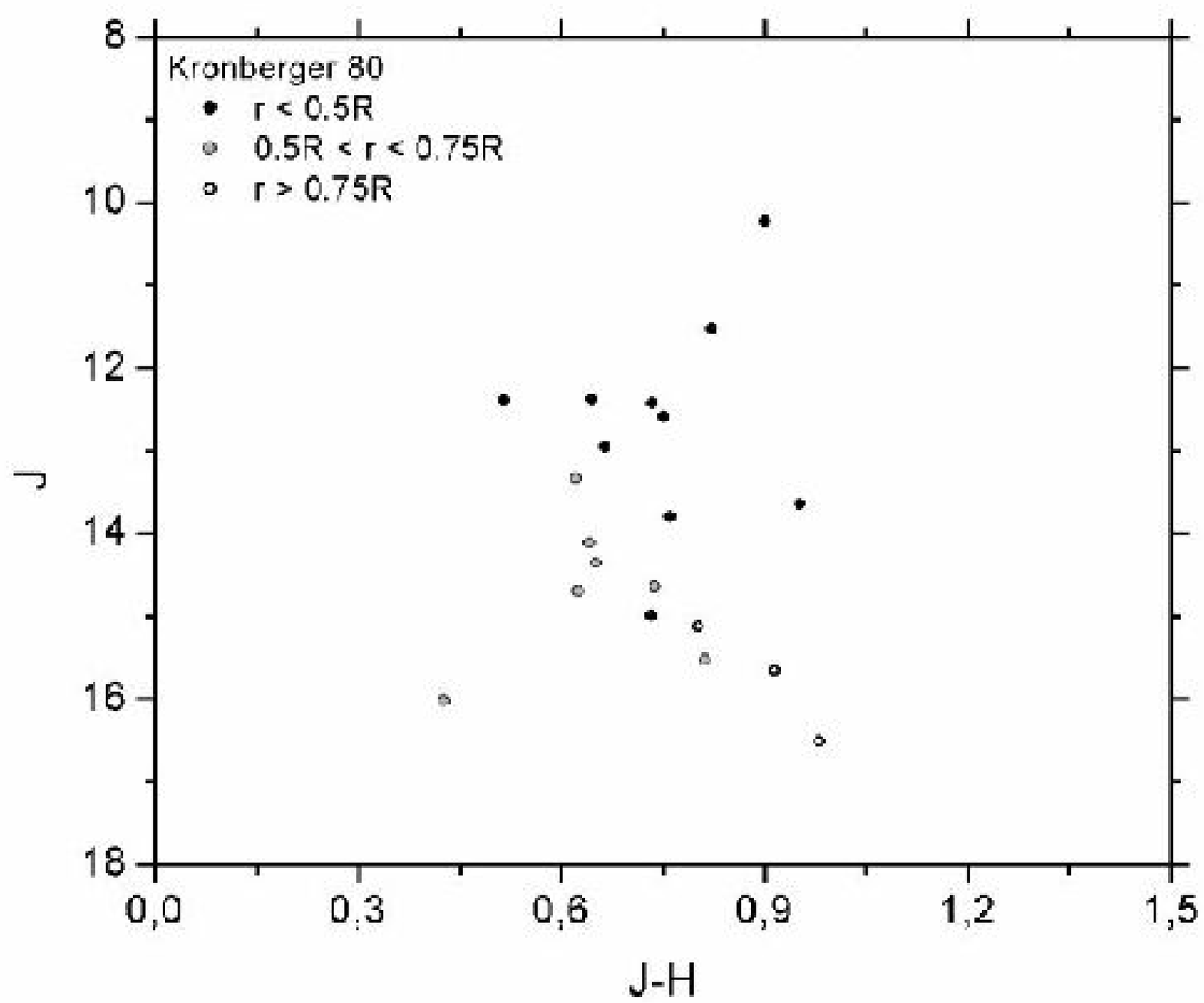}

\end{figure*}

\clearpage


\setcounter{figure}{9}
\begin{figure*}
\caption[]{Comparison of the CMDs of \object{Pismis~19} (left) and \object{Teutsch~79} (right). Both diagrams contain only those stars with $J$ and $H$ magnitudes derived either via aperture photometry ($rd_flg = 1$) or via point-spread function fitting ($rd_flg = 2$). Extraction radii of $1.5\arcmin$ (\object{Pismis~19}) and $1\arcmin$ (\object{Teutsch~79}) were taken.}
\label{fig10}

\includegraphics[width=8.75cm]{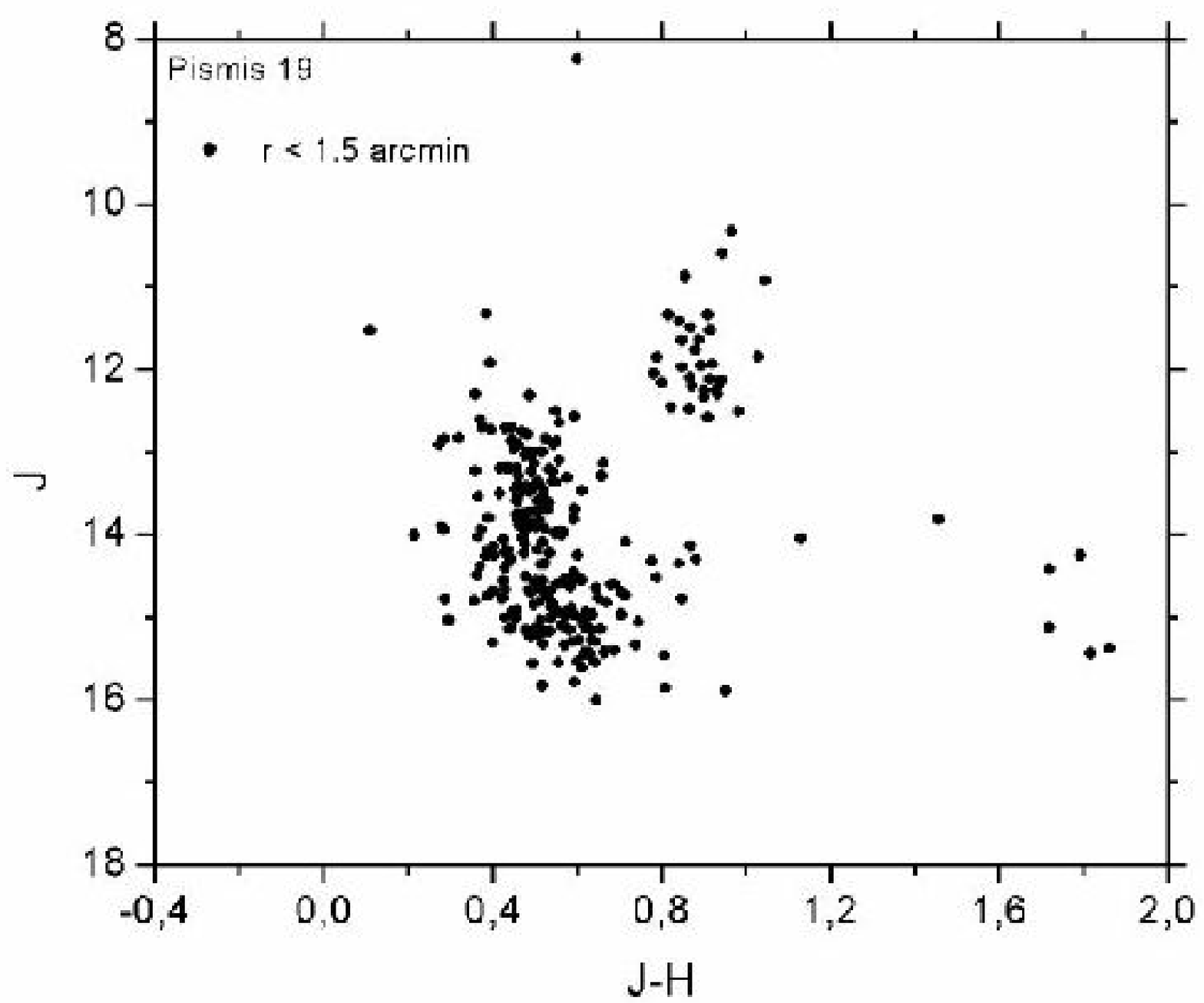}
\enskip
\includegraphics[width=8.75cm]{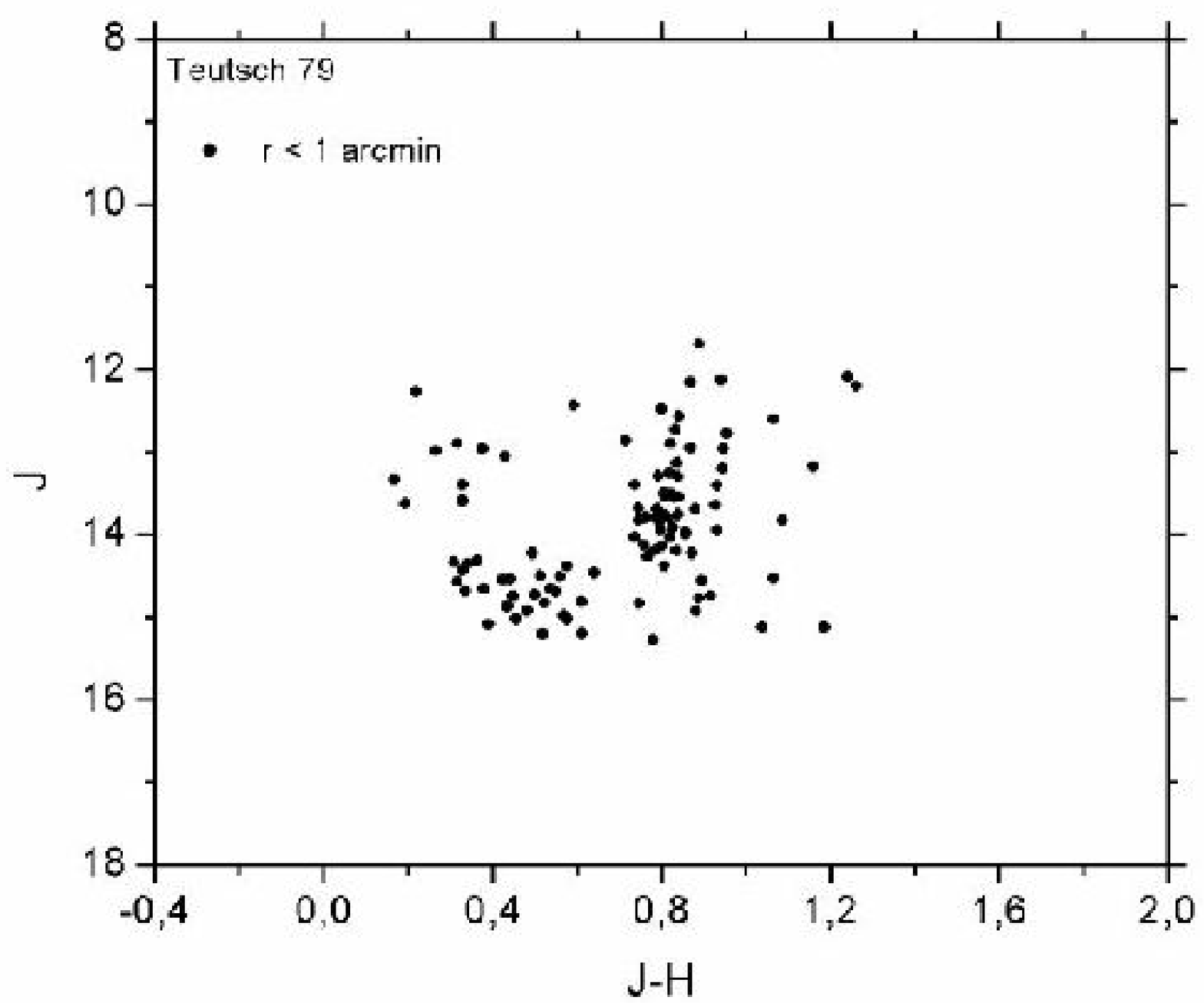}

\end{figure*}


\setcounter{figure}{10}
\begin{figure*}
\caption[]{Comparison of the CMDs of \object{NGC~2158} (left) and \object{Kronberger~81} (right). Both diagrams contain only those stars with $J$ and $H$ magnitudes derived either via aperture photometry ($rd_flg = 1$) or via point-spread function fitting ($rd_flg = 2$). In each case, an extraction radius of $1.75\arcmin$ was taken.}
\label{fig11}

\includegraphics[width=8.75cm]{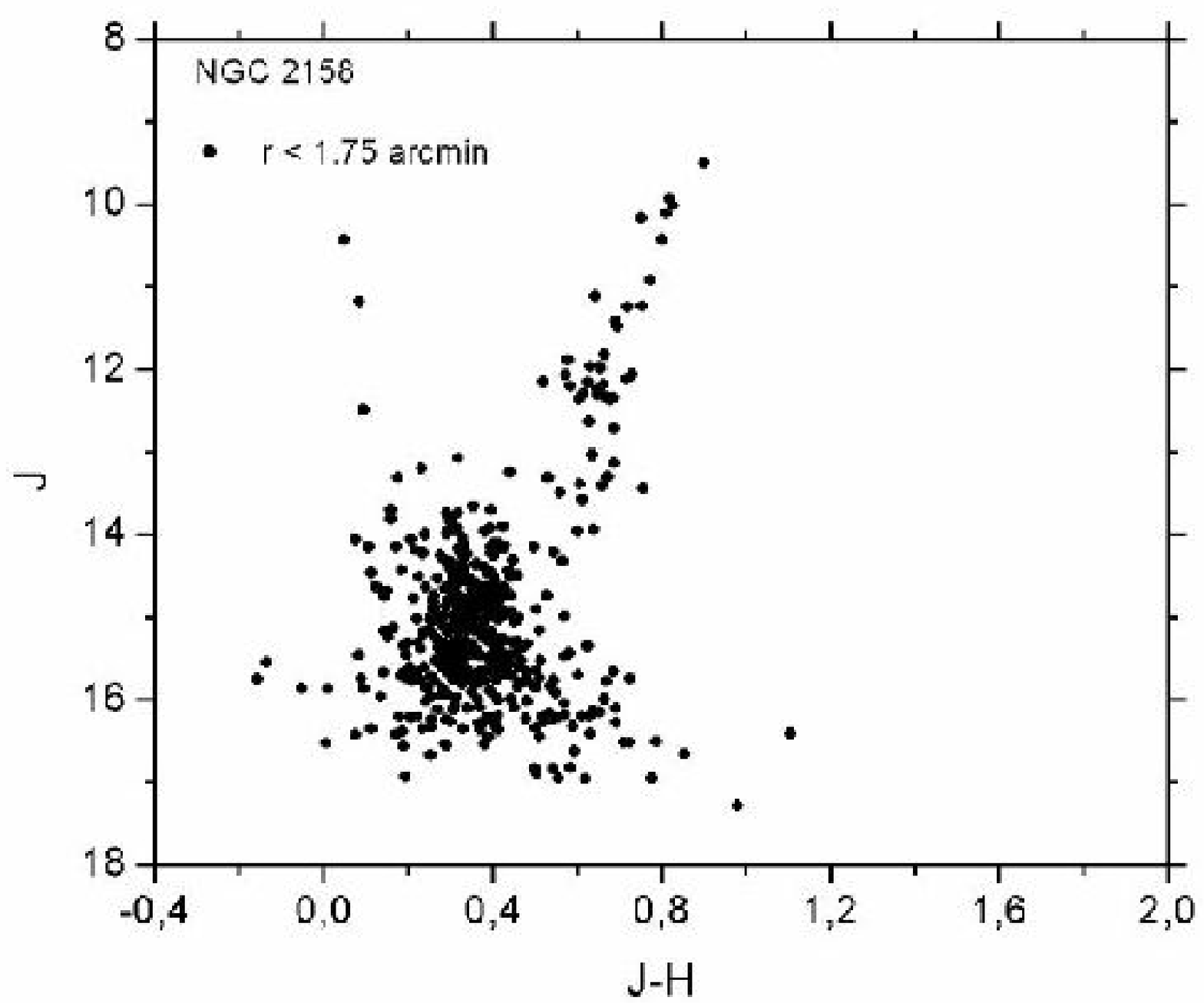}
\enskip
\includegraphics[width=8.75cm]{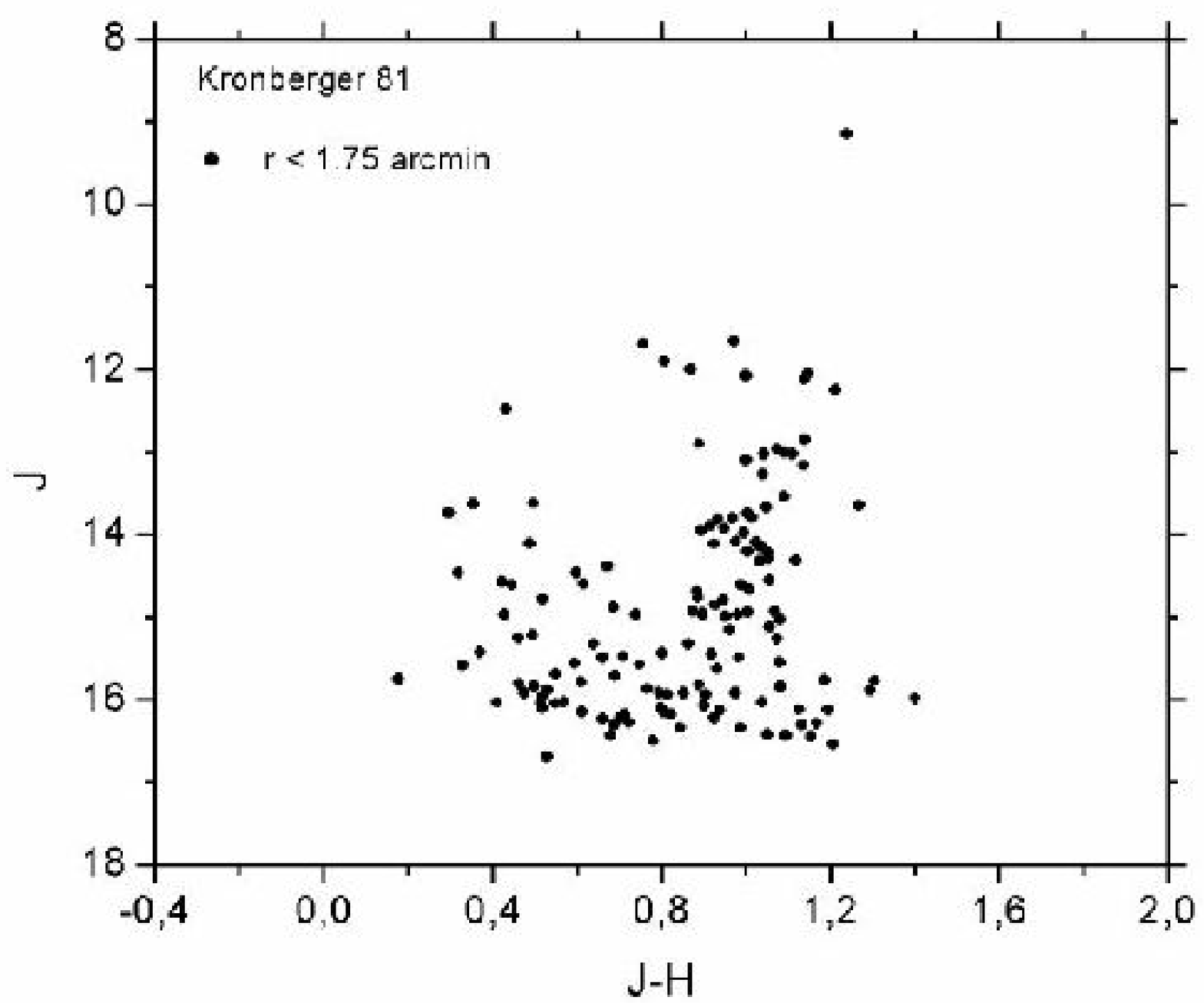}

\end{figure*}

\end{document}